\documentclass[journal]{IEEEtran}

\usepackage[dvips]{color}
\usepackage{graphicx}
\usepackage{amsmath}
 \usepackage{mdwlist}

\begin{document}

\title{A Survey on Wireless Security: Technical Challenges, Recent Advances and Future Trends}

\markboth{Proceedings of the IEEE (accepted to appear)}%
{Yulong Zou \MakeLowercase{\textit{et al.}}: A Survey on Wireless Security: Technical Challenges, Recent Advances and Future Trends}

\author{Yulong~Zou,~\IEEEmembership{Senior Member,~IEEE,}
        Jia~Zhu,
        Xianbin~Wang,~\IEEEmembership{Senior Member,~IEEE,}
        and Lajos~Hanzo,~\IEEEmembership{Fellow,~IEEE}

\thanks{Manuscript received May 29, 2014; revised October 6, 2014 and November 19, 2015; accepted April 21, 2016. This work was partially supported by the Priority Academic Program Development of Jiangsu Higher Education Institutions, the National Natural Science Foundation of China (Grant Nos. 61302104, 61401223 and 61522109), the Natural Science Foundation of Jiangsu Province (Grant Nos. BK20140887 and BK20150040), and the Key Project of Natural Science Research of Higher Education Institutions of Jiangsu Province (No. 15KJA510003).}

\thanks{Y. Zou and J. Zhu are with the School of Telecommunications and Information Engineering, Nanjing University of Posts and Telecommunications, Nanjing, China. email: \{yulong.zou, jiazhu\}@njupt.edu.cn.}
\thanks{X. Wang is with the Electrical and Computer Engineering Department, The University of Western Ontario, London, Ontario, Canada. email: xianbin.wang@uwo.ca.}
\thanks{L. Hanzo is with the School of Electronics and Computer Science, University of Southampton, Southampton, UK. email: \{lh@ecs.soton.ac.uk\}.}

}

\maketitle

\begin{abstract}
Due to the broadcast nature of radio propagation, the wireless air interface is open and accessible to both authorized and illegitimate users. This completely differs from a wired network, where communicating devices are physically connected through cables and a node without direct association is unable to access the network for illicit activities. The open communications environment makes wireless transmissions more vulnerable than wired communications to malicious attacks, including both the passive eavesdropping for data interception and the active jamming for disrupting legitimate transmissions. Therefore, this paper is motivated to examine the security vulnerabilities and threats imposed by the inherent open nature of wireless communications and to devise efficient defense mechanisms for improving the wireless network security. We first summarize the security requirements of wireless networks, including their authenticity, confidentiality, integrity and availability issues. Next, a comprehensive overview of security attacks encountered in wireless networks is presented in view of the network protocol architecture, where the potential security threats are discussed at each protocol layer. We also provide a survey of the existing security protocols and algorithms that are adopted in the existing wireless network standards, such as the Bluetooth, Wi-Fi, WiMAX, and the long-term evolution (LTE) systems. Then, we discuss the state-of-the-art in physical-layer security, which is an emerging technique of securing the open communications environment against eavesdropping attacks at the physical layer. Several physical-layer security techniques are reviewed and compared, including information-theoretic security, artificial noise aided security, security-oriented beamforming, diversity assisted security, and physical-layer key generation approaches. Since a jammer emitting radio signals can readily interfere with the legitimate wireless users, we also introduce the family of various jamming attacks and their counter-measures, including the constant jammer, intermittent jammer, reactive jammer, adaptive jammer and intelligent jammer. Additionally, we discuss the integration of physical-layer security into existing authentication and cryptography mechanisms for further securing wireless networks. Finally, some technical challenges which remain unresolved at the time of writing are summarized and the future trends in wireless security are discussed.

\end{abstract}


\IEEEpeerreviewmaketitle

\section*{NOMENCLATURE}

\begin{basedescript}{\desclabelstyle{\pushlabel}\desclabelwidth{5em}}
\item[3G] 3rd Generation
\item[AAA] Authentication, Authorization and Accounting
\item[AES] Advanced Encryption Standard
\item[AKA] Authentication and Key Agreement
\item[AP] Access Point
\item[ARQ] Automatic Repeat reQuest
\item[ASK] Authenticated Secret Key
\item[BS] Base Station
\item[CDMA] Code Division Multiple Access
\item[CK(s)] Ciphering Key(s)
\item[CSI] Channel State Information
\item[CSMA/CA] Carrier Sense Multiple Access with Collision Avoidance
\item[CST] Carrier Sensing Time
\item[CTS] Clear to Send
\item[DA] Destination Address
\item[DCF] Distributed Coordination Function
\item[DES] Data Encryption Standard
\item[DIFS] Distributed Inter-Frame Space
\item[DN] Destination Node
\item[DSSS] Direct-Sequence Spread Spectrum
\item[DoS] Denial of Service
\item[EPC] Evolved Packet Core
\item[E-UTRAN] Evolved-Universal Terrestrial Radio Access Network
\item[FHSS] Frequency-Hopping Spread Spectrum
\item[FTP] File Transfer Protocol
\item[GSVD] Generalized Singular Value Decomposition
\item[HSS] Home Subscriber Server
\item[HTTP] HyperText Transfer Protocol
\item[ICMP] Internet Control Message Protocol
\item[ICV] Integrity Check Value
\item[IK(s)] Integrity Key(s)
\item[IMSI] International Mobile Subscriber Identity
\item[IP] Internet Protocol
\item[IV] Initialization Vector
\item[LTE] Long Term Evolution
\item[MAC]  Medium Access Control
\item[MIC] Message Integrity Check
\item[MIMO] Multiple-Input Multiple-Output
\item[MISOME] Multiple-Input Single-Output Multiple-Eavesdropper
\item[MITM] Man In The Middle
\item[MME] Mobility Management Entity
\item[NIC] Network Interface Controller
\item[NP] Non-deterministic Polynomial
\item[OFDMA] Orthogonal Frequency-Division Multiple Access
\item[OSI] Open Systems Interconnection
\item[PER] Packet Error Rate
\item[PKM] Privacy and Key Management
\item[PN] Pseudo Noise
\item[PRNG] Pseudo-Random Number Generator
\item[QoS] Quality of Service
\item[RFCOMM] Radio Frequency COMMunications
\item[RSA] Rivest-Shamir-Adleman
\item[RSS] Received Signal Strength
\item[RTS] Request to Send
\item[SA] Source Address
\item[SIFS] Short Inter-Frame Space
\item[SINR] Signal-to-Interference-and-Noise Ratio
\item[SQL] Structured Query Language
\item[SMTP] Simple Mail Transfer Protocol
\item[SN] Source Node
\item[SNR] Signal-to-Noise Ratio
\item[SS] Subscriber Station
\item[SSL] Secure Sockets Layer
\item[TA] Transmitter Address
\item[TCP] Transmission Control Protocol
\item[TDMA] Time-Division Multiple Access
\item[TK] Temporal Key
\item[TKIP] Temporal Key Integrity Protocol
\item[TLS] Transport Layer Security
\item[TSC] TKIP Sequence Counter
\item[TTAK] TKIP-mixed Transmit Address and Key
\item[TTLS] Tunneled Transport Layer Security
\item[UDP] User Datagram Protocol
\item[UE(s)] User Equipment(s)
\item[UMTS] Universal Mobile Telecommunications System
\item[WEP] Wired Equivalent Privacy
\item[WiMAX] Worldwide Interoperability for Microwave Access
\item[WLAN] Wireless Local Area Network
\item[WMAN] Wireless Metropolitan Area Network
\item[WPA] Wi-Fi Protected Access
\item[WPA2] Wi-Fi Protected Access II
\item[WPAN] Wireless Personal Area Network

\end{basedescript}

\section{Introduction}

\IEEEPARstart {D}{uring} the past decades, wireless communications infrastructure and services have been proliferating with the goal of meeting rapidly increasing demands [1], [2]. According to the latest statistics released by the International Telecommunications Union in 2013 [3], the number of mobile subscribers has reached 6.8 billion worldwide and almost 40\% of the world's population is now using the Internet. Meanwhile, it has been reported in [4] that an increasing number of wireless devices are abused for illicit cyber-criminal activities, including malicious attacks, computer hacking, data forging, financial information theft, online bullying/stalking and so on. This causes the direct loss of about 83 billion Euros with an estimated 556 million users worldwide impacted by cyber-crime each year, according to the 2012 Norton cyber-crime report [4]. Hence, it is of paramount importance to improve wireless communications security to fight against cyber-criminal activities, especially because more and more people are using wireless networks (e.g., cellular networks and Wi-Fi) for online banking and personal emails, owing to the widespread use of smartphones.
\begin{figure}
  \centering
  {\includegraphics[scale=0.65]{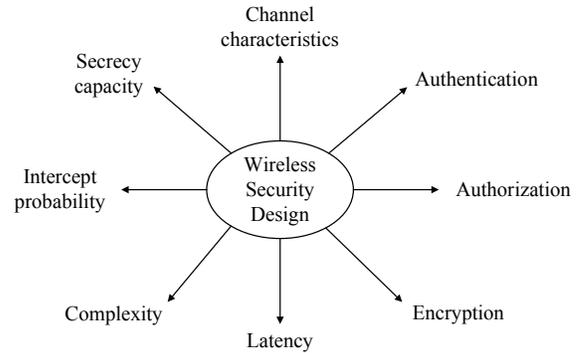}\\
  \caption{Wireless security methodologies and design factors.}\label{Fig1}}
\end{figure}

Wireless networks generally adopt the open systems interconnection (OSI) protocol architecture [5] comprising the application layer, transport layer, network layer [6], medium access control (MAC) layer [7] and physical layer [8], [9]. Security threats and vulnerabilities associated with these protocol layers are typically protected separately at each layer to meet the security requirements, including the authenticity, confidentiality, integrity and availability [10]. For example, cryptography is widely used for protecting the confidentiality of data transmission by preventing information disclosure to unauthorized users [11], [12]. Although cryptography improves the achievable communications confidentiality, it requires additional computational power and imposes latency [13], since a certain amount of time is required for both data encryption and decryption [14]. In order to guarantee the authenticity of a caller or receiver, existing wireless networks typically employ multiple authentication approaches simultaneously at different protocol layers, including MAC-layer authentication [15], network-layer authentication [16], [17] and transport-layer authentication [18]. To be specific, in the MAC layer, the MAC address of a user should be authenticated to prevent unauthorized access. In the network layer, the Wi-Fi protected access (WPA) and the Wi-Fi protected access II (WPA2) are two commonly used network-layer authentication protocols [19], [20]. Additionally, the transport-layer authentication includes the secure socket layer (SSL) and its successor, namely the transport layer security (TLS) protocols [21]-[23]. It becomes obvious that exploiting multiple authentication mechanisms at different protocol layers is capable of enhancing the wireless security, again, at the cost of high computational complexity and latency. As shown in Fig. 1, the main wireless security methodologies include the authentication, authorization and encryption, for which the diverse design factors e.g. the security level, implementation complexity and communication latency need to be balanced.

In wired networks, the communicating nodes are physically connected through cables. By contrast, wireless networks are extremely vulnerable owing to the broadcast nature of the wireless medium. Explicitly, wireless networks are prone to malicious attacks, including eavesdropping attack [24], denial-of-service (DoS) attack [25], spoofing attack [26], man-in-the-middle (MITM) attack [27], message falsification/injection attack [28], etc. For example, an unauthorized node in a wireless network is capable of inflicting intentional interferences with the objective of disrupting data communications between legitimate users. Furthermore, wireless communications sessions may be readily overheard by an eavesdropper, as long as the eavesdropper is within the transmit coverage area of the transmitting node. In order to maintain confidential transmission, existing systems typically employ cryptographic techniques for preventing eavesdroppers from intercepting data transmissions between legitimate users [29], [30]. Cryptographic techniques assume that the eavesdropper has limited computing power and rely upon the computational hardness of their underlying mathematical problems. The security of a cryptographic approach would be compromised, if an efficient method of solving its underlying hard mathematical problem was to be discovered [31], [32].

Recently, physical-layer security is emerging as a promising means of protecting wireless communications to achieve information-theoretic security against eavesdropping attacks. In [33], Wyner examined a discrete memoryless wiretap channel consisting of a source, a destination as well as an eavesdropper and proved that perfectly secure transmission can be achieved, provided that the channel capacity of the main link from the source to the destination is higher than that of the wiretap link from the source to the eavesdropper. In [34], Wyner's results were extended from the discrete memoryless wiretap channel to the Gaussian wiretap channel, where the notion of a so-called secrecy capacity was developed, which was shown to be equal to the difference between the channel capacity of the main link and that of the wiretap link. If the secrecy capacity falls below zero, the transmissions from the source to the destination become insecure and the eavesdropper would become capable of intercepting the source's transmissions [35], [36]. In order to improve the attainable transmission security, it is of importance to increase the secrecy capacity by exploiting sophisticated signal processing techniques, such as the artificial noise aided security [37]-[39], security-oriented beamforming [40], [41], security-oriented diversity approaches [42], [43] and so on.

In this paper, we are motivated to discuss diverse wireless attacks as well as the corresponding defense mechanisms and to explore a range of challenging open issues in wireless security research. The main contributions of this paper are summarized as follows. Firstly, a systematic review of security threats and vulnerabilities is presented at the different protocol layers, commencing from the physical layer up to the application layer. Secondly, we summarize the family of security protocols and algorithms used in the existing wireless networks, such as the Bluetooth, Wi-Fi, WiMAX and long-term evolution (LTE) standards. Thirdly, we discuss the emerging physical-layer security in wireless communications and highlight the class of information-theoretic security, artificial noise aided security, security-oriented beamforming, security-oriented diversity and physical-layer secret key generation techniques. Additionally, we provide a review on various wireless jammers (i.e., the constant jammer, intermittent jammer, reactive jammer, adaptive jammer and intelligent jammer) as well as their detection and prevention techniques. Finally, we outline some of open challenges in wireless security.

The remainder of this paper is organized as follows. Section II presents the security requirements of wireless networks, where the authenticity, confidentiality, integrity and availability of wireless services are discussed. In Section III, we analyze the security vulnerabilities and weaknesses of wireless networks at different protocol layers, including the application layer, transport layer, network layer, MAC layer and physical layer. Next, in Section IV, the security protocols and algorithms used in existing wireless networks, such as the Bluetooth, Wi-Fi, WiMAX and LTE standards, are discussed. Then, Section V presents the physical-layer security which is emerging as an effective paradigm conceived for improving the security of wireless communications against eavesdropping attacks by exploiting the physical-layer characteristics of wireless channels. In Section VI, we characterize the family of wireless jamming attacks and their counter-measures, {{while in Section VII, we discuss how physical-layer security may be invoked for efficiently complementing the existing suite of classic authentication and cryptography mechanisms.}} These discussions are followed by Section VIII, where some of the open challenges and future trends in wireless security are presented. Finally, Section IX provides our concluding remarks.

\section{Security Requirements in Wireless Networks}
Again, in wireless networks, the information is exchanged among authorized users, but this process is vulnerable to various
malicious threats owing to the broadcast nature of the wireless medium. The security requirements of wireless networks are specified for the sake of protecting the wireless transmissions against wireless attacks, such as eavesdropping attack, DoS attack, data falsification attack, node compromise attack and so on [44], [45]. For example, maintaining data confidentiality is a typical security requirement, which refers to the capability of restricting data access to authorized users only, while preventing eavesdroppers from intercepting the information.
Generally speaking, secure wireless communications should satisfy the requirements of authenticity, confidentiality, integrity and availability [46], as detailed below:

\begin{itemize}
\item
\textbf{Authenticity}: Authenticity refers to confirming the true identity of a network node to distinguish authorized users from unauthorized users. In wireless networks, a pair of communicating nodes should first perform mutual authentication before establishing a communications link for data transmission [47]. Typically, a network node is equipped with a wireless network interface card and has a unique medium access control (MAC) address, which can be used for authentication purposes. Again, in addition to MAC authentication, there are other wireless authentication methods, including network-layer authentication, transport-layer authentication and application-layer authentication.
\end{itemize}

\begin{itemize}
\item
\textbf{Confidentiality}: The confidentiality refers to limiting the data access to intended users only, while preventing the disclosure of the information to unauthorized entities [48]. Considering the symmetric key encryption technique as an example, the source node first encrypts the original data (often termed as plain-text) using an encryption algorithm with the aid of a secret key that is shared with the intended destination only. Next, the encrypted plain-text (referred to as cipher-text) is transmitted to the destination that then decrypts its received cipher-text using the secret key. Since the eavesdropper has no knowledge of the secret key, it is unable to interpret the plain-text based on the overheard cipher-text. Traditionally, the classic Diffie-Hellman key agreement protocol is used to achieve the key exchange between the source and destination and requires a trusted key management center [32]. Recently, physical-layer security emerges as a means of protecting the confidentiality of wireless transmission against eavesdropping attacks for achieving information-theoretic security [33], [49]. The details of physical-layer security will be discussed in Section V.
\end{itemize}

\begin{itemize}
\item
\textbf{Integrity}: The integrity of information transmitted in a wireless network should be accurate and reliable during its entire life-cycle representing the source-information without any falsification and modification by unauthorized users. The data integrity may be violated by so-called {{insider attacks}}, such as for example node compromise attacks [50]-[52]. More specifically, a legitimate node that is altered and compromised by an adversary, is termed as a compromised node. The compromised node may inflict damage upon the data integrity by launching malicious attacks, including message injection, false reporting, data modification and so on. In general, it is quiet challenging to detect the attacks by compromised nodes, since these compromised nodes running malicious codes still have valid identities. A promising solution to detect compromised nodes is to utilize the automatic code update and recovery process, which guarantees that the nodes are periodically patched and a compromised node may be detected, if the patch fails. The compromised nodes can be repaired and revoked through the so-called code recovery process.
\end{itemize}

\begin{itemize}
\item
\textbf{Availability}: The availability implies that the
authorized users are indeed capable of accessing a wireless network anytime and anywhere upon request. The violation of availability, referred to as denial of service, will result in the authorized users to become unable to access the wireless network, which in turn results in unsatisfactory user experience [53], [54]. {{For example, any unauthorized node is capable of launching DoS activities at the physical layer by maliciously generating interferences for disrupting the desired communications between legitimate users, which is also known as a jamming attack. In order to combat jamming attacks, existing wireless systems typically consider the employment of spread spectrum techniques, including direct-sequence spread spectrum (DSSS) [55], [56] and frequency-hopping spread spectrum (FHSS) solutions [57].}} To be specific, DSSS employs a pseudo-noise (PN) sequence to spread the spectrum of the original signal to a wide frequency bandwidth. In this way, the jamming attack operating without the knowledge of the PN sequence has to dissipate a much higher power for disrupting the legitimate transmission, which may not be feasible in practice due to its realistic power constraint. As an alternative, FHSS continuously changes the central frequency of the transmitted waveform using a certain frequency-hopping pattern, so that the jamming attacker cannot monitor and interrupt the legitimate transmissions.
\end{itemize}

\begin{table}
  \centering
  \caption{Summarization of Wireless Security Requirements.}
  {\includegraphics[scale=0.6]{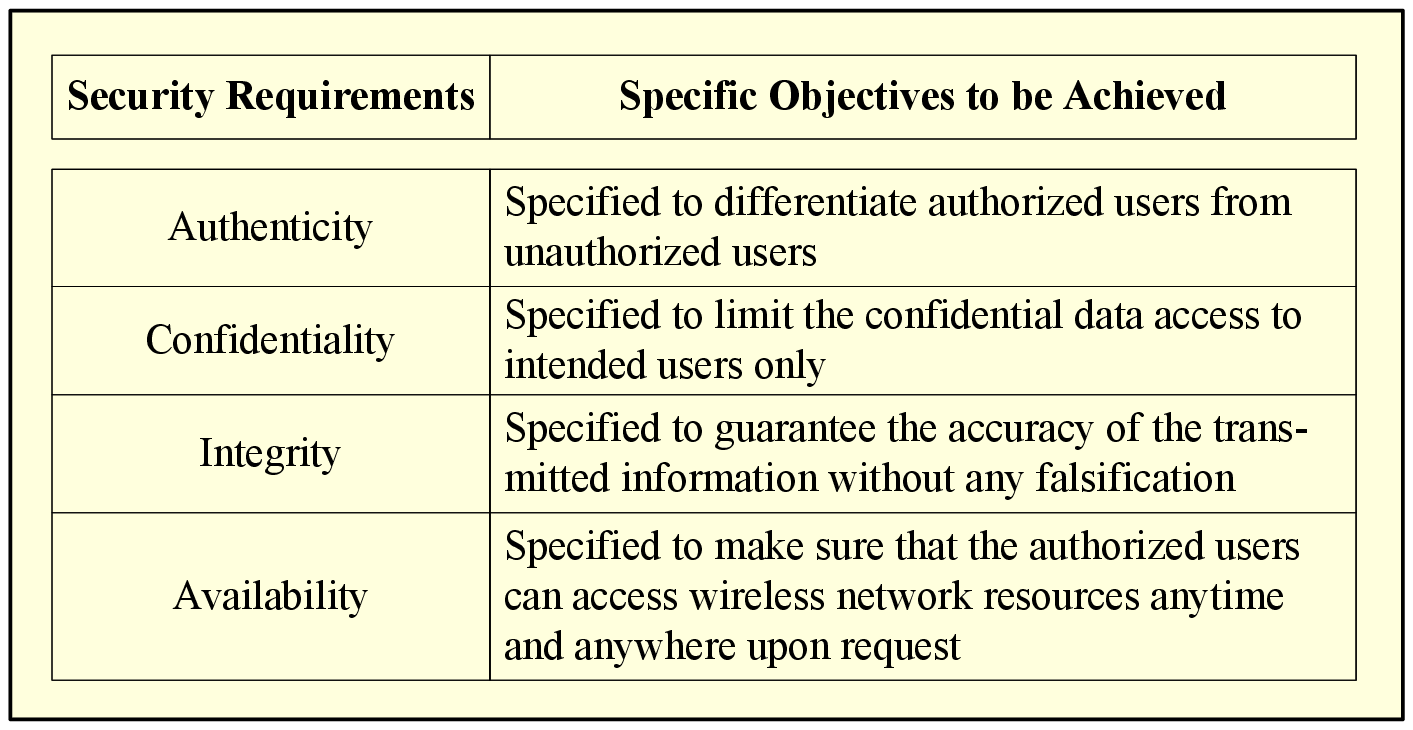}\label{Tab1}}
\end{table}

The above-mentioned authenticity, confidentiality, integrity and availability are summarized in Table I, which are commonly considered and implemented in the existing wireless networks, including the Bluetooth [58], Wi-Fi [59], Worldwide Interoperability for Microwave Access (WiMAX) [60], Long-Term Evolution (LTE) [61] standards and so on. {{In principle, wireless networks should be as secure as wired networks. This implies that the security requirements of wireless networks should be the same as those of wired networks, including the requirements of authenticity, confidentiality, integrity and availability. However, due to the broadcast nature of radio propagation, achieving these security requirements in wireless networks is more challenging than in wired networks. For example, the availability of wireless networks is extremely vulnerable, since a jamming attack imposing a radio signal can readily disrupt and block the wireless physical-layer communications. Hence, compared to wired networks, wireless systems typically employ an additional DSSS (or FHSS) technique in order to protect the wireless transmissions against jamming attacks.}}

\section{Security Vulnerabilities in Wireless Networks}

\begin{figure*}
  \centering
  {\includegraphics[scale=0.65]{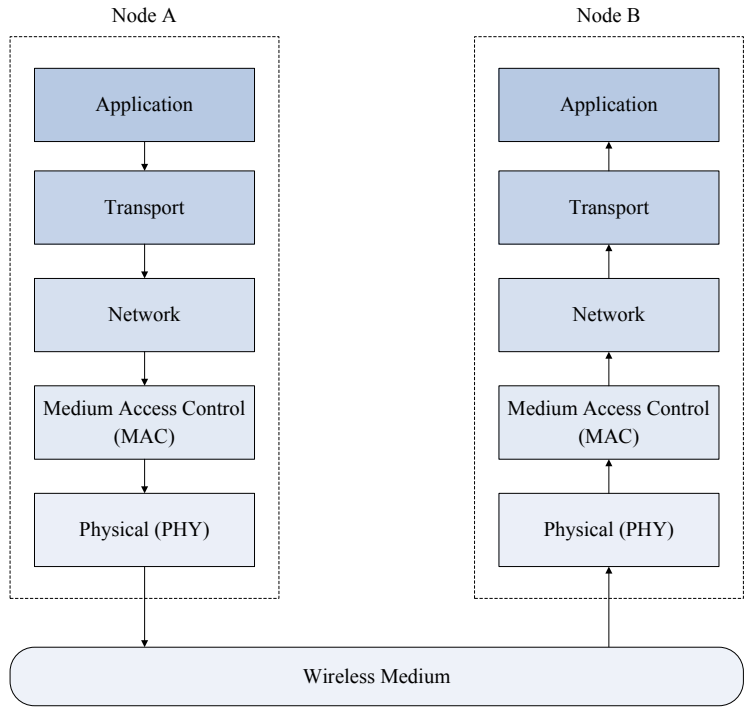}\\
  \caption{A generic wireless OSI layered protocol architecture consisting of the application layer, transport layer, network layer, MAC layer and physical layer.}\label{Fig2}}
\end{figure*}

\begin{figure}
  \centering
  {\includegraphics[scale=0.65]{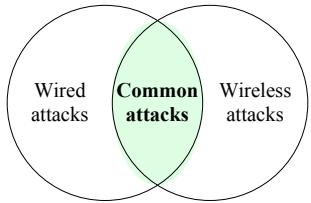}\\
  \caption{Relationship between the wired and wireless attacks.}\label{Fig3}}
\end{figure}

In this section, we present a systematic review of various security vulnerabilities and weaknesses encountered in wireless networks. Apart from their differences, wired and wireless networks also share some similarities. For example, they both adopt the OSI layered protocol architecture consisting of the physical layer, MAC layer, network layer, transport layer and application layer. As shown in Fig. 2, a network node (denoted by node A) employs these protocols for transmitting its data packets to another network node (i.e., node B). To be specific, the data packet at node A is first extended with the protocol overheads, including the application-layer overhead, transport-layer overhead, network-layer overhead, MAC overhead and physical-layer overhead. This results in an encapsulated packet. Then, the resultant data packet is transmitted via the wireless medium to node B, which will perform packet-decapsulation, commencing from the physical layer and proceeding upward to the application layer, in order to recover the original data packet. Note that the difference between the wired and wireless networks mainly lies in the PHY and MAC layers, while the application, transport and network layers of wireless networks are typically identical to those of wired networks. As a consequence, the wired and wireless networks share some common security vulnerabilities owing to their identical application, transport and network layers. Nevertheless, they also suffer from mutually exclusive attacks due to the fact that the wired and wireless networks have different PHY and MAC layers, as shown in Fig. 3.

\begin{table}
  \centering
  \caption{Main Protocols and Specifications of the Wireless OSI Layers.}
  {\includegraphics[scale=0.6]{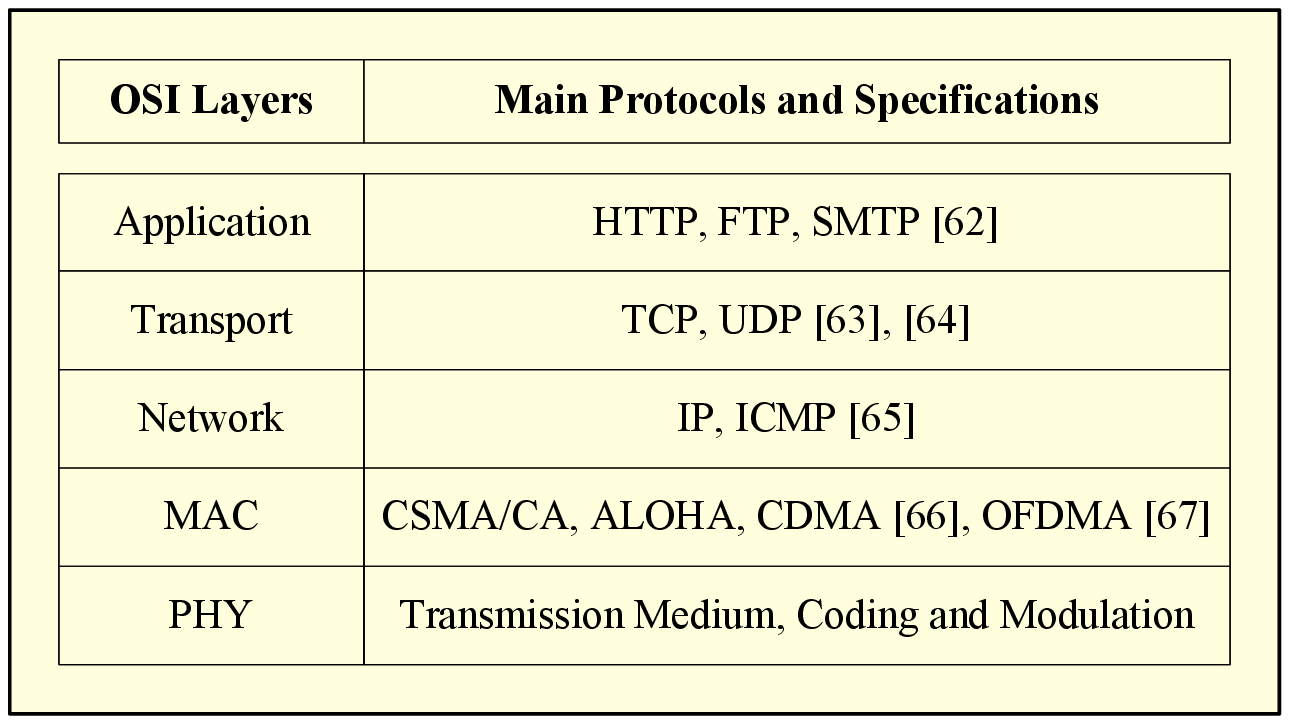}\label{Tab2}}
\end{table}

Table II shows the main protocols and specifications implemented at each of wireless OSI layers. For example, the application-layer supports the hypertext transfer protocol (HTTP) for the sake of delivering web services, while the file transfer protocol (FTP) is used for large-file-transfer, and the simple mail transfer protocol (SMTP) is invoked for electronic mail (e-mail) transmission and so on [62]. The commonly used transport-layer protocols include the transport control protocol (TCP) and the user datagram protocol (UDP) [63], [64]. The TCP ensures the reliable and ordered delivery of data packets, whereas UDP has no guarantee of such reliable and ordered delivery. In contrast to TCP, UDP has no handshaking dialogues and adopts a simpler transmission model, hence imposing a reduced protocol overhead. In the network layer, we also have different protocols, such as the Internet protocol (IP), which was conceived for delivering data packets based on IP addresses, and the Internet control message protocol (ICMP) designed for sending error messages for indicating, for example, that a requested service is unavailable or that a network node could not be reached [65]. Regarding the MAC layer, there are numerous different protocols adopted by various wireless networks, such as the carrier sense multiple access with collision avoidance (CSMA/CA) used in Wi-Fi networks, the slotted ALOHA employed in tactical satellite networks by military forces, code division multiple access (CDMA) involved in third-generation (3G) mobile networks [66] and orthogonal frequency-division multiple access (OFDMA) adopted in the long term evolution (LTE) and LTE-advanced networks [67]. Additionally, the physical layer specifies the physical characteristics of information transmission, including the transmission medium, modulation, line coding, multiplexing, circuit switching, pulse shaping, forward error correction, bit-interleaving and other channel coding operations, etc.

Every OSI layer has its own unique security challenges and issues, since different layers rely on different protocols, hence
exhibiting different security vulnerabilities [68]-[70]. Below we summarize the range of wireless attacks potentially encountered by various protocol layers.


\subsection{Physical-Layer Attacks}
\begin{table}
  \centering
  \caption{Main Types of Wireless Attacks at the PHY Layer.}
  {\includegraphics[scale=0.6]{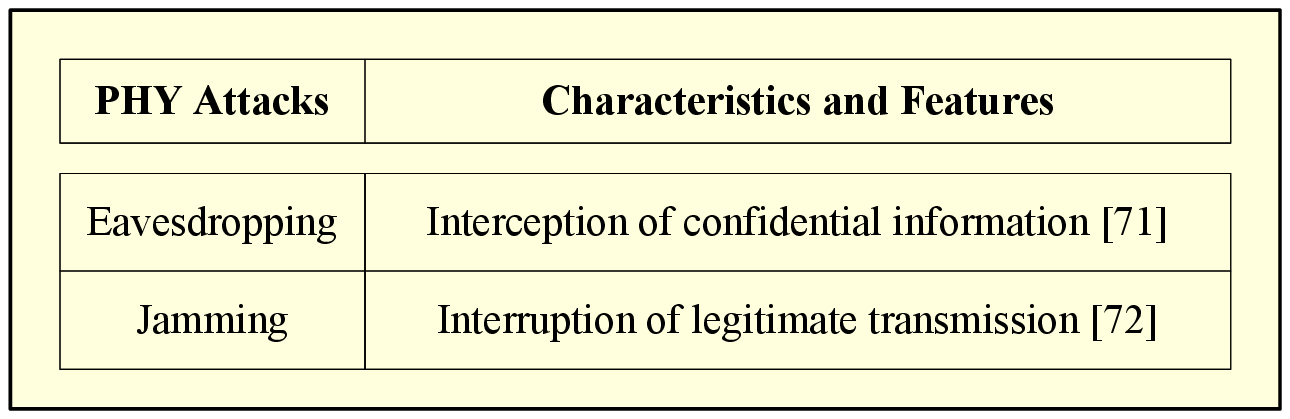}\label{Tab3}}
\end{table}

The physical layer is the lowest layer in the OSI protocol architecture, which is used for specifying the physical characteristics of signal transmission. Again, the broadcast nature of wireless communications makes its physical layer extremely vulnerable to eavesdropping and jamming attacks, which are two main types of wireless physical-layer attacks, as depicted in Table III. More specifically, the eavesdropping attack refers to an unauthorized user attempting to intercept the data transmission between legitimate users [71]. In wireless networks, as long as an eavesdropper lies in the transmit coverage area of the source node, the wireless communications session can be overheard by the eavesdropper. In order to maintain confidential transmission, typically cryptographic techniques relying on secret keys are adopted for preventing eavesdropping attacks from intercepting the data transmission. To be specific, the source node (SN) and destination node (DN) share a secret key and the so-called plain-text is first encrypted at SN, leading to the cipher-text, which is then transmitted to DN. In this case, even if an eavesdropper overhears the cipher-text transmission, it remains difficult to extract the plain-text from the cipher-text without the secret key.

Moreover, a malicious node in wireless networks can readily generate intentional interference for disrupting the data communications between legitimate users, which is referred to as a jamming attack (also known as DoS attack) [72]. The jammer aims for preventing authorized users from accessing wireless network resources and this impairs the network availability for the legitimate users. To this end, spread spectrum techniques are widely recognized as an effective means of defending against DoS attacks by spreading the transmit signal over a wider spectral bandwidth than its original frequency band. Again, the above-mentioned DSSS and FHSS techniques exhibit a high jamming-resistance at the physical layer.

\subsection{MAC-Layer Attacks}
\begin{table}
  \centering
  \caption{Main Types of Wireless Attacks at the MAC Layer.}
  {\includegraphics[scale=0.6]{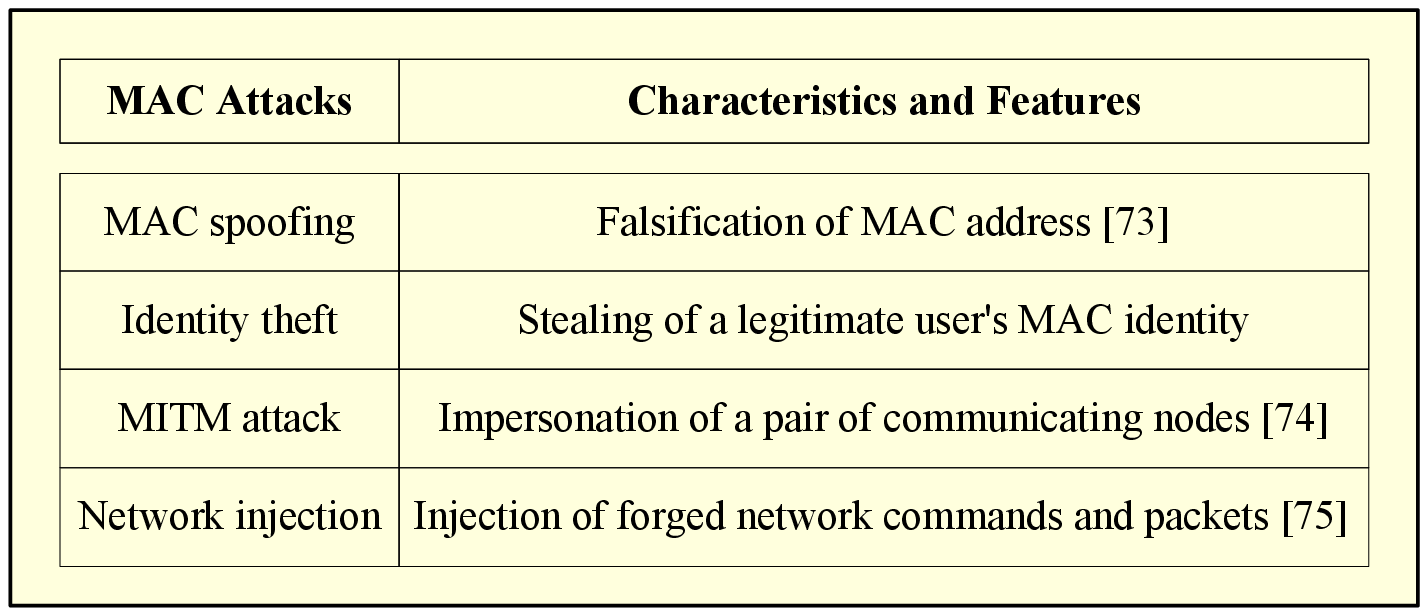}\label{Tab4}}
\end{table}
The MAC layer enables multiple network nodes to access a shared
medium with the aid of intelligent channel access control mechanisms
such as CSMA/CA, CDMA, OFDMA and so on. Typically, each network
node is equipped with a network interface controller (NIC) and has a
unique MAC address, which is used for user authentication. An
attacker that attempts to change its assigned MAC address with a malicious intention is termed as MAC spoofing, which is the primary technique of MAC attacks [73]. Although the MAC address is hard-coded into the NIC of a network node, it is still possible for a network node to spoof a MAC address and thus MAC spoofing enables the malicious node to hide its true identity or to impersonate another network node for the sake of carrying out illicit activities. Furthermore, a MAC attacker may overhear the network traffic and steal a legitimate node's MAC address by analyzing the overheard traffic, which is referred to as an identity-theft attack. An attacker attempting identity theft will pretend to be another legitimate network node and gain access to confidential information of the victim node.

In addition to the above-mentioned MAC spoofing and identity theft, the class of MAC-layer attacks also includes MITM attacks [74] and network injection [75]. Typically, a MITM attack refers to an attacker that first `sniffs' the network's traffic in order to intercept the MAC addresses of a pair of legitimate communicating nodes, then impersonates the two victims and finally establishes a connection with them. In this way, the MITM attacker acts as a relay between the pair of victims and makes them feel that they are communicating directly with each other over a private connection. In reality, their session was intercepted and controlled by the attacker. By contrast, the network injection attack aims for preventing the operation of networking devices, such as routers, switches, etc. by injecting forged network re-configuration commands. In this manner, if an overwhelming number of the forged networking commands are initiated, the entire network may become paralyzed, thus requiring rebooting or even reprogramming of all networking devices. The main types of wireless MAC attacks are summarized in Table IV.

\subsection{Network-Layer Attacks}
\begin{table}
  \centering
  \caption{Main Types of Wireless Attacks at the Network Layer.}
  {\includegraphics[scale=0.6]{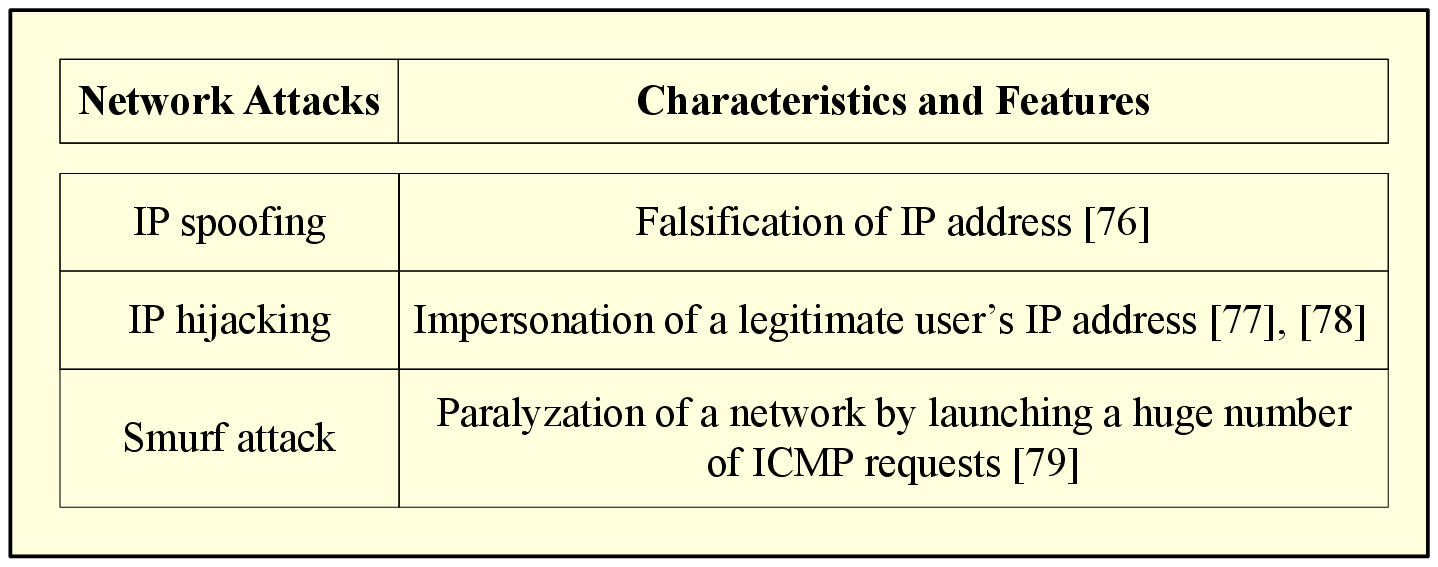}\label{Tab5}}
\end{table}
In the network layer, IP was designed as the principal protocol for delivering packets from a SN to a DN through intermediate routers based on their IP addresses. The network-layer attacks mainly aim for exploiting IP weaknesses, which include the IP spoofing and hijacking as well as the so-called Smurf attack [76]-[78], as illustrated in Table V. To be specific, IP spoofing is used for generating a forged IP address with the goal of hiding the true identity of the attacker or impersonating another network node for carrying out illicit activities. The network node that receives these packets associated with a forged source IP address will send its responses back to the forged IP address. This will waste significant network capacity and might even paralyze the network by flooding it with forged IP packets. IP hijacking is another illegitimate activity launched by hijackers for the sake of taking over another legitimate user's IP address. If the attacker succeeds in hijacking the IP address, it will be able to disconnect the legitimate user and create a new connection to the network by impersonating the legitimate user, hence gaining access to confidential information. There are some other forms of IP hijacking techniques, including prefix hijacking, route hijacking and border gateway protocol hijacking [78].

The Smurf attack is a DoS attack in the network layer, which intends to send a huge number of ICMP packets (with a spoofed source IP address) to a victim node or to a group of victims using an IP broadcast address [79]. Upon receiving the ICMP requests, the victims are required to send back ICMP responses, resulting in a significant amount of traffic in the victim network. When the Smurf attack launches a sufficiently high number of ICMP requests, the victim network will become overwhelmed and paralyzed by these ICMP requests and responses. To defend against Smurf attacks, a possible solution is to configure the individual users and routers by ensuring that they do not to constantly respond to ICMP requests. We may also consider the employment of firewalls, which can reject the malicious packets arriving from the forged source IP addresses.

\subsection{Transport-Layer Attacks}
\begin{table}
  \centering
  \caption{Main Types of Wireless Attacks at the Transport Layer.}
  {\includegraphics[scale=0.6]{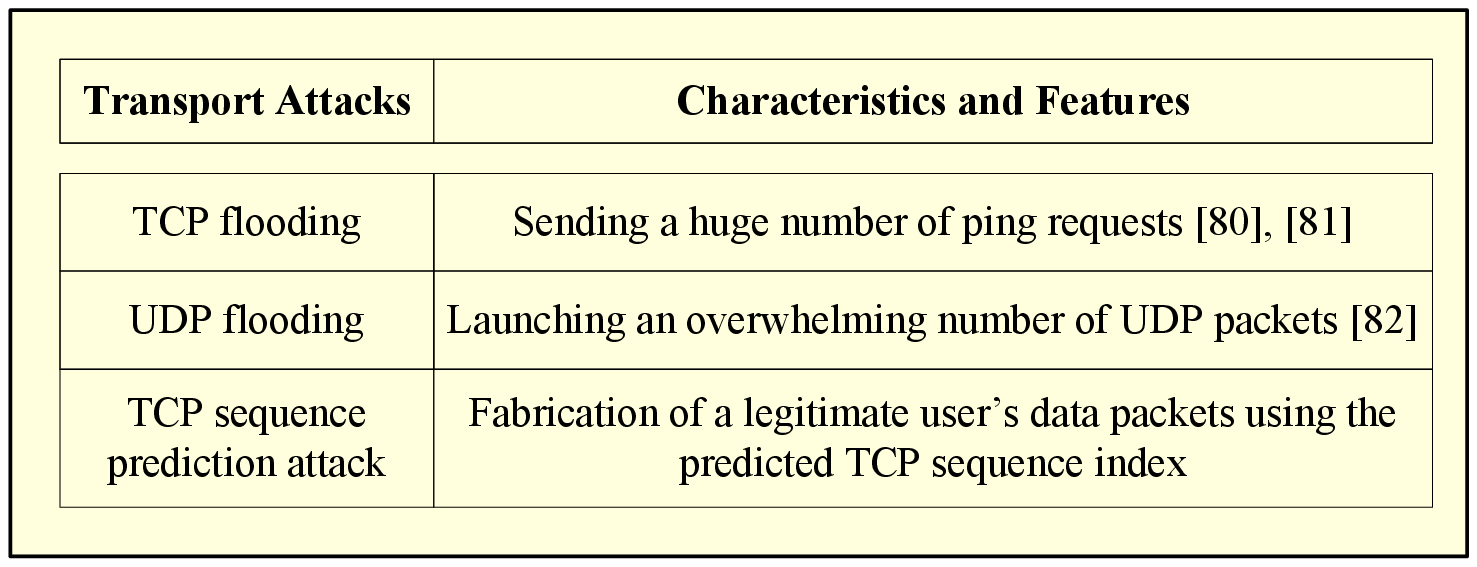}\label{Tab6}}
\end{table}
This subsection briefly summarizes the malicious activities in the transport layer, with an emphasis on the TCP and UDP attacks. To be specific, TCP is a connection-oriented transport protocol designed for supporting the reliable transmission of data packets, which is typically used for delivering e-mails and for transferring files from one network node to another. In contrast to TCP, UDP is a connectionless transport protocol associated with a reduced protocol overhead and latency, but as a price, it fails to guarantee reliable data delivery. It is often used by delay-sensitive applications which do not impose strict reliability requirements, such as IP television, voice over IP and online games. Both TCP and UDP suffer from security vulnerabilities including the TCP and UDP flooding as well as the TCP sequence number prediction attacks, as summarized in Table VI.

TCP attacks include TCP flooding attacks and sequence number
prediction attacks [80], [81]. The TCP flooding, which is also known as ping flooding, is a DoS attack in the transport layer, where the attacker sends an overwhelming number of ping requests, such as ICMP echo requests to a victim node, which then responds by sending ping replies, such as ICMP echo replies. This will flood both the input and output buffers of the victim node and it might even delay its connection to the target network, when the number of ping requests is sufficiently high. The TCP sequence prediction technique is another TCP attack that attempts to predict the sequence index of TCP packets of a transmitting node and then fabricates the TCP packets of the node. To be specific, the TCP sequence prediction attacker first guesses the TCP sequence index of a victim transmitter, then fabricates packets using the predicted TCP index, and finally sends its fabricated packets to a victim receiver. Naturally, the TCP sequence prediction attack will inflict damage upon the data integrity owing to the above-mentioned packet fabrication and injection.

The UDP is also prone to flooding attacks, which are imposed by sending an overwhelming number of UDP packets, instead of ping requests used in the TCP flood attack. Specifically, a UDP flood attacker transmits a large number of UDP packets to a victim node, which will be forced to send numerous reply packets [82]. In this way, the victim node will be overwhelmed by the malicious UDP packets and becomes unreachable by other legitimate nodes. Moreover, the UDP flooding attacker is capable of hiding itself from the legitimate nodes by using a spoofed IP address for generating malicious UDP packets. The negative impact of such UDP flooding attacks is mitigated by limiting the response rate of UDP packets. Furthermore, firewalls can be employed for defending against the UDP flooding attacks for filtering out malicious UDP packets.

\subsection{Application-Layer Attacks}
\begin{table}
  \centering
  \caption{Main Types of Wireless Attacks at the Application Layer.}
  {\includegraphics[scale=0.6]{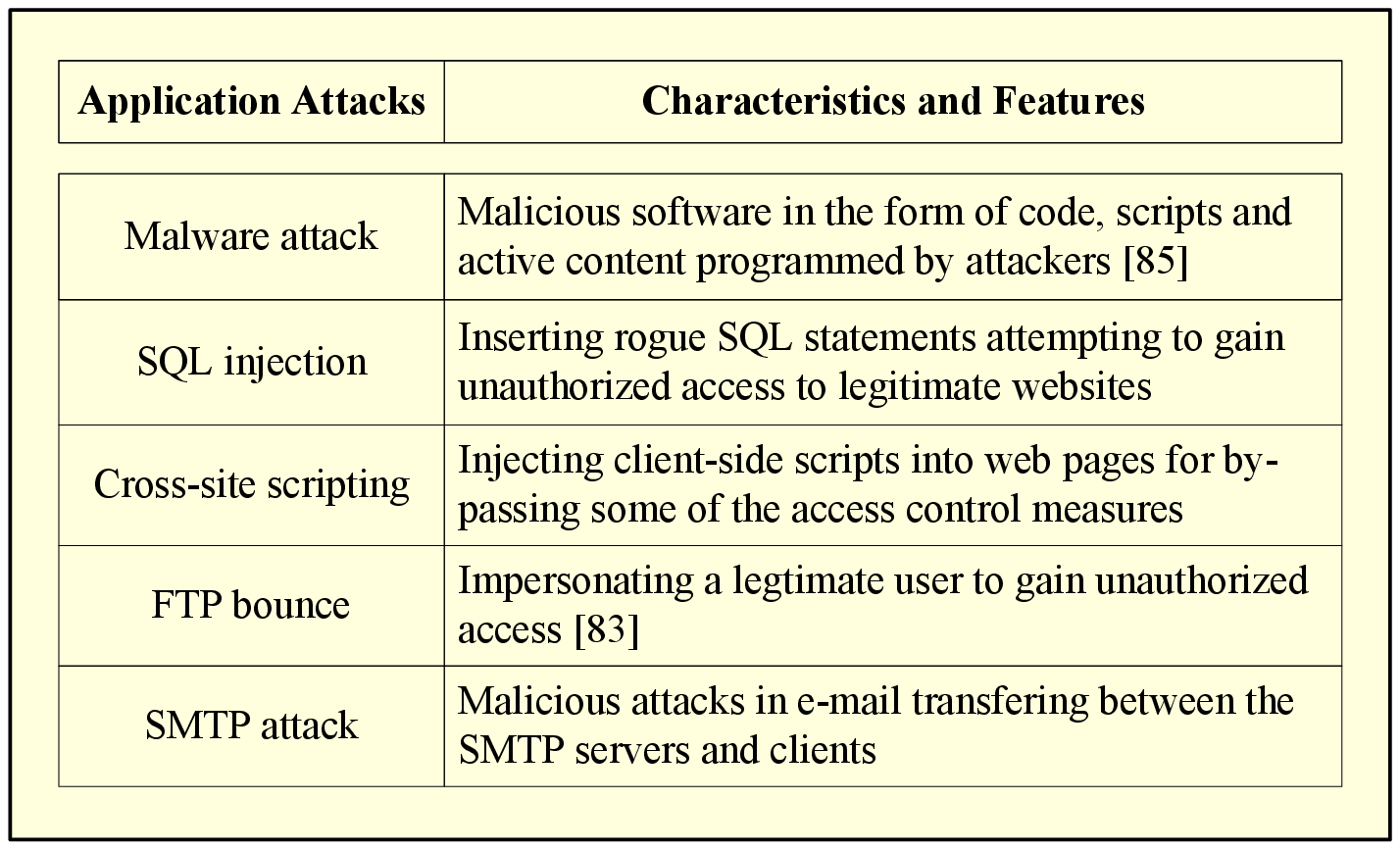}\label{Tab7}}
\end{table}

As mentioned above, the application layer supports HTTP [62] for web services, FTP [83] for file transfer and SMTP [84] for e-mail transmission. Each of these protocols is prone to security attacks. Logically, the application-layer attacks may hence be classified as HTTP attacks, FTP attacks and SMTP attacks. More specifically, HTTP is the application protocol designed for exchanging hypertext across the World Wide Web, which is subject to numerous security threats. The main HTTP attacks include the Malware attack (e.g., Trojan horse, viruses, worms, backdoors, keyloggers, etc.), structured query language (SQL) injection attack and cross-site scripting attack [85]. The terminology Malware refers to malicious software which is in the form of code, scripts and active content programmed by attackers attempting to disrupt legitimate transmissions or to intercept confidential information. The SQL injection is usually exploited to attack data-driven applications by inserting certain rogue SQL statements with an attempt to gain unauthorized access to legitimate websites. The last type of HTTP attacks to be mentioned is referred to as cross-site scripting attacks that typically occur in web applications and aim for bypassing some of the access control measures (e.g., the same-origin-policy) by injecting client-side scripts into web pages [85].

The FTP is used for large-file transfer from one network node to another, which also exhibits certain security vulnerabilities. The FTP bounce attacks and directory traversal attacks often occur in FTP applications [83]. The FTP bounce attack exploits the PORT command in order to request access to ports through another victim node, acting as a middle-man. We note however that most modern FTP servers are configured by default to refuse PORT commands in order to prevent FTP bounce attacks. The directory traversal attack attempts to gain unauthorized access to legitimate file systems by exploiting any potential security vulnerability during the validation of user-supplied input file names. In contrast to FTP, the SMTP is an application-layer protocol designed for transferring e-mails across the Internet, which, however, does not encrypt private information, such as the login username, the password and the messages themselves transmitted between the SMTP servers and clients, hence raising a serious privacy concern. Moreover, e-mails are frequent carriers of viruses and worms. Thus, the SMTP attacks include the password `sniffing', SMTP viruses and worms as well as e-mail spoofing [84]. Typically, antivirus software or firewalls (or both) are adopted for identifying and guarding against the aforementioned application-layer attacks. Table VII summarizes the aforementioned main attacks at the application layer.

\begin{figure}
  \centering
  {\includegraphics[scale=0.65]{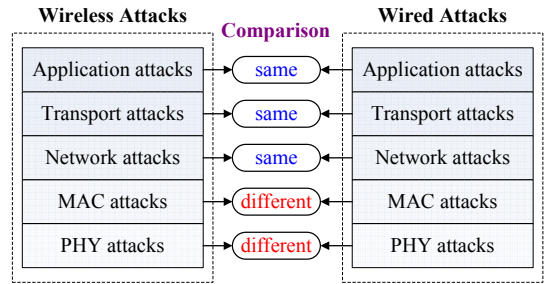}\\
  \caption{{{Comparison between the wireless and wired networks in terms of security attacks at different OSI layers.}}}\label{Fig4}}
\end{figure}

{{Finally, we summarize the similarities and differences between the wireless and wired networks in terms of their security attacks at the different OSI layers. As shown in Fig. 4, the application, transport and network-layer attacks of wireless networks are the same as those of wired networks, since the wireless and wired networks share common protocols at the application, transport and network layers. By contrast, wireless networks are different from wired networks in terms of the PHY and MAC attacks. In general, only the PHY and MAC layers are specified in wireless networking standards (e.g., Wi-Fi, Bluetooth, LTE, etc.). In wireless networks, conventional security protocols are defined at the MAC layer (sometimes at the logical-link-control layer) for establishing a trusted and confidential link, which will be summarized for different commercial wireless networks in Section IV. Additionally, the wireless PHY layer is completely different from its wireline based counterpart. Due to the broadcast nature of radio propagation, the wireless PHY layer is extremely vulnerable to both the eavesdropping and jamming attacks. To this end, physical-layer security is emerging as an effective means of securing wireless communications against eavesdropping, as it will be discussed in Section V. Next, Section VI will present various wireless jamming attacks and their counter-measures.}}



\section{Security Defense Protocols and Paradigms for Wireless Networks}
\begin{figure}
  \centering
  {\includegraphics[scale=0.55]{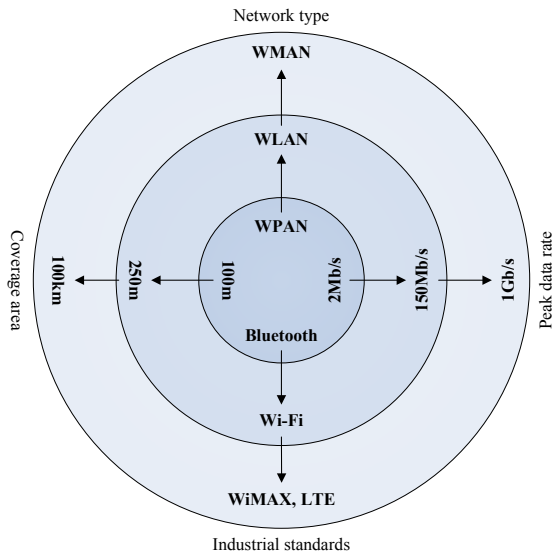}\\
  \caption{A family of wireless networks consisting of the wireless personal area network (WPAN), wireless local area network (WLAN) and wireless metropolitan area network (WMAN).}\label{Fig4}}
\end{figure}

This section is focused on the family of security protocols and paradigms that are used for improving the security of wireless networks. As compared to wired networks, the wireless networks have the advantage of avoiding the deployment of a costly cable based infrastructure. The stylized illustration of operational wireless networks is shown in Fig. 5, where the family of wireless personal area networks (WPAN), wireless local area networks (WLAN) and wireless metropolitan area networks (WMAN) are illustrated, which complement each other with the goal of providing users with ubiquitous broadband wireless services [86]. {{The objective of Fig. 5 is to provide a comparison amongst the WPAN, WLAN and WMAN techniques from different perspectives in terms of their industrial standards, coverage area and peak data rates.}} More specifically, a WPAN is typically used for interconnecting with personal devices (e.g., a keyboard, audio headset, printer, etc.) at a relatively low data rate and within a small coverage area. For example, Bluetooth is a common WPAN standard using short-range radio coverage in the industrial, scientific and medical band spanning the band 2400-2480MHz, which can provide a peak data rate of 2Mb\/s and a range of up to 100 meters (m) [87]. Fig. 5 also shows that a WLAN generally has a higher data rate and a wider coverage area than the WPAN, which is used for connecting wireless devices through an access point (AP) within a local coverage area. As an example, IEEE 802.11 (also known as Wi-Fi) consists of a series of industrial WLAN standards. Modern Wi-Fi standards are capable of supporting a peak data rate of 150Mb\/s and a maximum range of 250m [88]. Finally, a MAN is typically used for connecting a metropolitian city at a higher rate and over a lager coverage area than the WPAN and WLAN. For instance, in Fig. 5, we feature two types of industrial standards for WMAN, namely WiMAX and LTE [89], [90].

In the following, we will present an overview of the security protocols used in the aforementioned wireless standards (i.e., the Bluetooth, Wi-Fi, WiMAX and LTE) for protecting the authenticity, confidentiality, integrity and availability of legitimate
transmissions through the wireless propagation medium.

\subsection{Bluetooth}

\begin{figure}
  \centering
  {\includegraphics[scale=0.55]{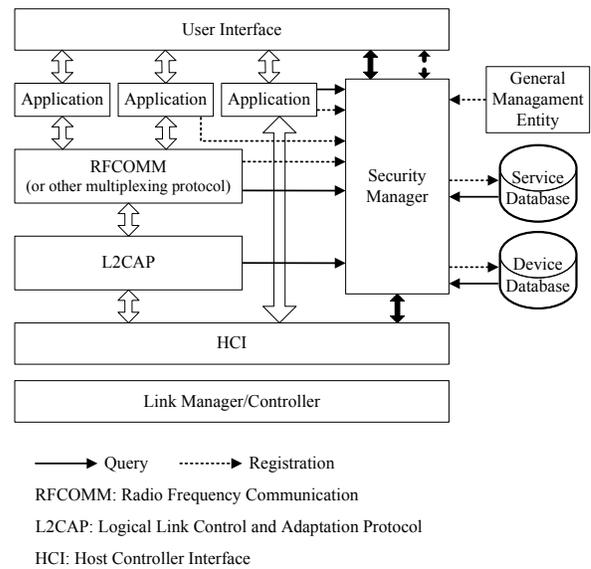}\\
  \caption{Bluetooth security architecture.}\label{Fig5}}
\end{figure}

Bluetooth is a short-range and low-power wireless networking standard, which has been widely implemented in computing and
communications devices as well as in peripherals, such as cell phones, keyboards, audio headsets, etc. However, Bluetooth devices are subject to a large number of wireless security threats and may easily become compromised. As a protection, Bluetooth introduces diverse security features and protocols for guaranteeing its transmissions against potentially serious attacks [91]. For security reasons, each Bluetooth device has four entities [92], including the Bluetooth device address (BD\_ADDR), private authentication key, private encryption key and a random number (RAND), which are used for authentication, authorization and encryption, respectively. More specifically, the BD\_ADDR contains 48 bits, which is unique for each Bluetooth device. The 128-bit private authentication key is used for authentication and the private encryption key that varies from 8 to 128 bits in length is used for encryption. In addition, RAND is a frequently changing 128-bit pseudo-random number generated by the Bluetooth device itself.

Fig. 6 illustrates the Bluetooth security architecture, where the key component is the security manager responsible for authentication, authorization and encryption [91]. As shown in Fig. 6, the service database and device database are mainly used for storing the security-related information on services and devices, respectively, which can be adjusted through the user interface. These databases can also be administrated by the general management entity. When a Bluetooth device receives an access request from another device, it will first query its security manager with the aid of its radio frequency communications (RFCOMM) or other multiplexing protocols. Then, the security manager has to respond to the query as to whether to allow the access or not by checking both the service database and device database. The generic access profile of Bluetooth defines three security modes:

(I) security mode 1 (non-secure), where no security procedure is initiated;

(II) security mode 2 (service-level enforced security), where the security procedure is initiated after establishing a link between the Bluetooth transmitter and receiver;

(III) security mode 3 (link level enforced security), where the security procedure is initiated before the link's establishment [91].


In Bluetooth systems, a device is classified into one of three categories: trusted/untrusted device, authenticated/unauthenticated device and unknown device. The trusted device category implies that the device has been authenticated and authorized as a trusted and fixed relationship, hence has unrestricted access to all services. By contrast, the untrusted device category refers to the fact that the device has indeed been authenticated successfully, but has no permanent fixed relationship, hence it is restricted to specific services. If a Bluetooth device is successfully authenticated, but has not completed any authorization process, it will be considered as an authenticated device. By definition, an unauthenticated device failed to authenticate and has a limited access to services. If a device has not passed any authentication and authorization process, it is classified as an unknown device and hence it is restricted to access services requiring the lowest privilege. Additionally, the Bluetooth services are also divided into the following three security levels: (1) authorization-level services, which can be accessed by trusted devices only; (2) authentication-level services, which require authentication, but no authorization. Hence, they remain inaccessible to the unauthenticated devices and unknown devices; and (3) open services, which are open to access by all devices. Below, we would like to discuss the detailed procedures of authentication, authorization and encryption in Bluetooth.
\begin{figure}
  \centering
  {\includegraphics[scale=0.65]{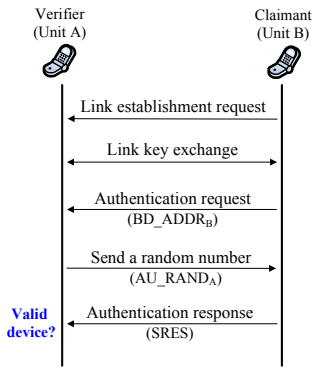}\\
  \caption{{{Bluetooth authentication process.}}\label{Fig7}}}
\end{figure}

The authentication represents the process of verifying the identity of Bluetooth devices based on the BD\_ADDR and link key. {{As shown in Fig. 7, the Bluetooth authentication adopts a ``challenge-response scheme" [93], where the verifier (Unit A) challenges the claimant (Unit B) which then responds for the sake of authentication. To be specific, the claimant first requests the verifier to establish a link and then exchanges a link key that is a 128-bit random number. Next, an authentication request with the claimant's address $\textrm{BD\_ADDR}_{\textrm{B}}$ is sent to the verifier, which returns a random number denoted by $\textrm{AU\_RAND}_{\textrm{A}}$. Then, both the verifier and claimant perform the same authentication function using the random number $\textrm{AU\_RAND}_{\textrm{A}}$, the claimant's address $\textrm{BD\_ADDR}_{\textrm{B}}$, and the link key to obtain their responses denoted by SRES' and SRES, respectively. Finally, the claimant sends its response SRES to the verifier, which will compare SRES with its own response SRES'. If SRES is identical to SRES', the authentication is confirmed. By contrast, a mismatch between SRES' and SRES represents an authentication failure.}}

\begin{figure}
  \centering
  {\includegraphics[scale=0.6]{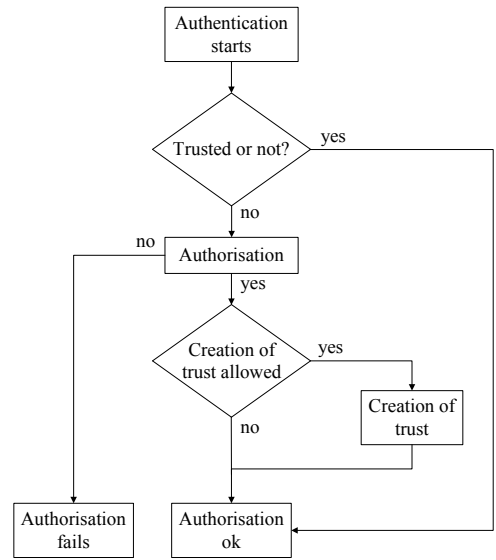}\\
  \caption{Flow chart of Bluetooth authorization.}\label{Fig7}}
\end{figure}
The authorization process is used for deciding whether a Bluetooth device has the right to access a certain service. Typically, trusted devices are allowed to access all services, however untrusted or unknown devices require authorization, before their access to services is granted. Fig. 8 shows a flow chart of the Bluetooth authorization process. Observe from Fig. 8 that the authorization process commences with checking the device database for deciding whether the Bluetooth device was authorized previously and considered trusted. If the Bluetooth device is trusted, the authorization is concluded. Otherwise, the authorization and the trust-creation will be performed sequentially. If the authorization fails, the access to certain services will be denied. Meanwhile, a successful authorization makes the corresponding Bluetooth devices trustworthy for accessing all services.

Additionally, encryption is employed in Bluetooth to protect the confidentiality of transmissions. The payload of a Bluetooth
data packet is encrypted by using a stream cipher, which consists of the payload key generator and key stream generator [93]. To be specific, first a payload key is generated with the aid of the link key and Bluetooth device address, which is then used for generating the key stream. Finally, the key stream and plain-text are added in modulo-2 in order to obtain the ciper text. It is pointed out that the payload key generator simply combines the input bits in an appropriate order and shifts them to four linear feedback shift registers to obtain the payload key. Moreover, the key stream bits are generated by using a method derived from the summation stream cipher generator by Massey and Rueppel [93].

\subsection{Wi-Fi}
The family of Wi-Fi networks mainly based on the IEEE 802.11 b/g standards has been explosively expanding. The most common security protocols in Wi-Fi are referred to as wired equivalent privacy (WEP) and Wi-Fi protected access (WPA) [94]. WEP was proposed in 1999 as a security measure for Wi-Fi networks to make wireless data transmissions as secure as in traditional wired networks. However, WEP has been shown to be a relatively weak security protocol, having numerous flaws. Hence, it can be `cracked' in a few minutes using a basic laptop computer. As an alternative, WPA was put forward in 2003 for replacing WEP, while the improved Wi-Fi protected access II (WPA2) constitutes an upgraded version of the WPA standard. Typically, WPA and WPA2 are more secure than WEP and thus they are widely used in modern Wi-Fi networks. Below, we detail the authentication and encryption processes of the WEP, WPA and WPA2 protocols.

\begin{figure}
  \centering
  {\includegraphics[scale=0.65]{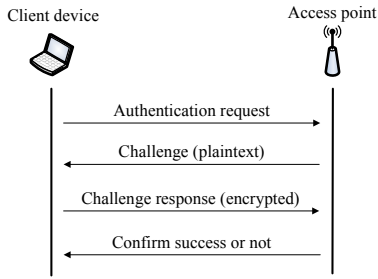}\\
  \caption{{{WEP authentication process.}}}\label{Fig8}}
\end{figure}
The WEP protocol consists of two main parts, namely the authentication part and encryption part, aiming for establishing access control by preventing unauthorized access without an appropriate WEP key and hence they achieve data privacy by encrypting the data streams with the aid of the WEP key. {{As shown in Fig. 9, the WEP authentication uses a four-step `challenge-response' handshake between a Wi-Fi client and an access point operating with the aid of a shared WEP key. To be specific, the client first sends an authentication request to the access point, which then replies with a plain-text challenge. After that, the client encrypts its received `challenge-text' using a pre-shared WEP key and sends the encrypted text to the access point. It then decrypts the received encrypted text with the aid of the pre-shared WEP key and attempts to compare the decrypted text to the original plain-text. If a match is found, the access point sends a successful authentication indicator to the client. Otherwise, the authentication is considered as failed.}}

Following the authentication, WEP activates the process of encrypting data streams using the simple rivest cipher 4 algorithm operating with the aid of the pre-shared WEP key [96]. Fig. 10 shows a block diagram of the WEP encryption, where first an initialization vector (IV) of 24 bits is concatenated to a 40-bit WEP key. This leads to a 64-bit seed for a pseudo-random number generator (PRNG), which is then used for generating the key stream. Additionally, an integrity check algorithm is performed such as a cyclic redundancy check on the plain-text in order to obtain an integrity check value (ICV), which can then be used for protecting the data transmission from malicious tampering. Then, the ICV is concatenated with the plain-text, which will be further combined with the aforementioned key stream in modulo-2 for generating the ciper-text. Although WEP carries out both the authentication and encryption functions, it still remains prone to security threats. For example, WEP fails to protect the information against forgery and replay attacks, hence an attacker may be capable of intentionally either modifying or replaying the data packets without the legitimate users becoming aware that data falsification and/or replay has taken place. Furthermore, the secret keys used in WEP may be `cracked' in a few minutes using a basic laptop computer [97]. Additionally, it is easy for an attacker to forge an authentication message in WEP, which makes it straightforward for unauthorized users to pretend to be legitimate users and hence to steal confidential information [98].

\begin{figure}
  \centering
  {\includegraphics[scale=0.65]{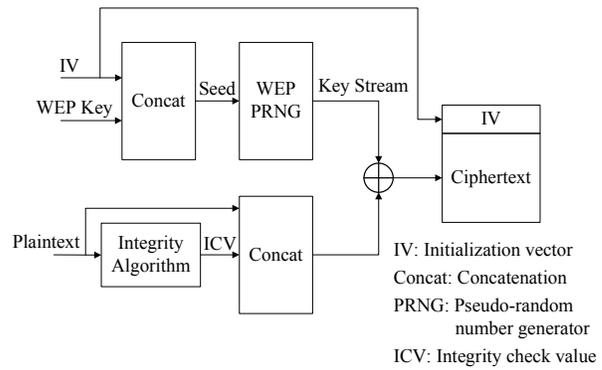}\\
  \caption{Block diagram of WEP encryption.}\label{Fig10}}
\end{figure}

As a remedy, WPA was proposed for addressing the aforementioned WEP security problems, which was achieved by Wi-Fi users without the need of changing their hardware. The WPA standard has two main types:

1) Personal WPA is mainly used in home without the employment of an authentication server, where a secret key is pre-shared between the client and access point, which is termed as WPA-PSK (pre-shared key);

2) Enterprise WPA used for enterprise networks, which requires an authentication server 802.1x for carrying out the security control in order to effectively guard against malicious attacks.

\begin{figure}
  \centering
  {\includegraphics[scale=0.63]{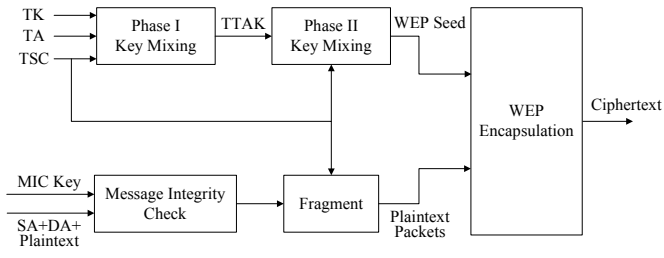}\\
  \caption{Illustration of TKIP encryption process.}\label{Fig11}}
\end{figure}

The main advantage of WPA over WEP is that WPA employs more powerful data encryption referred to as the temporal key integrity protocol (TKIP), which is assisted by a message integrity check (MIC) invoked for the sake of protecting the data integrity and confidentiality of Wi-Fi networks [99], [100]. Fig. 11 shows the TKIP encryption process, in which the transmitter address (TA), the temporal key (TK) and the TKIP sequence counter (TSC) constitute the inputs of the Phase I Key Mixing process, invoked in order to obtain a so-called TKIP-mixed transmit address and key (TTAK), which is then further processed along with the TSC in the Phase II Key Mixing stage for deriving the WEP seed, including a WEP IV and a base key. Furthermore, observe in Fig. 11 that the message integrity check (MIC) is performed both on the source address (SA), as well as on the destination address (DA) and the plain-text. The resultant MIC will then be appended to the plain-text, which is further fragmented into multiple packets, each assigned with a unique TSC. Finally, the WEP seed and plain-text packets are used for deriving the cipher-text by invoking the WEP encryption, as discussed earlier in Fig. 10, which is often implemented in the hardware of Wi-Fi devices. We note that even the WPA relying on the TKIP remains vulnerable to diverse practical attacks [101].

WiMAX (also known as IEEE 802.16) is a standard developed for wireless metropolitan area networks (WMAN) and the initial WiMAX system was designed for providing a peak data rate of 40 Mbps. In order to meet the requirements of the International Mobile Telecommunications-Advanced initiative, IEEE 802.16m was proposed as an updated version of the original WiMAX, which is capable of supporting a peak data rate of 1 Gbps for stationary reception and 100 Mbps for mobile reception [102]. As all other wireless systems, WiMAX also faces various wireless attacks and provides advanced features for enhancing the attainable transmission security. To be specific, a security sub-layer is introduced in the protocol stack of the WiMAX standard, as shown in Fig. 12 [103].

\subsection{WiMAX}
\begin{figure}
  \centering
  {\includegraphics[scale=0.65]{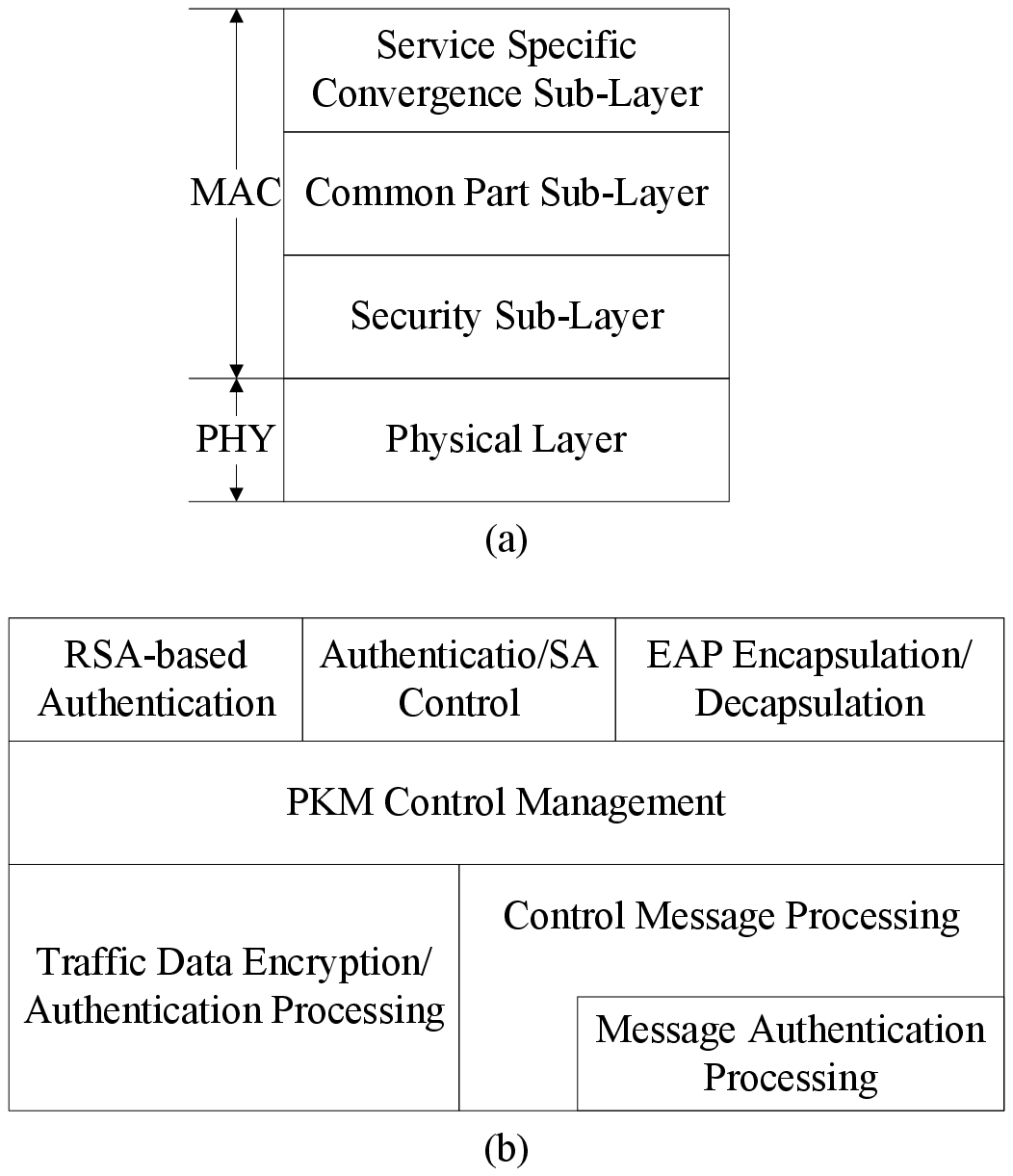}\\
  \caption{WiMAX protocol stack: (a) PHY-MAC illustration,
   and (b) security sub-layer specification.}\label{Fig12}}
\end{figure}

It is observed from Fig. 12(a) that the protocol stack of a WiMAX system defines two main layers, namely the physical (PHY) layer and medium access control (MAC) layer. Moreover, the MAC layer consists of three sub-layers, namely the service specific convergence sub-layer, the common part sub-layer and the security sub-layer. All the security issues and risks are considered and addressed in the security sub-layer. Fig. 12(b) shows the WiMAX security sub-layer, which will be responsible for authentication, authorization and encryption in WiMAX networks. The security sub-layer defines a so-called privacy and key management (PKM) protocol, which considers the employment of the X.509 digital certificate along with the Rivest-Shamir-Adleman (RSA) public-key algorithm and the advanced encryption standard (AES) algorithm for both user authentication as well as for key management and secure transmissions. The initial PKM version (PKMv1) as specified in early WiMAX standards (e.g., IEEE 802.16a/c) employs an unsophisticated one-way authentication mechanism and hence it is vulnerable to man-in-the-middle (MITM) attacks. To address this issue, an updated PKM version (PKMv2) was proposed in the more sophisticated WiMAX standard releases (e.g., IEEE 802.16e/m) [104], which relies on two-way authentication. The following discussions detail both the WiMAX authentication, as well as the authorization and encryption processes.

Authentication in WiMAX is achieved by the PKM protocol, which supports two basic authentication approaches, namely the RSA-based authentication and the extensible authentication protocol (EAP)-based authentication [105]. Fig. 13 shows the RSA-based authentication process, where a trusted certificate authority is responsible for issuing an X.509 digital certificate to each of the network nodes, including the subscriber station (SS) and the WiMAX base station (BS). An X.509 certificate contains both the public key and the MAC address of its associated network node. During the RSA-based authentication process shown in Fig. 13, when an SS receives an authentication request from a WiMAX BS, it sends its X.509 digital certificate to the BS, which then verifies whether the certificate is valid or not. If the certificate is valid, the SS is considered authenticated. By contrast, an invalid certificate implies that the SS fails to authenticate.

\begin{figure}
  \centering
  {\includegraphics[scale=0.65]{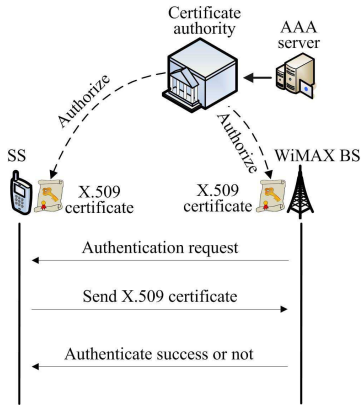}\\
  \caption{{{RSA-based authentication process.}}}\label{Fig13}}
\end{figure}

The EAP-based authentication process is illustrated in Fig. 14, where a WiMAX BS first sends an identity request to an SS who responds with its identity information. The WiMAX BS then forwards the SS' identity to an authentication, authorization and accounting (AAA) server over a secure networking protocol referred to as RADIUS. After that, the SS and AAA server start the authentication process, where three different EAP options are available depending on the SS and AAA server's capability, including the EAP-AKA (authentication and key agreement), EAP-TLS (transport layer security) and EAP-TTLS (tunneled transport layer security). Finally, the AAA server will indicate the success (or failure) of the authentication and notify the SS.

\begin{figure}
  \centering
  {\includegraphics[scale=0.65]{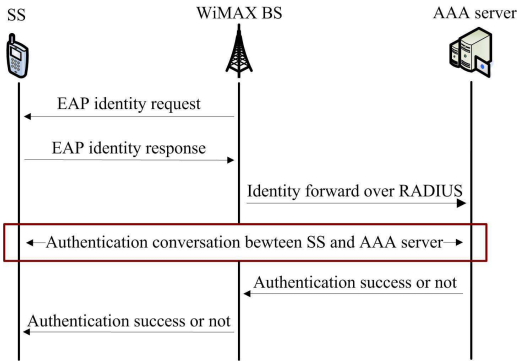}\\
  \caption{{{EAP-based authentication process.}}}\label{Fig14}}
\end{figure}

Additionally, the authorization process is necessary for deciding whether an authenticated SS has the right to access certain WiMAX services [106]. In the WiMAX authorization process, an SS first sends an authorization request message to the BS that contains both the SS' X.509 digital certificate, as well as the encryption algorithm and the cryptographic identity (ID). After receiving the authorization request, the BS validates the SS' request by interacting with an AAA server and then sends back an authorization reply to the SS. Once the positive authorization is confirmed, the SS will be allowed to access its intended services. Following user authentication and authorization, the SS is free to exchange data packets with the BS. In order to guarantee transmission confidentiality, WiMAX considers the employment of the AES algorithm for data encryption, which is much more secure than the data encryption standard (DES) algorithm [107]. Unlike the DES that uses the Feistel cipher design principles of [107], the AES cipher is based on a so-called substitution-permutation network and has a variable block size of 128, 192, or 256 bits [108]. This key-length specifies the number of transformation stages used for converting the plain-text into cipher-text. In WiMAX, the AES algorithm supports several different modes, including the cipher-block chaining mode, counter mode, and electronic codebook mode.

\subsection{LTE}
LTE is the most recent standard developed by the 3G partnership project for next-generation mobile networks designed for providing seamless coverage, high data rate and low latency [109]. It supports packet switching for seamless interworking with other wireless networks and also introduces many new elements, such as relay stations, home eNodeB (HeNB) concept, etc. A LTE network typically consists of an evolved packet core (EPC) and an evolved-universal terrestrial radio access network (E-UTRAN), as shown in Fig. 15 [90], [110]. The EPC is comprised of a mobility management entity (MME), a serving gateway, a packet data network gateway (PDN GW) and a home subscriber server (HSS). Moreover, the E-UTRAN includes a base station (also termed as eNodeB in LTE) and several user equipments (UEs). If channel conditions between the UEs and eNodeB are poor, a relay station may be activated for assisting their data communications. Furthermore, both in small offices and in residential environments, a HeNB may be installed for improving the indoor coverage by increasing both the capacity and reliability of the E-UTRAN. Although introducing these elements into LTE is capable of improving the network coverage and quality, it has its own new security vulnerabilities and threats.

\begin{figure}
  \centering
  {\includegraphics[scale=0.65]{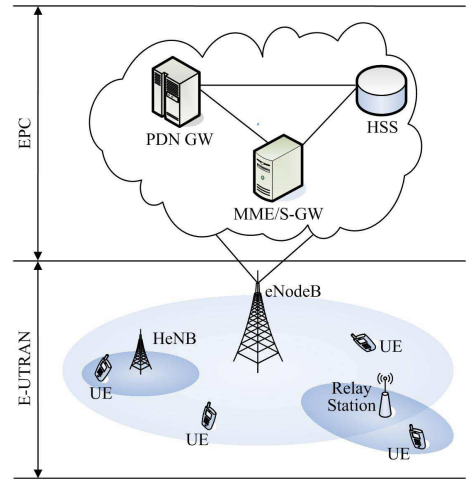}\\
  \caption{LTE network architecture.}\label{Fig15}}
\end{figure}

In order to facilitate secure packet exchange between the UEs and EPC, a so-called evolved packet system authentication and key agreement (EPS-AKA) protocol was proposed for defending LTE networks against various attacks, including redirection attacks, rogue base station attacks [111] and MITM attacks. A two-way authentication process was invoked between the UEs and EPC, which is adopted in the EPS-AKA protocol responsible for generating both the ciphering keys (CKs) and integrity keys (IKs) [112]. Both the CKs and IKs are used for data encryption and integrity check for enhancing the confidentiality and integrity of LTE transmissions. Fig. 16 shows this two-way authentication process of the LTE system using the EPS-AKA protocol, where an UE and a LTE network should validate each other's identity.

To be specific, the MME first sends a user identity verification request to the UE that then replies back with its unique international mobile subscriber identity (IMSI). Next, the MME sends an authentication data request to HSS, which consists of the UE's IMSI and the serving network's identity. Upon receiving the request, the HSS responds to the MME by sending back an EPS authentication vector containing the quantities $(\textrm{RAND, XRES, AUTN, KSI}_\textrm{ASME})$, where $\textrm{RAND}$ is an input parameter, while $\textrm{XRES}$ is an output of the authentication algorithm at the LTE network side. Furthermore, $\textrm{AUTN}$ indicates the identifier of the network authority, while $\textrm{KSI}_\textrm{ASME}$ is the key set identity of the access security management entity. Then, the MME sends an authentication request to the UE containing the $\textrm{RAND}$, $\textrm{AUTN}$ and $\textrm{KSI}_\textrm{ASME}$ quantities. As a result, the UE checks its received parameter $\textrm{AUTN}$ for authenticating the LTE network. If the network authentication is successful, the UE generates the response $\textrm{RES}$ and sends it to the MME, which compares $\textrm{XRES}$ with $\textrm{RES}$. If $\textrm{XRES}$ is the same as $\textrm{RES}$, this implies that the UE also passes the authentication.

\begin{figure}
  \centering
  {\includegraphics[scale=0.65]{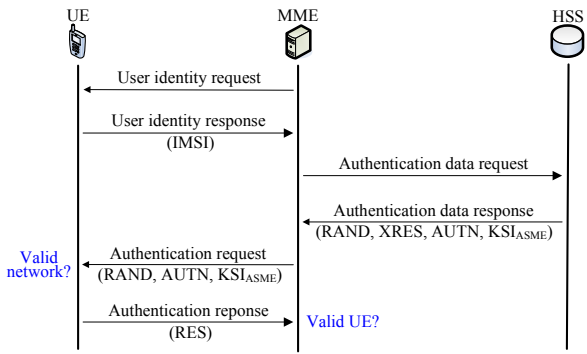}\\
  \caption{Two-way authentication in LTE by using the
  EPS-AKA protocol.}\label{Fig16}}
\end{figure}

In the universal mobile telecommunications system (UMTS), also known as the third generation mobile cellular system, KASUMI [113] is used as the ciphering algorithm for protecting the data confidentiality and integrity, which, however, has several security weaknesses and hence it is vulnerable to certain attacks, such as the related-key attack [113]. To this end, the LTE system adopts a more secure ciphering technique referred to as SNOW 3G [114] that is a block-based ciphering solution used as the heart of LTE confidentiality and integrity algorithms, which are referred to as the UEA2 and UIA2, respectively [114]. The SNOW 3G technique is referred to as a stream cipher having two main components, namely an internal state of 608 bits controlled by a 128-bit key and a 128-bit initialization vector (IV), which are utilized for generating the cipher-text by masking the plain-text. During the SNOW 3G operation process, we first perform key initialization to make the cipher synchronized to a clock signal and a 32-bit key stream word is produced in conjunction with every clock.

In summary, in the aforementioned subsections A-D, we have discussed the security protocols of Bluetooth, Wi-Fi, WiMAX as well as of LTE and observed that the existing wireless networks tend to rely on security mechanisms deployed at the upper OSI layers of Fig. 2 (e.g., MAC layer, network layer, transport layer, etc.) for both user authentication and data encryption. For example, the WEP and WPA constitute a pair of security protocols commonly used in Wi-Fi networks for guaranteeing the data confidentiality and integrity requirements, whereas WiMAX networks adopt the PKM protocol for achieving secure transmissions in the face of malicious attacks. By contrast, communication security at physical layer has been largely ignored in existing wireless security protocols. However, due to the broadcast nature of radio propagation, the physical layer of wireless transmission is extremely vulnerable to both eavesdropping and jamming attacks. This necessitates the development of physical-layer security as a complement to conventional upper-layer security protocols. The following section will introduce the physical-layer security paradigm conceived for facilitating secure wireless communications.

\section{Wireless Physical-Layer Security Against Eavesdropping}
\begin{figure}
  \centering
  {\includegraphics[scale=0.65]{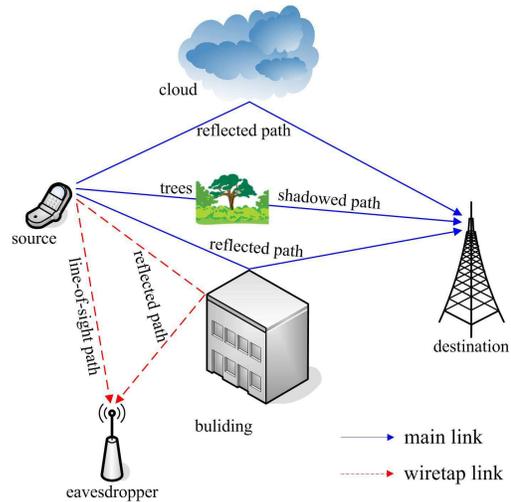}\\
  \caption{{{A wireless scenario transmitting from source to destination in multipath fading environments in the presence of an eavesdropper.}}}\label{Fig17}}
\end{figure}

In this section, we portray the field of wireless physical-layer security, which has been explored for the sake of enhancing the protection of wireless communications against eavesdropping attacks. Fig. 17 shows a wireless scenario transmitting from a source to a destination in the presence of an eavesdropper, where the main and wiretap links refer to the channels spanning from the source to the destination and to the eavesdropper, respectively. As shown in Fig. 17, when a radio signal is transmitted from the source, multiple differently delayed signals will be received at the destination via different propagation paths due to the signal reflection, diffraction and scattering experienced. Owing to the multipath effects, the differently delayed signal components sometimes add constructively, sometimes destructively. Hence, the attenuation of the signal that propagated through the space fluctuates in time, which is referred to as fading and it is usually modeled as a random process. The signal received at the destination may be attenuated significantly, especially when a deep fade is encountered, due to the shadowing in the presence of obstacles (e.g., trees) between the source and destination. Moreover, due to the broadcast nature of radio propagation, the source signal may be overheard by the eavesdropper, which also experiences a multipath fading process. There are three typical probability distribution models routinely used for characterizing the random wireless fading, including the Rayleigh fading [115], Rice fading [116] and Nakagami fading [117].


\begin{figure}
  \centering
  {\includegraphics[scale=0.65]{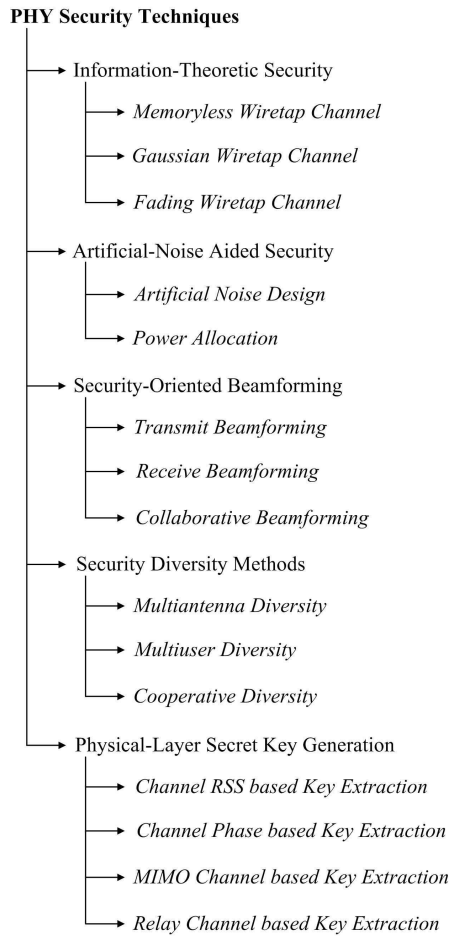}\\
  \caption{Classification of physical-layer security techniques.}\label{Fig18}}
\end{figure}

Recently, physical-layer security is emerging as a promising paradigm designed for improving the security of wireless transmissions by exploiting the physical characteristics of wireless channels [33], [34], [118]. More specifically, it was shown in [33] that reliable information-theoretic security can be achieved, when the wiretap channel spanning from the source to the eavesdropper is a degraded version of the main channel between the source and the destination. In [34], a so-called secrecy capacity was developed and shown as the difference between the capacity of the main channel and that of the wiretap channel, where a positive secrecy capacity means that reliable information-theoretic security is possible and vice versa. However, in contrast to wired channels that are typically time-invariant, wireless channels suffer from time-varying random fading, which results in a significant degradation of the wireless secrecy capacity [119], especially when a deep fade is encountered in the main channel due to shadowing by obstacles (e.g., buildings, trees, etc.) appearing between the source and destination. Hence considerable research efforts have been devoted to the development of various physical-layer security techniques, which can be classified into the following main research categories:

\begin{table*}
  \centering
  \caption{{Major Information-Theoretic Security Techniques.}}
  {\includegraphics[scale=0.75]{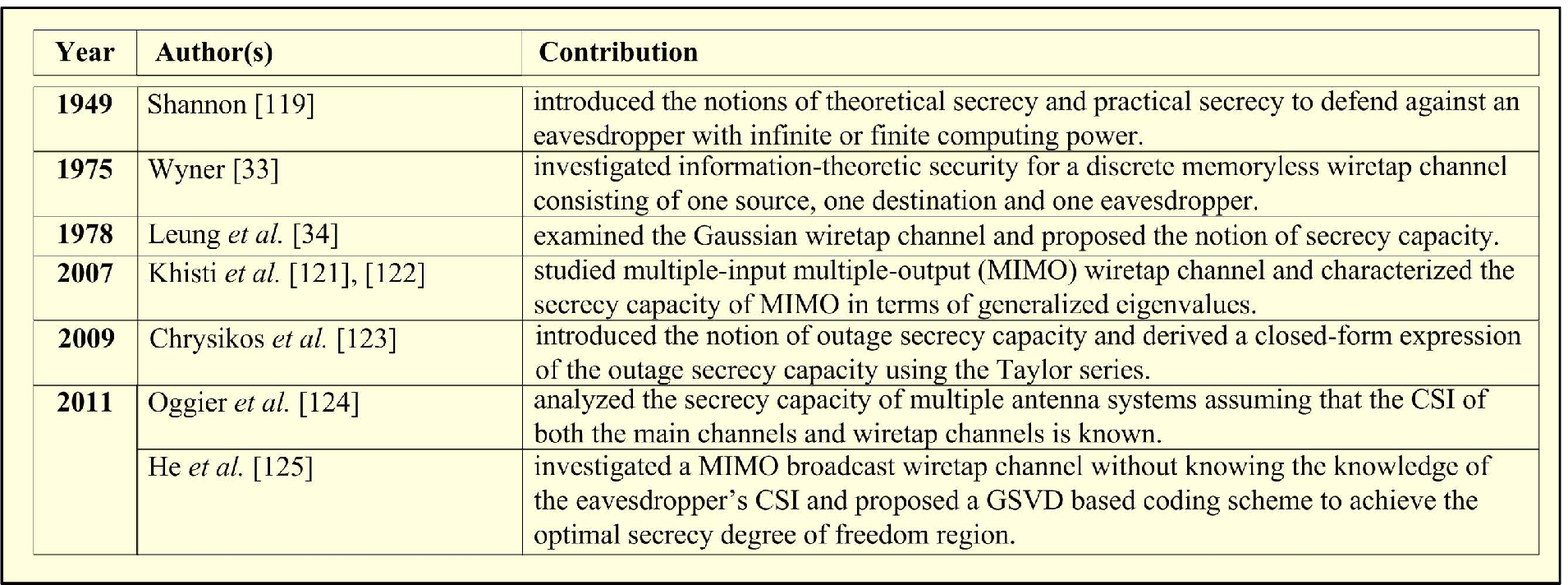}\label{Tab8}}
\end{table*}

1) information-theoretic security [119]-[125];

2) artificial noise aided security [126]-[130];

3) security-oriented beamforming techniques [131]-[136];

4) diversity-assisted security approaches [42], [137];

{{5) physical-layer secret key generation [147]-[161].}}

The aforementioned physical-layer security techniques are summarized in Fig. 18. In the following, we will detail these physical-layer security topics.

\subsection{Information-Theoretic Security}
Information-theoretic security examines fundamental limits of physical-layer security measures from an information-theoretic perspective. The concept of information-theoretic security was pioneered by Shannon in [119], where the basic theory of secrecy systems was developed with an emphasis on the mathematical structure and properties. To be specific, Shannon defined a secrecy system as a set of mathematical transformations of one space (the set of legitimate plain-text messages) into another space (the set of possible cryptograms), where each transformation corresponds to enciphering the information with the aid of a secret key. Moreover, the transformation is non-singular so that unique deciphering becomes possible, provided that the secret key is known. In [119], the notions of theoretical secrecy and practical secrecy were introduced, which was developed for the ease of guarding against eavesdropping attacks, when an adversary is assumed to have either infinite or more practically finite computing power. It was shown in [119] that a perfect secrecy system may be created, despite using a finite-length secret key, where the equivocation at the adversary does not approach zero, i.e. when the adversary is unable to obtain a unique solution to the cipher-text. {{To elaborate a little further, the equivocation is defined as a metric of quantifying how uncertain the adversary is of the original cipher-text after the act of message interception [119].}}

The secrecy system developed by Shannon in [119] is based on the employment of secret keys. However, the key management is challenging in certain wireless networks operating without a fixed infrastructure (e.g., wireless ad hoc networks) [32]. To this end, in [33], Wyner investigated the information-theoretic security without using secret keys and examined its performance limits for a discrete memoryless wiretap channel consisting of a source, a destination and an eavesdropper. It was shown in [33] that perfectly secure transmission can be achieved, provided that the channel capacity of the main link spanning from the source node (SN) to its destination node (DN) is higher than that of the wiretap link between the SN and eavesdropper. In other words, when the main channel conditions are better than the wiretap channel conditions, there exists a positive rate at which the SN and DN can reliably and securely exchange their information. In [34], Wyner's results were further extended to a Gaussian wiretap channel, where the notion of secrecy capacity was developed, which was obtained as the difference between the channel capacity of the main link and that of the wiretap link. If the secrecy rate is chosen below the secrecy capacity, reliable transmission from SN to DN can be achieved in perfect secrecy. In wireless networks, the secrecy capacity is severely {{degraded}} due to the time-varying fading effect of wireless channels. {{This is because fading attenuates the signal received at the legitimate destination, which reduces the capacity of the legitimate channel, thus resulting in a degradation of the secrecy capacity.}}

The family of multiple-input multiple-output (MIMO) systems is widely recognized as an effective means of mitigating the effects of wireless fading, which simultaneously increases the secrecy capacity in fading environments. In [120], Khisti \emph{et al.} investigated a so-called multiple-input single-output multiple-eavesdropper (MISOME) scenario, where both the source and eavesdropper are equipped with multiple antennas, whereas the intended destination has a single antenna. Assuming that the fading coefficients of all the associated wireless channels are fixed and known to all nodes (i.e., to the source, destination and eavesdropper), the secrecy capacity of the MISOME scenario can be characterized in terms of its generalized eigenvalues. Bearing in mind that the knowledge of the wiretap channel's impulse response is typically unavailable, Khisti \emph{et al.} [121] advocated the employment of a so-called masked beamforming scheme [118] for enhancing wireless physical-layer security, where the eavesdropper's channel knowledge is not relied upon for determining the transmit directions. It was shown that the masked beamforming scheme is capable of achieving a near-optimal security performance at sufficiently high signal-to-noise ratios (SNRs). Moreover, Khisti \emph{et al.} extended their results to time-varying wireless channels and developed both an upper and a lower bound on the secrecy capacity of the MISOME scenario operating in Rayleigh fading environments. {{In a nutshell, the work of Khisti \emph{et al.} [121] was mainly focused on characterizing the secrecy capacity of masked beamforming in an information-theoretic sense, which thus belongs to the family of information-theoretic security solutions.}}

As a further development, Khisti \emph{et al.} [122] examined the information-theoretic security achieved with the aid of multiple antennas in a more general scenario, where the source, destination and eavesdropper are assumed to have multiple antennas. They considered two cases: 1) the simplified and idealized deterministic case in which the channel state information (CSI) of both the main links and of the wiretap links are fixed and known to all the nodes; and 2) the more practical fading scenario, where the wireless channels experience time-varying Rayleigh fading and the source has the main channels' perfect CSI as well as the wiretap channels' statistical CSI knowledge. For the idealized deterministic case, they proposed the employment of the generalized-singular-value-decomposition (GSVD) based approach for increasing the secrecy capacity in high SNR regions. The GSVD scheme's performance was then further investigated in the fading scenario and the corresponding secrecy capacity was shown to approach zero if and only if the ratio of the number of eavesdropper antennas to source antennas was larger than two. Additionally, in [123], Chrysikos \emph{et al.} investigated the wireless information-theoretic security in terms of outage secrecy capacity, which is used for characterizing the maximum secrecy rate under a given outage probability requirement. A closed-form expression of the outage secrecy capacity was derived in [123] by using the first-order Taylor series for approximation of an exponential function.

\begin{table*}
  \centering
  \caption{{Major Artificial Noise Aided Security Techniques.}}
  {\includegraphics[scale=0.75]{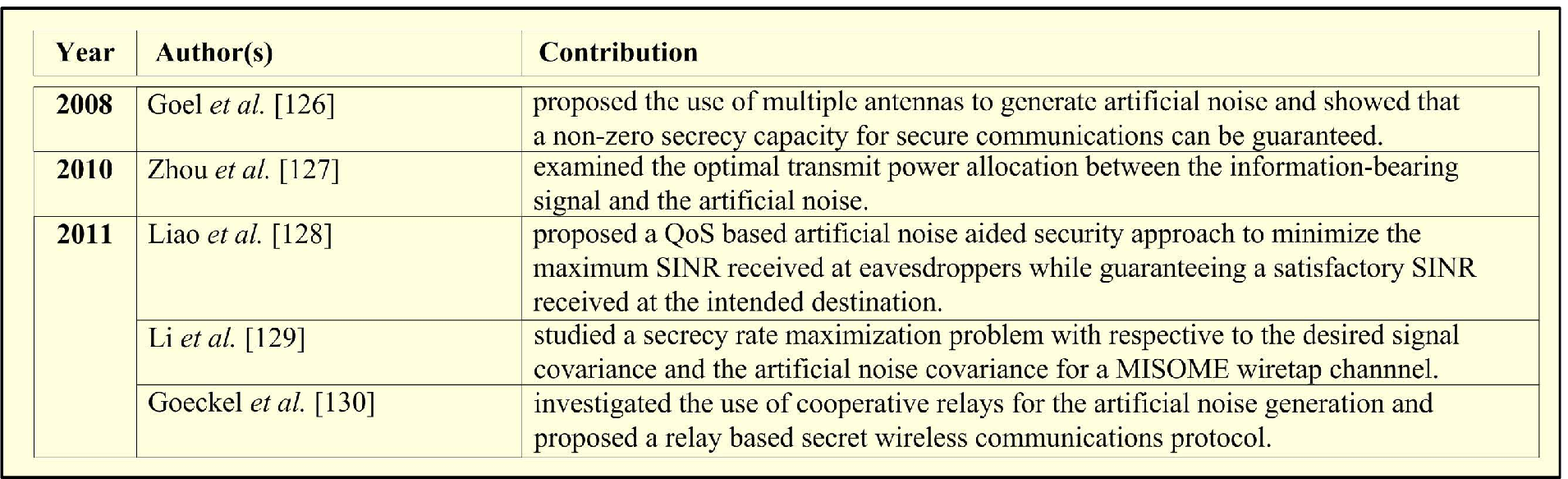}\label{Tab9}}
\end{table*}

The MIMO wiretap channel can also be regarded as a MIMO broadcast channel, where SN broadcasts its confidential information to both its legitimate DN and unintentionally also to an unauthorized eavesdropper. Perfect secrecy is achieved, when SN and DN can reliably communicate at a positive rate, while ensuring that the mutual information between the SN and eavesdropper becomes zero. In [124], Oggier and Hassibi analyzed the secrecy capacity of multiple antenna aided systems by converting the MIMO wiretap channel into a MIMO broadcast channel, where the number of antennas is arbitrary for both the transmitter and the pair of receivers (i.e., that of DN and of the eavesdropper). It was proven that through optimizing the transmit covariance matrix, the secrecy capacity of the MIMO wiretap channel is given by the difference between the capacity of the SN-to-DN channel and that of the SN-to-eavesdropper channel. It was pointed out that the secrecy capacity results obtained in [124] are based on the idealized simplifying assumption that SN knows the CSI of both the main channels and of the wiretap channels. This assumption is, however, invalid in practical scenarios, since the eavesdropper is passive and hence it remains an open challenge to estimate the eavesdropper's CSI. It is of substantial interest to study a more practical scenario, where SN only has statistical CSI knowledge of wiretap channels. To this end, the authors of [125] investigated a twin-receiver MIMO broadcast wiretap channel scenario, where the legitimate SN and DN are assumed to have no knowledge of the eavesdropper's CSI. A so-called `secrecy-degree-of-freedom-region' was developed for wireless transmission in the presence of an eavesdropper and a GSVD based scheme was proposed for achieving the optimal secrecy-degree-of-freedom-region. The major information-theoretic security techniques are summarized in Table VIII.

\subsection{Artificial Noise Aided Security}
The artificial noise aided security allows SN to generate specific interfering signals termed as artificial noise so that only the eavesdropper is affected adversely by the interfering signals, while the intended DN remains unaffected. This results in a reduction of the wiretap channel's capacity without affecting the desired channel's capacity and thus leads to an increased secrecy capacity, which was defined as the difference between the main channel's and the wiretap channel's capacity. Hence, a security improvement is achieved by using artificial noise. In [126], Goel and Negi considered a wireless network consisting of a SN, a DN and an eavesdropper for investigating the benefits of the artificial noise generation paradigm. More specifically, SN allocates a certain fraction of its transmit power for producing artificial noise, so that only the wiretap channel condition is degraded, while the desired wireless transmission from SN to DN remains unaffected by the artificial noise. To meet this requirement, Goel and Negi [126] proposed the employment of multiple antennas for generating artificial noise and demonstrated that the number of transmit antennas at SN has to be higher than that of the eavesdropper for ensuring that the artificial noise would not degrade the desired channel. It was shown that a non-zero secrecy capacity can be guaranteed for secure wireless communications by using artificial noise, even if the eavesdropper is closer to SN than DN.

\begin{table*}
  \centering
  \caption{{Major Security-Oriented Beamforming Techniques.}}
  {\includegraphics[scale=0.75]{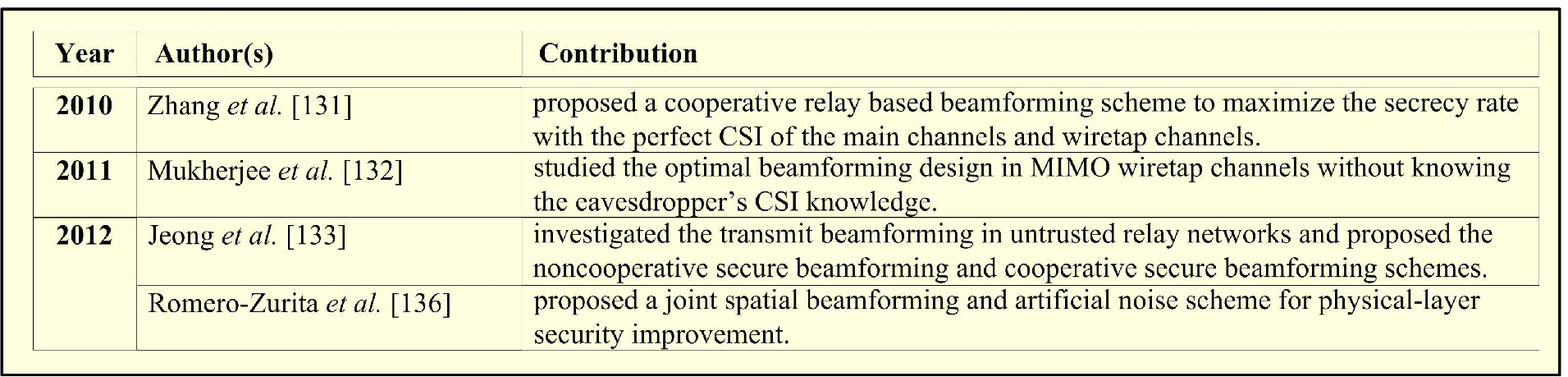}\label{Tab10}}
\end{table*}

Although the artificial noise aided security is capable of guaranteeing the secrecy of wireless transmission, this is achieved at the cost of wasting precious transmit power resources, since again, a certain amount of transmit power has to be allocated for generating the artificial noise. In [127], Zhou an McKay further examined the optimal transmit power sharing between the information-bearing signal and the artificial noise. They analyzed secure multiple-antenna communications relying on artificial noise and derived a closed-form secrecy capacity expression for fading environments, which was used as the objective function for quantifying the optimal power sharing between the information signal and artificial noise. The simple equal-power sharing was shown to be a near-optimal strategy, provided that the eavesdroppers do not collude with each other to jointly perform interception. Moreover, as the number of eavesdroppers increases, more power should be allocated for generating the artificial noise. In the presence of imperfect CSI, it was observed that assigning more power to the artificial noise for jamming the eavesdroppers is capable of achieving a better security performance than increasing the transmit power of the desired information signal.

However, the aforementioned artificial noise aided security work has been mainly focused on improving the secrecy capacity without considering the quality of service (QoS) requirements of the legitimate DN. Hence, in order to address this problem, a QoS based artificial noise aided security approach was presented in [128] for minimizing the maximum attainable signal-to-interference-and-noise ratio (SINR) encountered at the eavesdroppers, while simultaneously guaranteeing a satisfactory SINR at the intended DN. The optimization of the artificial noise distribution was formulated based on the CSIs of both the main channels and wiretap channels, which was shown to be a non-deterministic polynomial-time hard (NP-hard) problem. The classic semidefinite relaxation (SDR) technique [128] was used for approximating the solution of this NP-hard problem. Liao \emph{et al.} [128] demonstrated that the proposed QoS-based artificial noise aided security scheme is capable of efficiently guarding against eavesdropping attacks, especially in the presence of a large number of eavesdroppers. Li and Ma [129] proposed a robust artificial noise aided security scheme for a MISOME wiretap channel. Assuming that SN has perfect CSI knowledge of the main channels, but imperfect CSI knowledge of the wiretap channels, an optimization problem was formulated for the secrecy rate maximization with respect to both the desired signal's and the artificial noise's covariance, which is a semi-infinite optimization problem and can be solved with the aid of a simple one-dimensional search algorithm. It was shown that the proposed robust artificial noise design significantly outperforms conventional non-robust approaches in terms of its secrecy capacity.

In addition to relying on multiple antennas for artificial noise generation, cooperative relays may also be utilized for producing artificial noise to guard against eavesdropping attacks. In [130], the authors studied the employment of cooperative relays for artificial noise generation and proposed a secret wireless communications protocol, where a messaging relay was used for assisting the legitimate transmissions from SN to DN and a set of intervening relays were employed for generating the artificial noise invoked for jamming the eavesdroppers. The main focus of [130] was to quantify how many eavesdroppers can be tolerated without affecting the communications secrecy in a wireless network supporting a certain number of legitimate nodes. It was shown that if the eavesdroppers are uniformly distributed and their locations are unknown to the legitimate nodes, the tolerable number of eavesdroppers increases linearly with the number of legitimate nodes. The major artificial noise aided security techniques are summarized in Table IX.

\subsection{Security-Oriented Beamforming Techniques}
The family of security-oriented beamforming techniques allows SN to transmit its information signal in a particular direction to the legitimate DN, so that the signal received at an eavesdropper (that typically lies in a direction different from DN) experiences destructive interference and hence it becomes weak. Thus, the received signal strength (RSS) of DN would become much higher than that of the eavesdropper with the aid of security-oriented beamforming, leading to a beneficial secrecy capacity enhancement. In [131], the authors proposed the employment of cooperative relays to form a beamforming system relying on the idealized simplifying assumption of having the perfect CSI knowledge of all the main channels as well as of the wiretap channels and conceived a decode-and-forward relay based beamforming design for maximizing the secrecy rate under a fixed total transmit power constraint. The formulated problem was then solved by using the classic semi-definite programming and second-order cone programming techniques. It was shown in [131] that the proposed beamforming approach is capable of significantly increasing the secrecy capacity of wireless transmissions.

In [132], multiple antennas were used for beamforming in order to improve the attainable secrecy capacity of wireless transmissions from SN to DN in the presence of an eavesdropper. In contrast to the work presented in [131], where the perfect CSI knowledge of the wiretap channel was assumed, Mukherjee and Swindlehurst conceived the optimal beamforming designs in [132] without relying on the idealized simplifying assumption of knowing the eavesdropper's CSI, albeit the exact CSI of the main channel spanning from SN to DN was still assumed to be available. However, the perfect CSI of the main channel is typically unavailable at SN. To this end, Mukherjee and Swindlehurst further studied the impact of imperfect CSI on the attainable physical-layer security performance and presented a pair of robust beamforming schemes that are capable of mitigating the effect of channel estimation errors. It was shown that the proposed robust beamforming techniques perform well for moderate CSI estimation errors and hence achieve a higher secrecy capacity than the artificial noise aided security approaches.

In addition, the authors of [133] investigated the benefits of transmit beamforming in an amplify-and-forward relay network consisting of a SN, a RN and a DN, where the RN is indeed potentially capable of improving the SN-to-DN link, but it is also capable of launching a passive eavesdropping attack. Hence, a pair of secure beamforming schemes, namely a non-cooperative and a cooperative secure beamformers were proposed for maximizing the secrecy capacity of the SN-to-DN link. Extensive simulation results were provided for demonstrating that the secure beamforming schemes proposed are capable of outperforming conventional security approaches in terms of the attainable secrecy capacity. {{Moreover, in [134], a cross-layer approach exploiting the multiple simultaneous data streams of the family of operational IEEE 802.11 standards was devised by using zero-forcing beamforming, where a multi-antenna assisted AP was configured to utilize one of its data streams for communicating with the desired user, while the remaining data streams were exploited for actively interfering with the potential eavesdroppers. Extensive experimental evaluations were carried out in practical indoor WLAN environments, demonstrating that the proposed zero-forcing beamforming method consistently granted an SINR for the desired user, which was 15dB higher than that of the eavesdropper.}}

Naturally, this beamforming technique may also be combined with the artificial noise based approach for the sake of further enhancing the physical-layer security of wireless transmissions against eavesdropping attacks. Hence, in [135], the authors examined a joint beamforming and artificial noise aided design for conceiving secure wireless communications from SN to DN in the presence of multiple eavesdroppers. The beamforming weights and artificial noise covariance were jointly optimized by minimizing the total transmit power under a specific target secrecy rate constraint. To elaborate a little further, this joint beamforming and artificial noise aided design problem was solved by using a two-level optimization approach, where the classic semidefinite relaxation method and the golden section based method [135] were invoked for the inner-level optimization and the outer-level optimization, respectively. Numerical results illustrated that the joint beamforming and artificial noise aided scheme significantly improves the attainable secrecy capacity of wireless transmission as compared to the conventional security-oriented beamforming approaches. In [136], Romero-Zurita \emph{et al.} studied the joint employment of spatial beamforming and artificial noise generation for enhancing the attainable physical-layer security of a MISO channel in the presence of multiple eavesdroppers, where no CSI knowledge was assumed for the wiretap channel. The optimal power sharing between the information signal and artificial noise was examined under a specific guaranteed secrecy probability requirement. By combining the beamforming and artificial noise techniques, both the security and reliability of wireless transmissions were substantially improved. The major security-oriented beamforming techniques are summarized in Table X.

\begin{figure}
  \centering
  {\includegraphics[scale=0.65]{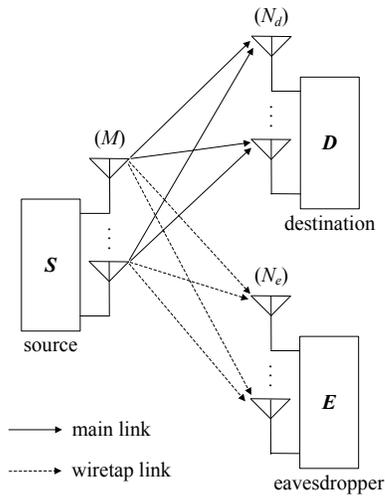}\\
  \caption{A MIMO wireless system consisting of a SN and a DN in the presence of an eavesdropper, where $M$, $N_d$ and $N_e$ represent the number of antennas at SN, DN and eavesdropper, respectively.}\label{Fig19}}
\end{figure}

\subsection{Diversity Assisted Security Approaches}
This subsection is focused on the portrayal of diversity techniques invoked for the sake of improving the physical-layer security of wireless transmissions [137]. In contrast to the artificial noise aided approaches, which dissipate additional power assigned to the artificial noise generation, the diversity-aided security paradigm is capable of enhancing the wireless security without any additional power. Traditionally, diversity techniques have been used for improving the attainable transmission reliability, but they also have a substantial potential in terms of enhancing the wireless security against eavesdropping attacks. Below we will discuss several diversity-aided security approaches, including multiple-antenna aided diversity, multiuser diversity and cooperative diversity.

Multiple-antenna aided transmit diversity has been shown to constitute an effective means of combatting the fading effect, hence also increasing the secrecy capacity of wireless transmissions [138], [139]. As shown in Fig. 19, provided that SN has multiple antennas, the optimal antenna can be activated for transmitting the desired signal, depending on whether the CSI of the main channel and of the wiretap channel is available. To be specific, if the CSI of both the main channel and of the wiretap channel is known at SN, the specific transmit antenna associated with the highest secrecy capacity can be chosen as the optimal antenna to transmit the desired signal, which has the potential of significantly improving the secrecy capacity of wireless transmissions. If only the main channel's CSI is available, we can choose a transmit antenna associated with the highest main channel capacity to transmit the desired signal. Since the transmit antenna selection is exclusively based on the main channel's CSI and the wiretap channel is typically independent of the main channel, the main channel's capacity will be increased with the aid of transmit antenna selection, while no capacity improvement can be achieved for the wiretap channel. This finally results in an increase of the secrecy capacity, as an explicit benefit of transmit antenna selection.

\begin{figure}
  \centering
  {\includegraphics[scale=0.65]{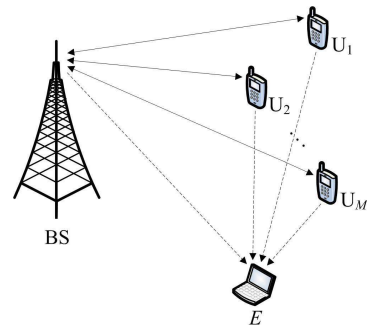}\\
  \caption{A multiuser diversity system consisting of a BS and $M$ users in the presence of an eavesdropper.}\label{Fig20}}
\end{figure}

The multiuser diversity of Fig. 20 also constitutes an effective means of improving the physical-layer security in the face of eavesdropping attacks. Considering that a base station (BS) serves multiple users in a cellular network, an orthogonal multiple access mechanism, such as the orthogonal frequency-division multiple access (OFDMA) of LTE [140] or code-division multiple access (CDMA) of 3G systems [141], enables the multiple users to communicate with the BS. Considering the OFDMA as an example, given a slot or subband of OFDM subcarriers, we should determine which particular user is assigned to access this specific subband for data transmission. More specifically, a user is enabled with the aid of multiuser scheduling to access the OFDM subband and then starts transmitting its signal to the BS. Meanwhile, due to the broadcast nature of wireless medium, an eavesdropper may intercept the source message. In order to effectively protect the wireless transmission against eavesdropping attacks, the multiuser scheduling should be designed for minimizing the capacity of the wiretap channel, while maximizing the capacity of the main channel [142]. This action requires the CSI of both the main channel and of the wiretap channel. If only the main channel's CSI is available, the multiuser scheduling can be designed for maximizing the main channel's capacity without the wiretap channel's CSI knowledge. It is worth mentioning that the multiuser scheduling is capable of significantly improving the main channel's capacity, while the wiretap channel's capacity remains the same, which results in a secrecy capacity improvement with the aid of multiuser diversity, even if the CSI of the wiretap channel is unknown.

\begin{figure}
  \centering
  {\includegraphics[scale=0.65]{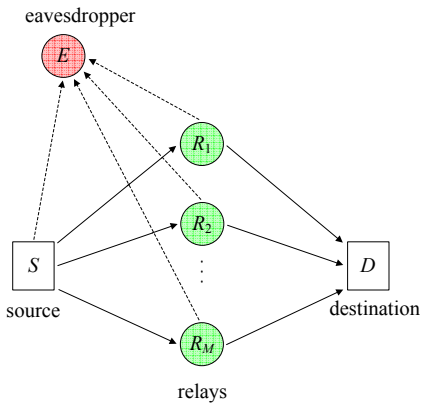}\\
  \caption{A cooperative diversity system consisting of a SN, $M$ relays and a DN in the presence of an eavesdropper.}\label{Fig21}}
\end{figure}

As an alternative, cooperative diversity [143], [144] also has a great potential in terms of protecting the wireless transmissions against eavesdropping attacks. When considering a wireless network consisting of a single SN, multiple RNs and a DN as shown in Fig. 21, the multiple relays can be exploited for assisting the signal transmission from SN to DN. In order to prevent the eavesdropper from intercepting the source signal from a security perspective, the best relay selection emerges as a means of improving the security of wireless transmissions against eavesdropping attacks [145]. Specifically, a RN having the highest secrecy capacity (or the highest main channel capacity if only the main channel's CSI is known) is selected to assist the SN's transmission to the intended DN. By using the best RN selection, a beneficial cooperative diversity gain can be achieved for the sake of increasing the secrecy capacity, which explicitly demonstrates the advantages of wireless physical-layer security.

\subsection{Physical-Layer Secret Key Generation}
In this subsection, we present the family of wireless secret key generation techniques by exploiting the physical-layer characteristics of radio propagation, including the amplitude and phase of wireless fading [146]. To be specific, as shown in Fig. 22, a pair of legitimate transceivers, namely Alice and Bob are connected through a reciprocal wireless channel, where the fading gain of the main channel spanning from Alice to Bob, denoted by $h_{ab}$, is identical to that from Bob to Alice, namely $h_{ba}$. Since Alice and Bob can directly estimate $h_{ba}$ and $h_{ab}$, respectively, using classic channel estimation methods [147], [148], they may exploit their estimated CSIs $\hat h_{ba}$ and $\hat h_{ab}$ for the secret key generation and agreement process. By contrast, a third party (e.g., Eve) based at a different location experiences independent wiretap channels of $h_{ae}$ and $h_{be}$, which are uncorrelated with the CSIs $h_{ab}$ and $h_{ba}$ of the main legitimate channel between Alice and Bob, as seen in Fig. 22. Since Alice and Bob both estimate the main channel by themselves without exchanging their estimated CSIs $\hat h_{ab}$ and $\hat h_{ba}$ over the air, it is impossible for Eve to acquire the main channel's CSI for deriving and duplicating the secret keys. The secret key extraction and agreement process based on the physical characteristics of the main channel is capable of achieving reliable information-theoretic security without resorting to a fixed key management infrastructure [149].

The research of physical-layer key generation and agreement can be traced back to the middle of the 1990s [150], [151], where the feasibility of generating secret keys based on the wireless channel's CSI was shown to achieve reliable information-theoretic security without devising any practical key extraction algorithms. To this end, a RSS based secret key extraction algorithm was proposed in [152] by exploiting RSS measurements of the main channel in order to generate secret bits for an IEEE 802.11 network in an indoor wireless environment. In [153], Jana \emph{et al.} further investigated the key generation rate of RSS based secret key extraction in diverse wireless environments. It was shown in [153] that it is possible to generate secret bits at a sufficiently high rate based on the wireless channel variations in highly dynamic mobile scenarios. However, in static environments, where the network devices are fixed, the rate of bits generated is too low to be suitable for a secret key, which is due to the lack of random variations in the wireless channels.

{{To this end, Gollakota and Katabi [154] proposed the so-called iJam approach, which processed the desired transmit signal in a specific manner that still allows the legitimate receiver to decode the desired signal, but prevents the potential eavesdroppers from decoding it. The iJam scheme renders the secret key generation both fast and independent of the wireless channel variations. Furthermore, a testbed was also developed in [154] for implementing the iJam technique using USRP2 radios and the IEEE 802.11 specifications. The associated experimental results demonstrated that the iJam scheme was indeed capable of generating the physical-layer secret keys faster than conventional approaches. To be specific, the iJam scheme generated secret keys at a rate of 3-18 Kb/s without any measurable disagreement probability, whereas the conventional approaches exhibited a maximum generation rate of 44 bits/s in conjunction with a 4\% bit disagreement probability between the legitimate transmitter and receiver. More recently, an extension of the RSS based key extraction from a twin-device system to a multi-device network was studied in [155], where a collaborative key generation scheme was proposed for multiple devices by exploiting the RSS measurements and then experimentally validating it in both indoor and outdoor environments.}}

\begin{figure}
  \centering
  {\includegraphics[scale=0.55]{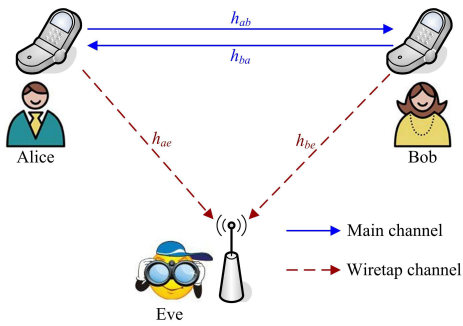}\\
  \caption{{{A wireless system consisting of two legitimate transceivers (Alice and Bob) in the presence of an eavesdropper (Eve).}}}\label{Fig22}}
\end{figure}

\begin{table*}
  \centering
  \caption{{{Major Physical-Layer Secret Key Generation Techniques.}}}
  {\includegraphics[scale=0.75]{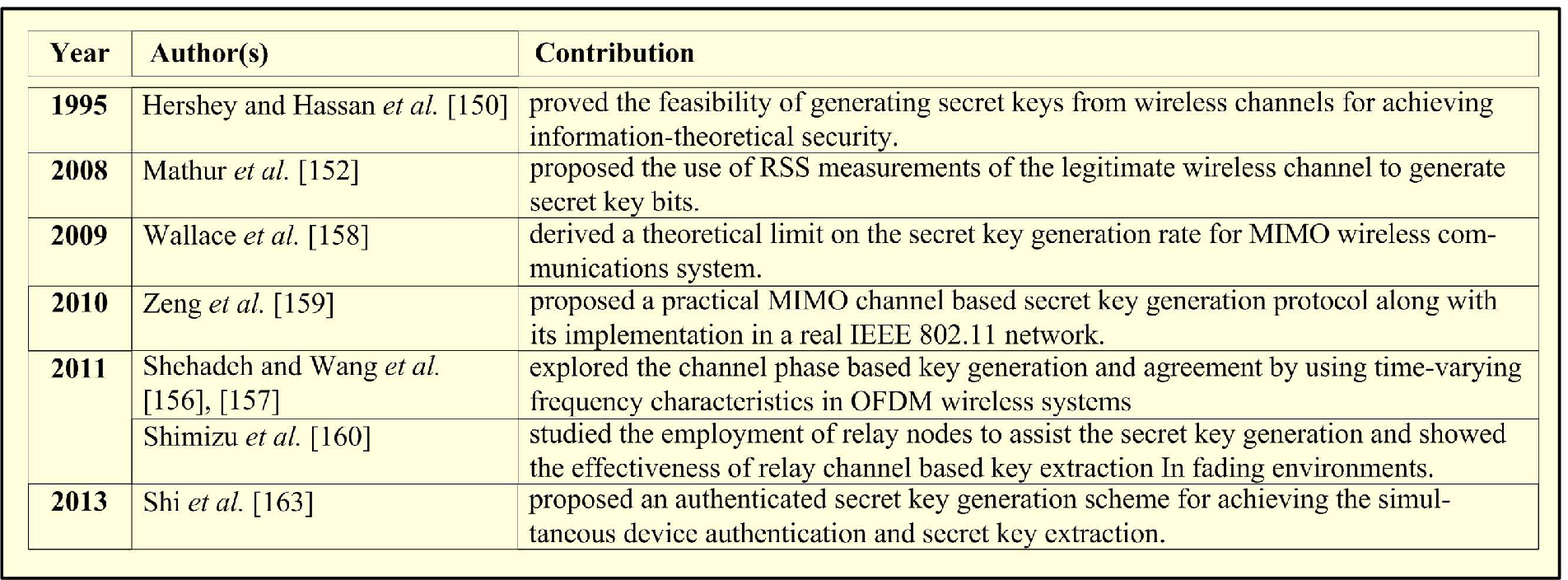}\label{Tab11}}
\end{table*}

Although it is feasible to exploit the RSS for wireless secret key extraction and agreement, the RSS based methods have a low key generation rate, which limits their applications in practical wireless systems. In order to alleviate this problem, the channel phase may also be considered as an alternative means of assisting the generation of secret keys, which is capable of beneficially exploiting the phase measurements across different carriers and thus enhances the secret key generation rate. In [156], Shehadeh \emph{et al.} proposed a channel phase based key agreement scheme, which generates secret bits from the time-varying frequency-domain characteristics in an OFDM based wireless system. More specifically, the OFDM system's subcarrier phase correction process was studied in the context of secret key generation, showing that the employment of higher fast Fourier transform (FFT) sizes is potentially capable of improving the secret bit generation rate. Additionally, the authors of [157] employed multiple randomized channel phases for conceiving an efficient key generation scheme, which was evaluated through both analytical and simulation studies. This solution was found to be highly flexible in the context of multiuser wireless networks and increased the secret key generation rate by orders of magnitude. {{It has to be pointed out that exploiting the phase measurements across multiple OFDM subcarriers is beneficial in terms of increasing the attainable key generation rate. However, the channel phase extracted by a pair of legitimate devices is unlikely to be reciprocal due to the different hardware characteristics of the different devices. This non-reciprocity embedded in the phase measurements results in a high disagreement rate for the legitimate devices during the generation of secret keys.}}

As an alternative, multiple-input multiple-output (MIMO) techniques used by the legitimate transceivers are capable of significantly increasing the channel's randomness, which can be exploited for secret key generation and agreement, leading to the concept of MIMO based key generation. In [158], a theoretical characterization of the MIMO based key generation was explored in terms of deriving a performance limit on the number of secret key bits generated per random channel realization, assuming that the main channel and the wiretap channel are Rayleigh distributed. As a further development, Zeng \emph{et al.} proposed a practical multiple-antenna based secret key generation protocol in [159], which was implemented for an IEEE 802.11 network in both indoor and outdoor mobile environments. It was also shown in [160] that even if an eavesdropper is capable of increasing the number of its antennas, it cannot infer more information about the secret keys generated from the main channel. {{However, the secrecy improvement of MIMO based secret key generation is achieved at the cost of an increased system complexity, since more computing and memory resources are required for estimating the MIMO channel, as the number of transmit/receive antennas increases.}}  In order to further improve the reliability and efficiency of secret key generation, the employment of relay nodes was investigated in [160] for assisting the secret key generation in two different scenarios, namely in conjunction with a single-antenna aided relay and a multiple-antenna assisted relay, respectively. It was demonstrated in [160] that the relay channel based key generation method is capable of substantially improving the key generation rate in Rayleigh fading environments. {{Although the relay nodes can be exploited for enhancing the key generation rate, they may become compromised by an adversary aiming for launching malicious activities. Hence, it is of interest to explore the security issues associated with untrusted relays as well as the corresponding countermeasures.}}

{{It is worth mentioning that the success of the aforementioned physical-layer key generation solutions rely on the assumption that the main channel between the transmitter and legitimate receiver is reciprocal and uncorrelated with the wiretap channel experienced at an eavesdropper located more than half-a-wavelength away from the legitimate receiver. However, this assumption has not been rigorously evaluated in the open literature and indeed, it maybe invalid in some practical scenarios, which do not experience extensive multipath scattering. It was shown in [161] that in reality a strong correlation may be encountered between the main channel and the wiretap channel, even when the eavesdropper is located significantly more than half-a-wavelength away from the legitimate receiver. In [161], the authors demonstrated that a so-called passive inference attacker is potentially capable of exploiting this correlation for inferring a part of the secret keys extracted between a pair of legitimate devices. Additionally, in [162], Eberz \emph{et al.} presented a practical man-in-the-middle (MITM) attack against the physical-layer key generation and showed that the MITM attack can be readily launched by impersonating the legitimate transmitter and receiver as well as by injecting the eavesdropper's data packets. It was demonstrated in [162] that the MITM attack is capable of imposing intentional sabotaging of the physical-layer key generation by inflicting a high key disagreement rate, whilst additionally inferring up to 47\% of the secret keys generated between the legitimate devices.}}

{{In order to mitigate the effects of MITM attacks, Shi \emph{et al.} [163] examined the potential benefits of simultaneous device authentication and secret key extraction based on the wireless physical-layer characteristics, where an authenticated secret key generation (ASK) scheme was proposed by exploiting the heterogeneous channel characteristics in the context of wireless body area networks. Specifically, in case of simple routine body movements, the variations of wireless channels between line-of-sight on-body devices are relatively insignificant, while the wireless channels between the non-line-of-sight devices fluctuate quiet significantly. The ASK scheme exploits the relatively static channels for reliable device authentication and the dynamically fluctuating channels for secret key generation. Extensive experiments were conducted by using low-end commercial-off-the-shelf sensors, demonstrating that the ASK scheme is capable of effectively authenticating body devices, while simultaneously generating secret keys at a high rate. More importantly, the ASK is resilient to MITM attacks, since it performs the authentication and key generation simultaneously. Consequently, it becomes difficult for an MITM attacker to promptly pass through the authentication phase and to get involved in the resultant key generation phase. The major physical-layer secret key generation techniques are summarized in Table XI at a glance.}}

\section{Wireless Jamming Attacks and Their Counter-measures}
As mentioned earlier, due to the shared nature of radio propagation, wireless transmissions are vulnerable to both the eavesdropping and jamming attacks. In the previous section, we have presented a comprehensive overview of how physical-layer security may be exploited for guarding against eavesdropping. Let us now focus our attention on the family of wireless jamming attacks and their counter-measures in this section. In wireless networks, a jamming attack can be simply launched by emitting unwanted radio signals to disrupt the transmissions between a pair of legitimate nodes.

The objective of a jamming attacker (also referred to as jammer) is to interfere with either the transmission or the reception (or both) of legitimate wireless communications. For example, a jammer may continuously transmit its signal over a shared wireless channel so that legitimate nodes always find the channel busy and keep deferring their data transmissions. This, however, is energy-inefficient, since the jammer has to transmit constantly. To improve its energy efficiency, a jammer may opt for transmitting an interfering signal only when it detects that a legitimate transmitter is sending data. There are many different types of wireless jammers, which may be classified into the following five categories [164]:

1) \emph{constant jammer}, where a jamming signal is continuously transmitted;

2) \emph{intermittent jammer}, where a jamming signal is emitted from time to time;

3) \emph{reactive jammer}, where a jamming signal is only imposed, when the legitimate transmission is detected to be active;

4) \emph{adaptive jammer}, where a jamming signal is tailored to the level of received power at the legitimate receiver;

5) \emph{intelligent jammer}, where weaknesses of the upper-layer protocols are exploited for blocking the legitimate transmission.

Clearly, the first four types of jammers all exploit the shared nature of the wireless medium and can be regarded as wireless physical-layer jamming attacks. By contrast, the intelligent jammer attempts to capitalize on the vulnerabilities of the upper-layer protocols [165], including the MAC, network, transport and application layers. Typically, the network, transport and application layers are defined in the TCP/IP protocols and not specified in wireless standards (e.g., Bluetooth, WLAN, etc.), which are responsible for the PHY and MAC specifications only. The jammers targeting the network, transport and application layers essentially constitute DoS attacks (e.g., Smurf attack, TCP/UDP flooding, Malware attack, etc.), which have been summarized in Section III-C to Section III-E. Let us now discuss the aforementioned five main types of wireless jamming attacks and their counter-measures in a little more detail.

\subsection{Constant Jammer}
Again, the constant jammer continuously transmits a jamming signal over the shared wireless medium. The jamming signal can have an arbitrary waveform associated with a limited bandwidth and constrained power, including but not limited to pseudo-random noise, modulated Gaussian waveforms, or any other signals. The effect of a constant jammer is twofold. On the one hand, it increases the interference and noise level for the sake of degrading the signal reception quality at a legitimate receiver. On the other hand, it also makes a legitimate transmitter always find the wireless channel busy, which keeps preventing the legitimate transmitter from gaining access to the channel. Hence, the constant jammer is capable of disrupting the legitimate communications, regardless of the specific wireless system. However, the constant jammer is energy-inefficient, since it has to continuously transmit a jamming signal.

The basic idea behind detecting the presence of a constant jammer is to identify an abnormal signal received at a legitimate receiver [164], [166]. There are certain statistical tests that can be exploited for the detection of the constant jammer, such as the received signal strength (RSS), carrier sensing time (CST), and packet error rate (PER), etc. To be specific, the RSS test is based on a natural measurement used for detecting the presence of a constant jammer, since the signal strength received at a legitimate node would be directly affected by the presence of a jamming signal. The RSS detector first accumulates the energy of the signal received during an observation time period and then compares the accumulated energy to a predefined threshold to decide as to whether a constant jammer is present or absent. If the accumulated energy is higher than the threshold, implying that a jamming signal may be present, then the presence of a constant jammer is confirmed. As an alternative, the CST can also be used as a measurement for deciding whether a constant jammer is preventing the legitimate transmission or not, since the CST distribution will be affected by the jammer. More specifically, the presence of a jamming signal may render the wireless channel constantly busy and hence might lead to an unusually high CST, which can be used for jammer detection.

Additionally, the PER is defined as the number of unsuccessfully decoded data packets divided by the total number of received packets, which can also be used for detecting the presence of a jamming signal, since the legitimate communications will be severely corrupted by the constant jammer, leading to an unduly high PER. Normally, the legitimate wireless communications links operating in the absence of a jammer should have a relatively low PER (e.g., lower than $0.1$). Indeed, it was shown in [166] that even in a highly congested network, the PER is unlikely to exceed $0.2$. By contrast, in the presence of an effective jammer, the legitimate data transmissions will be overwhelmed by the jamming signal and background noise. This would result in an excessive PER, close to one [166], which indicates that indeed, the PER may be deemed to be an effective measurement for detecting the presence of a constant jammer. Conversely, an ineffective jammer, which only slightly affects the PER, fails to inflict a significant damage upon the legitimate wireless system and thus may not have to be detected for invoking further counter-measures.

Once the presence of a jammer is detected, it is necessary to decide upon how to defend the legitimate transmissions against jamming attacks. Frequency hopping is a well-known classic anti-jamming technique [167]-[169], which rapidly changes the carrier frequency with the aid of a pseudo-random sequence known to both the transmitter and receiver. The frequency hopping regime can be either proactive or reactive. In proactive frequency hopping, the transmitter will proactively perform pseudo-random channel switching, regardless of the presence or absence of the jammer. Hence, proactive hopping does not have to detect the presence of a jammer. By contrast, reactive frequency hopping starts switching to a different channel only when the presence of a jamming signal is detected. Compared to proactive hopping, reactive hopping has the advantage of requiring a reduced number of frequency hops for achieving a certain level of secrecy. Overall, frequency hopping is highly resistant to jamming attacks, unless of course the jammer has explicit knowledge of the pseudo-random hopping pattern. Typically, cryptographic techniques are used for generating the pseudo-random hopping pattern under the control of a secret key that is pre-shared by the legitimate transmitter and receiver.

\subsection{{{Intermittent Jammer}}}
An attacker, which transmits a jamming signal from time to time for the sake of interfering with the legitimate communications, is referred to as an intermittent jammer [170]. The intermittent jammer transmits for a certain time and then sleeps for the remaining time. Typically, increasing the sleeping time would save more energy for the jammer, which of course comes at the cost of a performance degradation in terms of the jamming effectiveness, since less time becomes available for transmitting the jamming signal. The jammer can strike a trade-off between the jamming effectiveness and energy savings by appropriately adjusting the transmit time and sleeping time. Hence, compared to the constant jammer, the intermittent jammer generally reduces the energy consumption, which is attractive for energy-constrained jammers.

Similarly to the constant jammer, the presence of an intermittent jammer will affect the same statistical measurements of the legitimate transmissions, including the RSS, CST and PER, which thus can be used for its detection. After detecting an intermittent jammer, again frequency hopping may be activated for protecting the legitimate transmissions. More specifically, when a legitimate node is deemed to be jammed, it switches to another channel and communicates with its destination over the newly established link.

\subsection{Reactive Jammer}
The reactive jammer starts to transmit its jamming signal only when it detects that the legitimate node is transmitting data packets [171], [172]. This type of jammer first senses the wireless channel and upon detecting that the channel is busy, implying that the legitimate user is active, it transmits a jamming signal for the sake of corrupting the data reception at the legitimate receiver. The success of a reactive jammer depends on its sensing accuracy concerning the legitimate user's status. For example, when the legitimate signal received at a reactive jammer is weak (e.g., due to fast fading and shadowing effects) and hence cannot be detected, the reactive jammer then becomes ineffective in jamming the legitimate transmissions. In contrast to both constant and intermittent jammers that attempt to block the wireless channel regardless of the legitimate traffic activity on the channel, the reactive jammer remains quiet when the channel is idle and starts emitting its jamming signal only when the channel is deemed to be busy. This implies that the reactive jammer is more energy-efficient than both the constant and the intermittent jammers.

The detection of the presence of a reactive jammer is typically harder than that of the constant and intermittent jammers. As discussed above, the constant and intermittent jammers both intend to interfere with the reception of a legitimate data packet as well as to hinder the transmission of the legitimate packets by maliciously seizing the wireless channel. By contrast, a reactive jammer inflicts less damage, since it corrupts the reception without affecting the legitimate transmitter's activity to gain access to the wireless channel. This means that the CST becomes an ineffective measurement for detecting the reactive jammer. Since the reception of legitimate wireless communications will be affected in the presence of a reactive jammer, we can still consider the employment of RSS and PER based techniques for the detection of the reactive jammer. Generally, an abnormal increase of the RSS and/or a surprisingly high PER indicate the presence of a reactive jammer.

An effective technique of preventing a reactive jammer from disrupting communications is to assist the legitimate user in becoming undetectable, because then the jammer remains silent. Direct-sequence spread spectrum (DSSS) [173] techniques spread the radio signal over a very wide frequency bandwidth, so that the signal has a low power spectral density (PSD), which may even be below the background noise level. This makes it difficult for a reactive jammer to differentiate the DSSS modulated legitimate signal from the background noise. In this way, the reactive jammer may become unable to track the legitimate traffic activity and thus cannot disrupt the legitimate transmissions. Additionally, the aforementioned frequency hopping technique is also effective in guarding against a reactive jamming attack, as long as the hopping rate is sufficiently high (e.g., faster than the jammer reacts).

\subsection{Adaptive Jammer}
The adaptive jammer refers to an attacker who can adjust its jamming power to any specific level required for disrupting the legitimate receiver [174]. More specifically, in wireless communication systems, the RSS depends on the time-varying fading. If the main channel spanning from the transmitter to the legitimate receiver is relatively good and the signal arriving at the legitimate receiver is sufficiently strong, the adaptive jammer may have to increase its jamming power for the sake of corrupting the legitimate reception. One the other hand, if the main channel itself is experienced an outage due to a deep fade, then naturally, the legitimate receiver is unable to succeed in decoding its received signal even in the absence of a jammer. In this extreme case, no jamming power is needed for the adaptive jammer. Hence, compared to the constant, intermittent and reactive jammers, the adaptive jammer is the most energy-efficient jamming attacker, which can achieve the highest energy efficiency when aiming for disrupting the legitimate transmissions. It can be observed that the adaptive jammer should have the RSS knowledge of the legitimate receiver for adapting its jamming power, which, however, is challenging for a jammer to obtain in practice, since the main channel's RSS varies in time and it is unknown to the jammer. This limits the application of the adaptive jammer in practical wireless systems. The adaptive jammer usually serves as an idealized optimum jamming attacker for benchmarking purposes.

The detection of an adaptive jammer is challenging in the sense that it will dynamically adjust its jamming power to conceal its existence. Similarly to the reactive jammer, the adaptive jammer transmits nothing if the legitimate transmission is deemed to be inactive, implying that the CST technique is ineffective for detecting the adaptive jammer. Although the RSS and PER based solutions can be employed for detecting the presence or absence of an adaptive jammer, the separate employment of the two individual statistics may be insufficient. As a consequence, Xu \emph{et al.} proposed a so-called consistency check method in [166], which relies on the joint use of the RSS and PER measurements. To be specific, if both the RSS and PER are unexpectedly high, it is most likely that there is a jammer, which results in a high RSS due to the presence of a jamming signal that interferes with the legitimate reception, leading to a high PER. If we encounter a low RSS and a high PER, this implies that the main channel is poor. Moreover, the joint occurrence case of a high RSS and low PER indicates that the legitimate transmissions perform well. Finally, it is unlikely in general, that both the RSS and PER are low simultaneously.

As mentioned above, the adaptive jammer is an idealized adversary who is assumed to have knowledge of the legitimate signal characteristics, including the RSS, carrier frequency and bandwidth, waveform, and so on. In order to guard against such a sophisticated jamming attacker, a simple but effective defense strategy is to evade the adversary. Hence in [166], Xu \emph{et al.} proposed a pair of evasion methods to defend against jamming attacks, namely the channel surfing and the spatial retreating solutions. To be specific, in channel surfing, the legitimate transmitter and receiver are allowed to change their jammed channel to a new channel operating at the link layer, which is a different philosophy from that of frequency hopping operating at the physical layer. The spatial retreating technique enables a jammed wireless node to move away and escape from the jammed area to avoid the jamming signal. In case of spatial retreating, it is crucial to accurately determine the position of jammers, which enables the victims to move away from the jammed area. To this end, in [175], Liu \emph{et al.} proposed an error-minimizing framework for accurately localizing multiple jammers by relying on a direct measurement of the jamming signal strength, demonstrating its advantage over conventional methods in terms of its localization accuracy.

\subsection{Intelligent Jammer}
The jamming attackers discussed so far belong to the family of physical-layer jammers operating without taking into account any upper-layer protocol specifications. By contrast, an intelligent jammer is assumed to have a good understanding of the upper-layer protocols and attempts to jam the vitally critical network control packets (rather than data packets) by exploiting the associated protocol vulnerabilities. This subsection is mainly focused on the jamming of MAC control packets. For example, let us consider the MAC jamming of the IEEE 802.11 standard (also known as Wi-Fi) that is widely used for WLANs [176]. The IEEE 802.11 MAC procedure is referred to as the distributed coordination function (DCF) [177], which is shown in Fig. 23. To be specific, if a source node senses the channel to be idle, it waits for a time period termed as the distributed inter-frame space (DIFS) and then sends a request-to-send (RTS) control frame to an access point (AP). After succeeding in decoding the RTS frame and waiting for a time period called the short inter-frame space (SIFS), the AP will send a clear-to-send (CTS) control frame, which indicates that the AP is ready to receive data packets. Finally, the source node waits for a SIFS time duration and starts transmitting a data packet to the AP, which will send an acknowledgement (ACK) frame after a SIFS time interval to confirm that it successfully decoded the data packet.

\begin{figure}
  \centering
  {\includegraphics[scale=0.65]{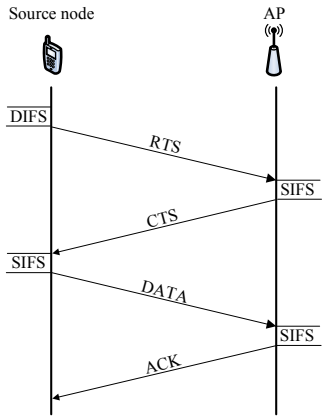}\\
  \caption{{{IEEE 802.11 DCF process.}}}\label{Fig23}}
\end{figure}

In order to block the legitimate communications between the source node and the AP of Fig. 23, an intelligent jammer can simply corrupt the RTS/CTS control frames, rather than data packets, which minimizes its energy consumption. There are several different types of intelligent jamming attackers, including the RTS jammer, CTS jammer and ACK jammer. More specifically, an RTS jammer senses the channel to be idle for a DIFS time period, and then transmits a jamming signal for corrupting a possible RTS packet. By contrast, a CTS jammer attempts to detect the presence of an RTS frame and upon detecting the RTS arrival, it waits for the RTS period plus a SIFS time interval before sending a jamming pulse for disrupting the CTS frame. The CTS jamming strategy will result in a zero throughput for the legitimate transmissions, since no data packets will be transmitted by the source node without successfully receiving a CTS frame. Similarly to the CTS jamming, an ACK jammer also senses the wireless medium. Upon detecting the presence of a packet, it waits for a SIFS time interval at the end of the data packet transmission and then jams the wireless channel, leading to the corruption of an ACK frame. If the source node constantly fails to receive the ACK, it will finally give up transmitting data packets to the AP.

The aforementioned intelligent jammers can be detected by tracing the traffic of MAC control packets to identify abnormal events in terms of sending and/or receiving RTS, CTS and ACK frames. For example, if the AP (or source node) consistently fails to send and receive the RTS, CTS or ACK, it may indicate the presence of an intelligent jammer. As mentioned earlier, an intelligent jammer takes advantage of specific upper-layer protocol parameters for significantly degrading the network performance. In order to defend against such an intelligent jammer, a protocol hopping approach, as a generalization of the physical-layer frequency hopping, was proposed in [178], which allows legitimate nodes to hop across various protocol parameters that the jammer may exploit. A game-theoretic framework was formulated in [178] for modeling the interactions between an intelligent jammer and the protocol functions, which was shown to achieve an improved robustness against intelligent jamming attacks.

\begin{table*}
  \centering
  \caption{Characteristics of Different Jamming Attacks.}
  {\includegraphics[scale=0.65]{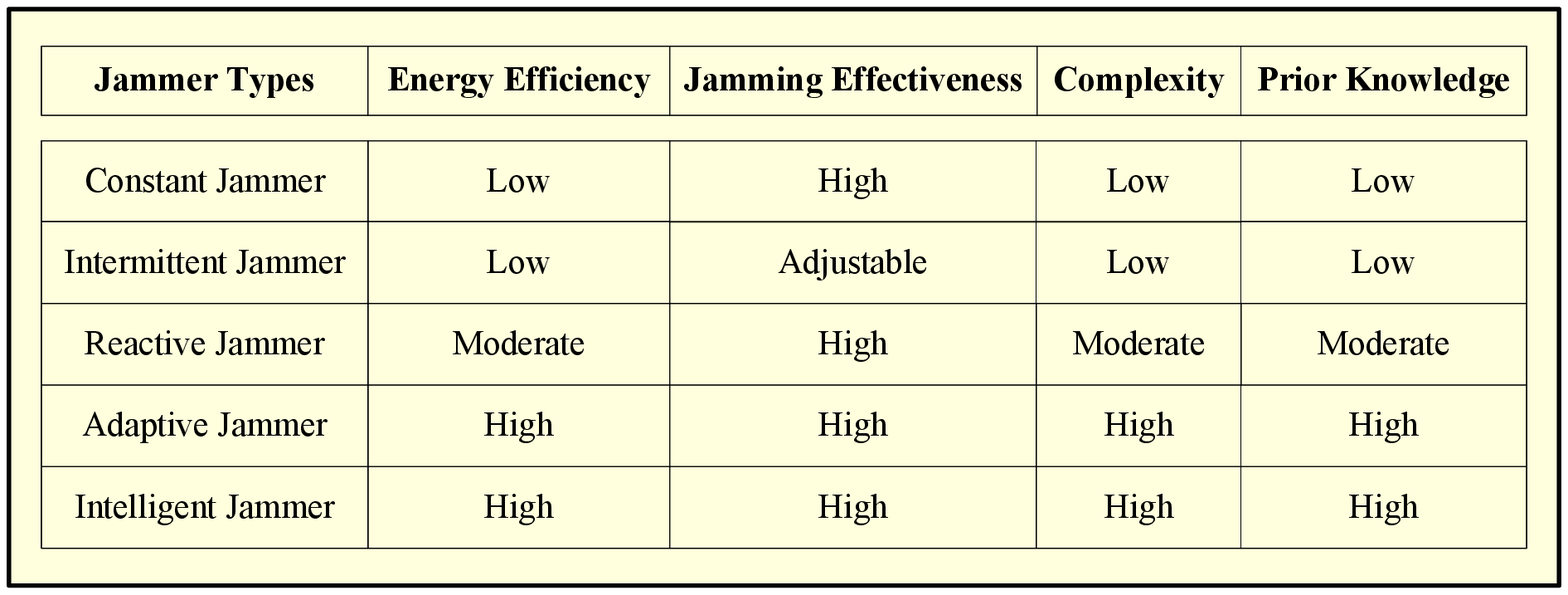}\label{Tab12}}
\end{table*}

Finally, Table XII summarizes the characteristics of the constant, intermittent, reactive, adaptive and intelligent jammers in terms of their energy efficiency, jamming effectiveness, implementation complexity and prior knowledge requirements. As shown in Table XII, the constant and intermittent jammers have a low implementation complexity and required no prior knowledge for effectively jamming the legitimate communications, but their energy efficiency is poor. By contrast, the adaptive and intelligent jammers achieve a high energy efficiency and jamming effectiveness, which however, require some prior knowledge (e.g., the legitimate main channel quality, the protocol parameters, etc.) and have a high complexity. As an alternative, the reactive jammer exhibits a high jamming effectiveness and at the same time, achieves a moderate performance in terms of its energy efficiency, implementation complexity and prior knowledge requirements.

\section{{Integration of Physical-Layer Security into Existing Wireless Networks}}
{{As discussed earlier, the authentication and encryption constitute of a pair of salient techniques adopted in existing wireless security architectures for satisfying the stringent authenticity and integrity requirements of wireless networks (e.g., Wi-Fi, LTE, etc.). Meanwhile, physical-layer security has emerged as a new means of enhancing the security of wireless communications, which is typically considered as a complement to the existing classic authentication and cryptography mechanisms, rather than replacing them [179]. Recently, there has witnessed a growing research efforts devoted to the integration of physical-layer security into the existing body of classic wireless authentication and cryptography [179]-[193]. Below we present an in-depth discussion on the physical-layer authentication and cryptography solutions conceived for wireless networks.}}

{{Authentication constitutes an essential security requirement designed for reliably differentiating authorized nodes from unauthorized ones in wireless networks. Conventionally, the MAC address of a network node has been used for authentication, which is however, vulnerable to MAC spoofing attacks and can be arbitrarily changed for the sake of impersonating another network node. To this end, an increasing research attention has been devoted to the physical-layer authentication [180]-[190] of wireless networks, where either the hardware properties of radio-frequency (RF) based devices (also known as device fingerprints) or the propagation characteristics of wireless channels (e.g. the time-varying fading) have been employed for authentication. This line of work is based on the premise that both the device fingerprints and the wireless channels are unique and non-forgeable by an adversary. To elaborate a little further, the random manufacturing imperfections lead to the fact that a pair of RF devices, even produced by the same manufacturing and packaging process, would have different hardware specifications, as exemplified by their clock timing deviations and carrier frequency offsets (CFO), which can be invoked as unique fingerprints for device identification. Additionally, as mentioned in Section V-E, an adversary located at least at a distance of half-a-wavelength away from the legitimate receiver experiences an independent fading channel. This would make it a challenge for the adversary to predict and mimic the wireless channel between the legitimate users, which can thus be used as a unique link-specific signature for physical-layer authentication.}}

{{Specifically, in [180], a clock timing based hardware fingerprinting approach was proposed for differentiating the authorized devices from spoofing attackers in Wi-Fi networks, which is passive and non-invasive, hence requiring no extra cooperation from the fingerprintee hosts. It was shown by extensive experiments that the clock timing based fingerprint identification is accurate and very effective in differentiating between Wi-Fi devices. Later on, Brik \emph{et al.} [181] considered the joint use of multiple distinctive radiometric signatures, including the frequency error, synchronization frame correlation, I/Q offset, magnitude error, and phase error, which were inferred from the modulated symbols for identifying different IEEE 802.11 network nodes. This technique was termed as PARADIS [181]. Quantitatively, the experimental results demonstrated that by applying sophisticated machine-learning algorithms, PARADIS was capable of differentiating the legitimate nodes with a probability of at least 99\% in a set of more than 130 network nodes equipped with identical 802.11 NICs in the presence of background noise and wireless fading.}}

{{As a design alternative, a CFO based physical-layer authentication scheme was proposed in [182], where the expected CFO was estimated with the aid of Kalman filtering fed with previous CFO estimates. Then, the expected CFO was compared to the current CFO estimate in order to determine, whether the received radio signal obeys a consistent CFO pattern. To be specific, if the difference between the expected CFO and current CFO estimate was higher than a predefined threshold, it indicated the presence of an unknown wireless device. Moreover, the threshold value was adaptively adjusted based on both the background noise level and on the Kalman prediction based errors for the sake of further improving the authentication accuracy. Additionally, a SDR based prototype platform was developed in [182] for validating the feasibility of the proposed CFO based wireless device authentication in the face of multipath fading channels.}}

{{In addition to the device fingerprint based authentication solutions [180]-[182], the wireless channel is also considered as an effective metric for device authentication [183]-[187]. Specifically, in [183], Xiao \emph{et al.} studied the employment of channel probing and responses for determining whether an unauthorized user is attempting to invade a wireless network. The reliability of the proposed CSI based authentication scheme was analyzed in the face of complex Gaussian noise environments. The simulation results relying on the ray-tracing tool WiSE validated the efficiency of the CSI based authentication approach under a range of realistic practical channel conditions. However, this approach is vulnerable to the so-called mimicry attacker, which is able to forge a CSI signature, as long as the attacker roughly knows the radio signal at the legitimate receiver's location. In order to guard against the mimicry attack, a time-synchronized link signature was presented in [184] by integrating the timing factor into the wireless physical-layer features. The provided experimental results showed that the proposed time-synchronized physical-layer authentication is indeed capable of mitigating the mimicry attack with a high probability.}}

{{More recently, in [185], the angle-of-arrival (AoA) information was exploited as a highly sensitive physical-layer signature for uniquely identifying each client in IEEE 802.11 networks. To be specific, a multi-antenna AP was relied upon, in order to estimate all the directions a client's radio signals arrive from. Once spotting a suspicious transmission, the AP and the client initiate an AoA signature-based authentication protocol for mitigating the attack. It was shown in [185] that the proposed AoA signature-based authentication scheme was capable of preventing 100\% of Wi-Fi spoofing attacks, whilst maintaining a false alarm probability of just 0.6\%. As a further development, Du \emph{et al.} [186] examined the extension of physical-layer authentication from single-hop communication networks to dual-hop scenarios by proposing a pair of physical-layer authentication mechanisms, namely the PHY-CRAMR and PHY-AUR techniques for wireless networks operating in the presence of an untrusted relay. The security performance of the PHY-CRAMR and PHY-AUR techniques was analyzed by relying on extensive simulations, showing that both schemes are capable of achieving a high successful authentication probability and a low false alarm rate, especially at sufficiently high SNRs.}}

{{It is worth mentioning that all the aforementioned constitutions [180]-[186] have primarily been focused on exploiting the device fingerprints or channel characteristics by relying on their intrinsic randomness. However, these stochastic features are beyond our control. As a consequence, in [187]-[189] Yu \emph{et al.} explored the benefits of a sophisticated deliberate fingerprint embedding mechanism for physical-layer `challenge-response' authentication, which facilitated striking flexible performance trade-offs by design. More precisely, a stealthy fingerprint was superimposed onto the data in the deliberate fingerprinting mechanism, whilst additionally both the data and an authentication message were transmitted separately by relying on conventional tag-based authentication methods [190]. Naturally, the authentication message used in conventional tag-based methods reduces the spectral efficiency, whilst at the same time, being exposed to eavesdropping. By contrast, a deliberately embedded fingerprint can be designed by ensuring that it has high spectral efficiency and remains impervious to eavesdropping. It was shown in [187]-[189] that a compelling trade-off between the stealth, security and robustness can be struck by the deliberate fingerprint embedding based approach in wireless fading environments.}}

{{Having presented a range of physical-layer authentication techniques [180]-[190], let us now consider the integration of physical-layer security with classic cryptographic approaches [179], [191]-[193]. Traditionally, the cryptographic techniques relying on secret keys have been employed for protecting the communication confidentiality. However, the distribution and management of secret keys remains quite a challenging task in wireless networks. To this end, Abdallah \emph{et al.} [179] have investigated the subject of physical-layer cryptography by exploiting the existing automatic repeat request (ARQ) protocol for achieving the reliable exchange of secret keys between the legitimate users without any information leakage to passive eavesdroppers. Specifically, in [191], the secret bits were distributed across the ARQ packets and only the 1-bit ACK/NACK feedback from the legitimate receiver was exploited for key sharing. It was shown in [191] that a useful non-zero secrecy rate can be achieved even when the wiretap channel spanning from the source to the eavesdropper has a better condition than the legitimate main channel.}}

{{Additionally, Xiao \emph{et al.} studied the benefits of ARQ mechanisms in terms of generating so-called \emph{dynamic secrets} by taking advantage of the inevitable information loss in error-prone wireless communications, where the dynamic secrets are constantly extracted from the communication process with the aid of hash functions \footnote{A hash function is any function that is capable of converting an input data of variable size to an output data of fixed size.}. It was shown in [192] that the dynamic secret mechanism is complementary to the family of existing security protocols and it has the benefit of being time-variant, hence remains hard to reveal. However, in [191] and [192], the ARQ feedback was assumed to be perfectly received and decoded without errors, which may not be practical due to the presence of hostile channel impairments. In order to make these investigations more realistic, the authors of [193] modeled the practical ARQ feedback channel as a correlated erasure channel and evaluated both the secrecy outage probability and the secrecy capacity of ARQ aided physical-layer cryptography. It was shown in [193] that a significant secrecy improvement can be achieved even when the eavesdropper's channel conditions are unknown to the legitimate users.}}

\begin{figure}
  \centering
  {\includegraphics[scale=0.65]{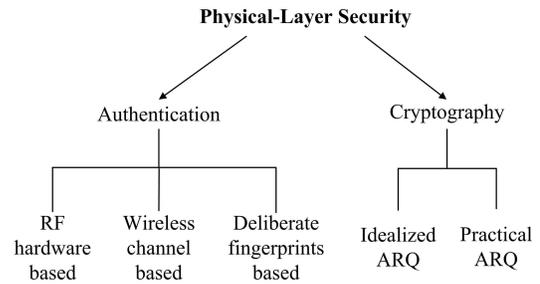}\\
  \caption{{{{{Classification of physical-layer authentication and cryptography.}}}}}\label{Fig23}}
\end{figure}

{{In summary, we have presented an in-depth review concerning the integration of physical-layer security with classic wireless security mechanisms, including physical-layer authentication and cryptography. As shown in Fig. 24, the physical-layer authentication techniques may be classified into three categories, namely the RF hardware properties based, wireless channel characteristics based, and deliberate fingerprint based authentication approaches [180]-[189]. Meanwhile, in the subject of physical-layer cryptography, the existing research efforts [191]-[193] have mainly been focused on exploiting the classic ARQ protocol for securing the exchange of secret keys between legitimate users, where even the realistic practical ARQ feedback associated with transmission errors has been considered.}}

\section{Open Challenges and Future Work}
This section presents a range of challenging open issues and future directions for wireless security research. As mentioned in the previous sections, extensive research efforts have been devoted to this subject, but numerous challenges and issues still remain open at the time of writing.

\subsection{Mixed Attacks in Wireless Networks}
Most of the physical-layer security research [119]-[193] only addressed the eavesdropping attacks, but has neglected the joint consideration of different types of wireless attacks, such as eavesdropping and denial-of-service (DoS) attacks. It will be of particularly importance to explore new techniques of jointly defending against multiple types of wireless attacks, which may be termed as mixed wireless attacks. In order to effectively guard against mixed attacks including both eavesdropping and DoS attacks, we should aim for minimizing the detrimental impact of interference inflicted by DoS attacks on the legitimate transmission. The security defense mechanism should not only consider the CSI of the interfering ink spanning from the DoS attacker to the legitimate receiver, but ideally should also take into account the CSI of the wiretap link between the legitimate transmitter and the eavesdropper, in addition to the CSI of the main link from the legitimate transmitter to the legitimate receiver. It will be of interest to investigate the security defense mechanisms in different scenarios in the presence of both full and partial knowledge of the CSI of the main link as well as of the interfering link and that of the wiretap link. The full CSI based scenario will provide a theoretical performance upper bound as a guide for developing new signal processing algorithms to guard against mixed attacks. Moreover, considering the fact that the eavesdropper remains silent and the CSI of the wiretap channel is typically unknown, it is of practical interest to conceive security protocols for the scenario, where the eavesdropper's CSI is
unavailable.

\subsection{Joint Optimization of Security, Reliability and Throughput}
Security, reliability and throughput constitute the main driving factors for the research and development of wireless networks [194]. In conventional wireless systems, the mechanisms assuring security, reliability and throughput are designed individually and separately, which is however potentially suboptimal, since the three factors are coupled and affect each other [195]. For example, the reliability and throughput of the main link can be improved by increasing the source's transmit power, which however also increases the capacity of the wiretap channel spanning from the source to the
eavesdropper and raises the risk that the eavesdropper succeeds in intercepting the source message through the wiretap link. Similarly,
although increasing the data rate at SN improves the security level by reducing the probability of an intercept event, it comes at the expense of a degradation in transmission reliability, since the outage probability of the main link increases for higher data rates. Therefore, it is necessary to investigate the joint optimization of security, reliability and throughput for the sake of maintaining secure, reliable and high-rate wireless communications, which is an open challenge to be solved in the future. The goal of the joint optimization is to maximize the wireless security performance under the target reliability and throughput requirements. For example, convex optimization and game theory may be considered for formulating and solving the security-reliability-throughput tradeoff in wireless networks.

\subsection{Cross-Layer Wireless Security Design and Analysis}
Presently, cross-layer aided security design is in its infancy. The goal of wireless cross-layer aided security design is to enable efficient information exchange among different protocol layers for the sake of improving the level of wireless security with minimal network overhead. In general, wireless networks adopt the layered OSI protocol architecture that consists of the physical layer, MAC layer, network layer, transport layer and application layer. Traditionally, the aforementioned protocol layers have been protected separately in order to meet their individual communications security requirements, including their authenticity, integrity and confidentiality [10]. However, these traditional layered security mechanisms are potentially inefficient, since each protocol layer introduces additional computational complexity and latency. For example, in order to meet the authenticity requirements, the existing wireless networks typically adopt multiple authentication approaches at different layers, including MAC-layer authentication, network-layer authentication and transport-layer authentication. The employment of multiple separate authentication mechanisms at different protocol layers improves the security level of wireless networks, which, however, comes at the expense of a high complexity and latency. As a consequence, it will be of high interest to explore the benefits of cross-layer aided wireless security in order to guard against the above-mentioned mixed wireless attacks. Intuitively, the physical-layer characteristics and properties of wireless channels may also be further exploited by the upper-layer security algorithms, including the user authentication, secret key generation and data protection algorithms. It is anticipated that the cross-layer security framework will further improve the wireless security at a reduced cost, as compared to the traditional layered security mechanisms.

\subsection{Physical-Layer Security for the Emerging 5G Systems}
Given the proliferation of smart devices and the increasing demand for multimedia communications, the amount of mobile traffic has substantially grown in recent years and it may soon exceed the capacity of the operational fourth-generation (4G) mobile communications systems [196]. To meet this challenging requirement, substantial efforts have been devoted to the research and development of the fifth-generation (5G) mobile systems [197]-[200] relying on advanced wireless technologies, such as the massive MIMOs and millimeter wave (mmWave) solutions. Meanwhile, it is expected that a strict security requirement is desired for the 5G systems, since more and more sensitive information (e.g., financial data, personal emails and files) will be transmitted wirelessly [196]-[198]. To this end, physical-layer security as a beneficial complement to conventional security mechanisms will have a great potential in the context of 5G systems. For instance, by deploying a large number of antennas in 5G systems, the aforementioned artificial noise and beamforming assisted techniques can be readily utilized for improving the transmission performance of legitimate users, while degrading the reception quality of eavesdroppers. However, the application of massive MIMOs for enhancing the physical-layer security also has its own challenges to be addressed, such as the deleterious effects of pilot contamination, power allocation and channel reciprocity [196]. Therefore, it is of high importance to explore the opportunities and challenges of combining the physical-layer security techniques with 5G enabling technologies, such as massive MIMOs and mmWave solutions.

\subsection{Field Experiments for Physical-Layer Security Investigations}
As discussed above, there are various physical-layer security schemes including the artificial noise, beamforming and diversity aided security enhancement approaches, which have been shown to be effective in terms of improving both the secrecy capacity and the secrecy outage probability of wireless communications [126]-[146]. However, their security benefits have so far only been shown theoretically relying on idealized simplifying assumptions, such as the availability of perfect CSI knowledge [201], [202]. By considering the artificial noise based method as an example, the accurate CSI of the main channel is required for the appropriate design of an artificial noise, so that the legitimate receiver remains unaffected by the noise, while the eavesdropper is interfered with. However, regardless of the specific channel estimation methods used [148], [149], estimation errors always contaminate the estimation of the CSI, hence the perfect CSI estimation cannot be achieved in practical wireless systems. Given an inaccurate CSI, it is impossible to devise an artificial noise that only interferes with the eavesdropper without affecting the legitimate receiver. Typically, the less accurate the CSI of the main channel, the more interference is received at the legitimate receiver, hence resulting in a degradation of the wireless physical-layer security. Similarly, the CSI estimation errors would also cause a performance degradation for the beamforming- and diversity-aided security approaches. However, it remains unclear to what extent the CSI estimation error affects the attainable physical-layer security performance in terms of the secrecy capacity and secrecy outage probability. It will be of great benefit to conduct field experiments for the sake of verifying the efficiency of various physical-layer security approaches in real wireless communications systems in the presence of both jamming and eavesdropping attacks.

\section{Conclusions}
In this paper, we have presented a survey of the wireless security challenges and defense mechanisms conceived for protecting the authenticity, confidentiality, integrity and availability of wireless transmissions against malicious attacks. We have discussed the range of wireless attacks and security threats potentially experienced at different protocol layers from the application layer to the physical layer, which are classified into application-layer and transport-layer attacks, network-layer, MAC-layer as well as physical-layer attacks. Then, existing security paradigms and protocols conceived for guarding against the different protocol layers' attacks have been reviewed in the context of several widely deployed wireless networks, including the Bluetooth, Wi-Fi, WiMAX and LTE. Bearing in mind that wireless transmissions are highly vulnerable to eavesdropping attacks owing to the broadcast nature of radio propagation, we have also discussed the state-of-the-art in physical-layer security, which is emerging as a promising paradigm of defending the wireless transmissions against eavesdropping attacks by exploiting the physical-layer characteristics of wireless channels. More specifically, several physical-layer security techniques, including information-theoretic security, artificial noise aided security, security-oriented beamforming, diversity-assisted security and physical-layer secret key generation approaches have been presented as well as compared. Additionally, we have summarized various types of wireless jamming attacks along with their detection and prevention techniques. {{Finally, we have also discussed the integration of physical-layer security into classic wireless authentication and cryptography, as well as}} highlighted a range of open challenges to be addressed:

\begin{itemize}
\item
\emph{Mixed wireless attacks}, where new theories and techniques have to be explored for jointly defending the system against multiple types of wireless attacks;

\item
\emph{Joint optimization of security, reliability and throughput}, where an efficient wireless transmission mechanism has to be developed by maximizing the security performance under specific target reliability and throughput requirements;

\item
\emph{Cross-layer wireless security design}, where a cross-layer security framework has to be investigated for the sake of improving the wireless security at a reduced security overhead and latency as compared to the conventional layered security mechanisms, where the OSI protocol layers are protected separately.

\item
{{\emph{5G physical-layer security}, where the combination of physical-layer security with 5G enabling technologies, such as massive MIMOs and mmWave solutions has to be explored for the sake of meeting the strict security requirements imposed by the emerging 5G communication systems.}}

\item
{{\emph{Field experiments}, where the efficiency of various physical-layer security approaches has to be verified with the aid of field tests in real wireless communications systems without the idealized simplifying assumptions that are routinely used in theoretical studies.}}

\end{itemize}

Based on the solutions presented throughout this paper, we provide some general guidelines for wireless communications security design:

\begin{itemize}
\item
Wireless networks are based on the layered OSI protocol architecture that consists of the application layer, transport layer, network layer, MAC layer and physical layer. Each layer shall be protected in order to meet the security requirements of wireless networks. Bearing in mind the fact that the different layers support different protocols and exhibit different security vulnerabilities, the security mechanisms invoked by the different wireless protocols should be customized so as to guard against malicious attacks as efficiently as possible.

\item
The security paradigms, such as user authentication and data encryption used in conventional wireless networks are typically designed separately at the different protocol layers, which, however, results in high latency and overhead. To this end, cross-layer security would be a candidate for protecting wireless networks against various attacks. To be specific, the physical-layer characteristics of wireless channels may be potentially considered and exploited for designing or customizing the upper-layer security algorithms, including the identity authentication, key generation and so on.

\item
The secrecy capacity of wireless communications in the presence of eavesdroppers may be severely degraded due to the time-varying multipath fading effects, which may be significantly mitigated by exploiting a range of diversity-aided techniques, such as time diversity, frequency diversity and spatial diversity. For example, spatial diversity can be achieved for the sake of attaining the wireless secrecy capacity improvements by using multiple antennas at the legitimate transmitter and/or the legitimate receiver.

\item
The multiuser scheduling may be employed for the sake of achieving the multiuser diversity gain to improve the wireless secrecy capacity. Additionally, multiuser MIMO may be invoked for further improvements in secrecy capacity, which combines the benefits of the multiuser diversity as well as the MIMO diversity and multiplexing gains.

\item
Artificial noise generation techniques may be used for improving the wireless physical-layer security against eavesdropping attacks by ensuring that only the eavesdropping attackers are adversely affected by the artificial noise, while the legitimate receiver is unaffected. In order to maximize the security benefits of using the artificial noise assisted method, the power sharing between the desired information-bearing signals and the artificial noise should be given careful attention.

\item
It is worth mentioning that additional power resources are dissipated in generating the artificial noise to confuse the eavesdropper. Given a fixed total transmit power, increasing the artificial noise power is capable of deteriorating the eavesdropper's channel condition, it, however, comes at the cost of performance degradation of the legitimate receiver, since less transmit power is available for the desired signal transmission. Hence, the power allocation between the artificial noise and desired signal should be carefully considered for the sake of optimizing the wireless physical-layer security.

\item
Beamforming approaches may also be invoked for improving the wireless security design, which enables the legitimate transmitter to send its information signal in a particular direction to the legitimate receiver by ensuring that the signal received at the legitimate receiver experiences constructive interference, whereas that received at an eavesdropper experiences destructive interference. Moreover, combining the beamforming and the artificial noise aided techniques would further enhance the wireless physical-layer security against eavesdropping attacks.

\item
The security benefits of the artificial noise generation and beamforming techniques are typically maximized at the cost of a throughput or reliability degradation. The conventional mechanisms assuring the security, reliability and throughput are designed separately, which are not optimized jointly. It is therefore suggested to consider the joint optimization of security, reliability and throughput for secure wireless communications. For example, the joint optimization problem may be addressed by maximizing the wireless security performance under the target reliability and throughput requirements.

\item
CSIs of the main channel and/or the wiretap channel are essential in assuring the wireless physical-layer security against eavesdropping attacks. Both the artificial noise and beamforming aided security approaches rely on the CSIs. The accuracy of estimated CSIs has a significant impact on the physical-layer security performance (e.g., the secrecy capacity). It is thus suggested to employ the pilot-based channel estimation approaches, rather than the semi-blind or blind channel estimation, for the sake of obtaining accurate CSIs.

\item
Performing the accurate channel estimation increases the complexity of the wireless transceiver, especially in fast-fading channels, where the CSI has to be estimated more frequently and the CSI feedback rate has to be increased, resulting in higher transmission overhead in terms of both bandwidth and power. Hence, some balanced system design principles are suggested, where the wireless secrecy capacity may be sacrificed with the intention of reducing the CSI estimation complexity and feedback overhead.

\item
{{Physical-layer key generation and agreement techniques are capable of generating secret keys based on the random variations of wireless fading channels for securing wireless networks without the need for a fixed key management infrastructure. However, in static environments, where the wireless nodes are stationary, the channel fading would fluctuate slowly, resulting in a limited number of secret bits to be generated. In these cases, we may consider the employment of MIMO-aided and relay-assisted methods for enhancing the channel's randomness for the sake of improving the secret key generation rate.}}

\item
{{Wireless communications can be disrupted by a jammer at the physical layer by transmitting an interfering signal. Although the FHSS technique is capable of effectively guarding against some of the known physical-layer jamming attacks, the frequency hopping pattern agreement between the legitimate transceivers is challenging in wireless networks. It is therefore advisable to combine FHSS with physical-layer security by exploiting the characteristics of wireless channels for the frequency hopping pattern agreement.}}

\end{itemize}

\ifCLASSOPTIONcaptionsoff
  \newpage
\fi

\appendices

\begin{IEEEbiography}[{\includegraphics[width=1in,height=1.25in]{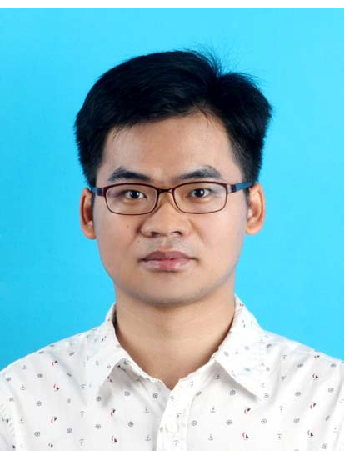}}]{Yulong Zou} (SM'13) is a Full Professor and Doctoral Supervisor at the Nanjing University of Posts and Telecommunications (NUPT), Nanjing, China. He received the B.Eng. degree in information engineering from NUPT, Nanjing, China, in July 2006, the first Ph.D. degree in electrical engineering from the Stevens Institute of Technology, New Jersey, USA, in May 2012, and the second Ph.D. degree in signal and information processing from NUPT, Nanjing, China, in July 2012.

His research interests span a wide range of topics in wireless communications and signal processing, including the cooperative communications, cognitive radio, wireless security, and energy-efficient communications. Dr. Zou was awarded the 9th IEEE Communications Society Asia-Pacific Best Young Researcher in 2014 and a co-receipt of the Best Paper Award at the 80th IEEE Vehicular Technology Conference in 2014. He is currently serving as an editor for the IEEE Communications Surveys \& Tutorials, IET Communications, and China Communications. In addition, he has acted as TPC members for various IEEE sponsored conferences, e.g., IEEE ICC/GLOBECOM/WCNC/VTC/ICCC, etc.

\end{IEEEbiography}

\begin{IEEEbiography}[{\includegraphics[width=1in,height=1.25in]{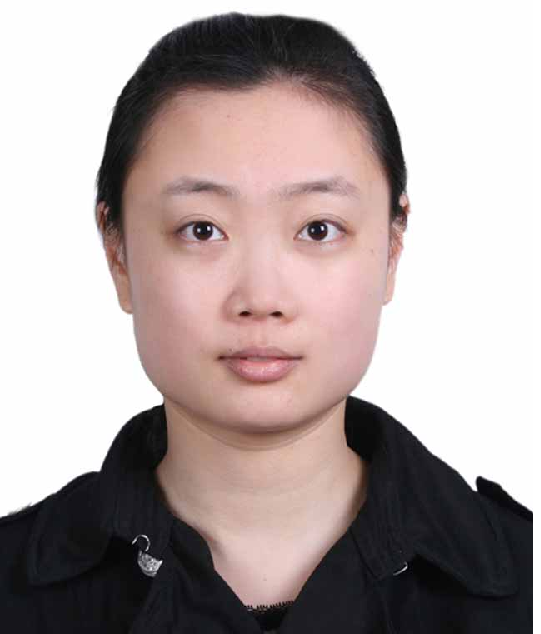}}]{Jia Zhu} is an Associate Professor at the Nanjing University of Posts and Telecommunications (NUPT), Nanjing, China. She received the B.Eng. degree in Computer Science and Technology from the Hohai University, Nanjing, China, in July 2005, and the Ph.D. degree in Signal and Information Processing from the Nanjing University of Posts and Telecommunications, Nanjing, China, in April 2010. From June 2010 to June 2012, she was a Postdoctoral Research Fellow at the Stevens Institute of Technology, New Jersey, the United States. Since November 2012, she has been with the Telecommunication and Information School of NUPT, Nanjing, China. Her general research interests include the cognitive radio, physical-layer security, and communications theory.
\end{IEEEbiography}

\begin{IEEEbiography}[{\includegraphics[width=1in,height=1.25in]{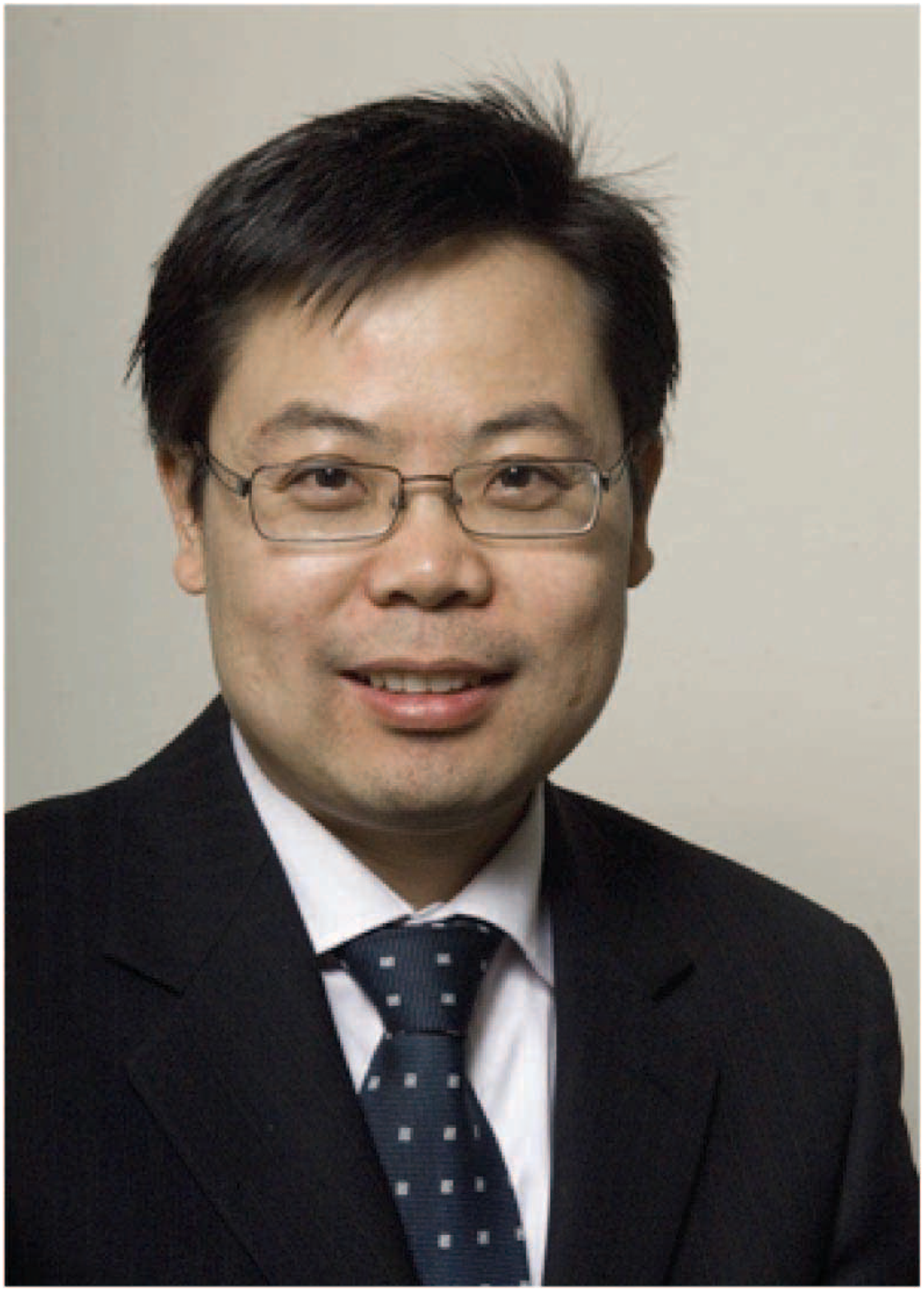}}]
{Xianbin Wang}(S'98-M'99-SM'06) is a Full Professor at The University of Western Ontario and a Canada Research Chair in Wireless Communications. He received his Ph.D. degree in electrical and computer engineering from National University of Singapore in 2001.

Prior to joining Western, he was with Communications Research Centre Canada as Research Scientist/Senior Research Scientist between July 2002 and Dec. 2007. From Jan. 2001 to July 2002, he was a system designer at STMicroelectronics, where he was responsible for system design for DSL and Gigabit Ethernet chipsets. He was with Institute for Infocomm Research, Singapore (formerly known as Centre for Wireless Communications), as a Senior R \& D engineer in 2000. His primary research area is wireless communications and related applications, including adaptive communications, wireless security, and wireless infrastructure based position location. Dr. Wang has over 150 peer-reviewed journal and conference papers on various communications system design issues, in addition to 23 granted and pending patents and several standard contributions.

Dr. Wang is an IEEE Distinguished Lecturer and a Senior Member of IEEE. He was the recipient of three IEEE Best Paper Awards. He currently serves as an Associate Editor for IEEE Wireless Communications Letters, IEEE Transactions on Vehicular Technology and IEEE Transactions on Broadcasting. He was also an editor for IEEE Transactions on Wireless Communications between 2007 and 2011. Dr. Wang was involved in a number of IEEE conferences including GLOBECOM, ICC, WCNC, VTC, and ICME, on different roles, such as symposium chair, track chair, TPC and session chair.

\end{IEEEbiography}

\begin{IEEEbiography}[{\includegraphics[width=1in,height=1.25in]{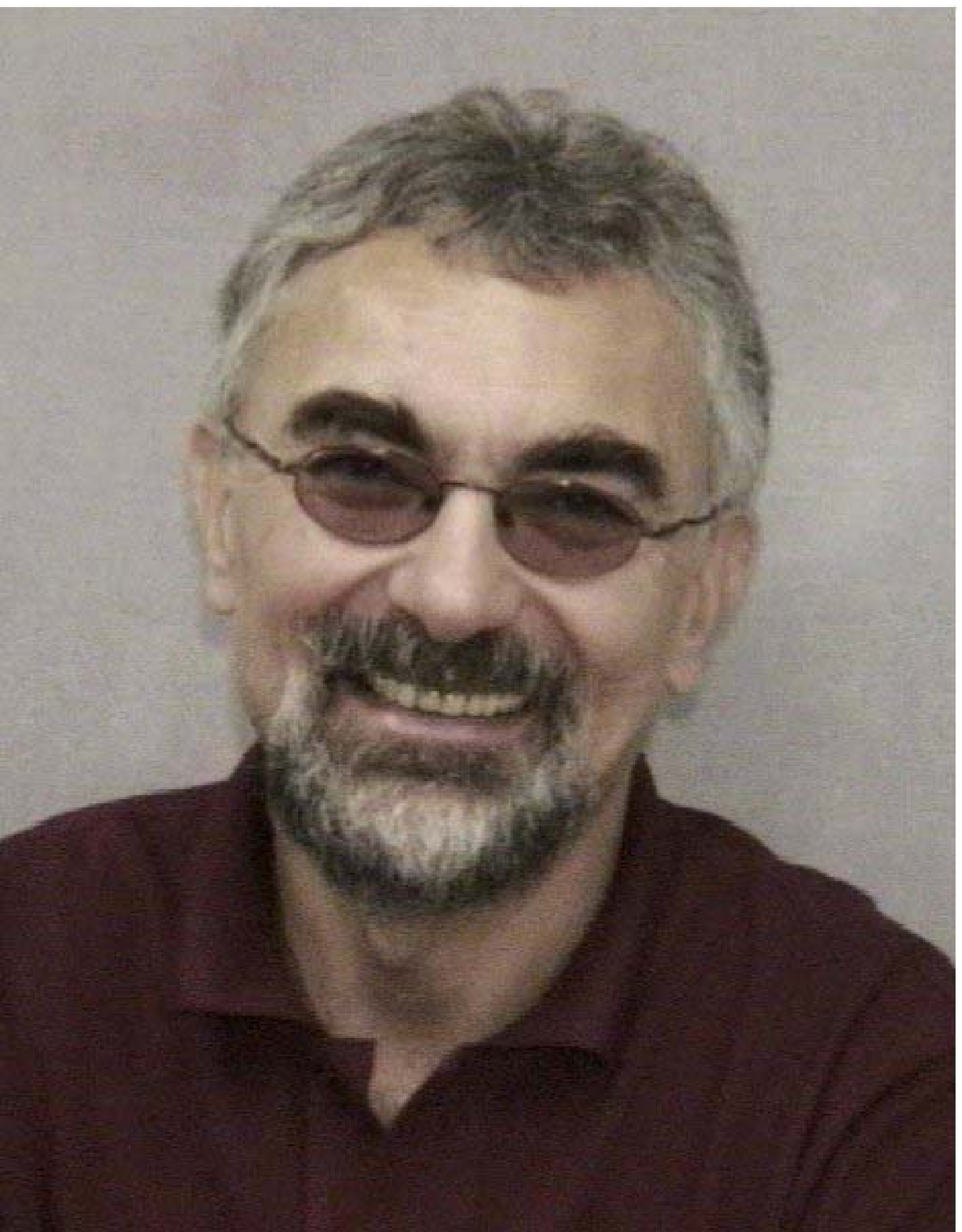}}]{Lajos Hanzo} received his Masters degree in electronics in 1976 and his Doctorate in 1983 from the Technical University of Budapest. In 2010 he was awarded the university's highest honour, namely the Honorary Doctorate ``Doctor Honaris Causa". Since 1986 he has been with the University of Southampton, UK and in 2004 he was awarded the Doctor of Sciences (DSc) degree. During his 38-year career in telecommunications he has held various research and academic posts in Hungary, Germany and the UK. Since 1986 he has been a member of academic staff in the School of Electronics and Computer Science, University of Southampton, UK, where he currently holds the Chair in Telecommunications and he is head of the Communications Research Area. During 2009 - 2012 he was also a Chaired Professor at Tsinghua University, Beijing, China and the EIC of the IEEE Press.

Lajos Hanzo has co-authored 20 John Wiley/IEEE Press books totalling about 10 000 pages on mobile radio communications, and published 1400+ research papers and book chapters at IEEE Xplore. Lajos has 19 000+ citations. He has also organised and chaired major IEEE conferences, such as WCNC'2006, WCNC'2009, VTC'2011, ICC'2013, presented Tutorial/overview lectures at international conferences. He presented a number of named lectures and keynotes.

Lajos is also an IEEE Distinguished Lecturer, a Wolfson Fellow of the Royal Society, as well as a Fellow of both the IEEE and the IEE/IET, Fellow of the Royal Academy of Engineering (FREng). He is acting as a Governor of the IEEE VTS. He has been awarded a number of distinctions, most recently the IEEE Wireless Technical Committee Achievement Award (2007), the IET Sir Monti Finniston Achievement Award across all disciplines of engineering (2008), a Honorary Doctorate of the Technical University of Budapest (2010) and the IEEE Radio Communications Technical Committee Achievement Award (2013). His most recent paper awards are: WCNC'2007 in Hong Kong, ICC'2009 Dresden and ICC'2010 Cape Town, WCNC'2013 Shanghai. For further research please visit www-mobile.ecs.soton.ac.uk.
\end{IEEEbiography}


\begin{thebibliography}{11}

\bibitem{IEEEhowto:1}
O. Aliu, A. Imran, M. Imran, and B. Evans, ``A survey of self
organisation in future cellular networks," \emph{IEEE Communications
Surveys \& Tutorials}, vol. 15, no. 1, pp. 336-361, February 2013.

\bibitem{IEEEhowto:2}
H. ElSawy, E. Hossain, and M. Haenggi, ``Stochastic geometry for
modeling, analysis, and design of multi-tier and cognitive cellular
wireless networks: A Survey," \emph{IEEE Communications Surveys \&
Tutorials}, vol. PP, no. 99, pp. 1-24, June 2013.

\bibitem{IEEEhowto:3}
ITU, ``The world in 2013: ICT facts and figures," January 2013,
available on-line at
http://www.itu.int/en/ITU-D/Statistics/Documents/facts/ICTFactsFigures2013.pdf

\bibitem{IEEEhowto:4}
Symantec Norton Department, ``The 2012 Norton cybercrime report,"
September 2012, available on-line at
http://www.norton.com/2012cybercrimereport.

\bibitem{IEEEhowto:5}
M. M. Rashid, E. Hossain, and V. K. Bhargava, ``Cross-layer analysis
of downlink V-BLAST MIMO transmission exploiting multiuser
diversity," \emph{IEEE Transactions on Wireless Communications},
vol. 8, no. 9, pp. 4568-4579, September 2009.

\bibitem{IEEEhowto:6}
F. Foukalas, V. Gazis, N. Alonistioti, ``Cross-layer design
proposals for wireless mobile networks: A survey and taxonomy,"
\emph{IEEE Communications Surveys \& Tutorials}, vol. 10, no. 1, pp.
70-85, April 2008.

\bibitem{IEEEhowto:7}
R. Jurdak, C. Lopes, and P. Baldi, ``A survey, classification and
comparative analysis of medium access control protocols for ad hoc
networks," \emph{IEEE Communications Surveys \& Tutorials}, vol. 6,
no. 1, pp. 2-16, April 2004.

\bibitem{IEEEhowto:8}
M, Takai, J, Martin, and R. Bagrodia, ``Effects of wireless physical
layer modeling in mobile ad hoc networks," \emph{Proceedings of the
2nd ACM International Symposium on Mobile ad Hoc Networking \&
Computing (MobiHoc)}, Carlifonia, USA, September 2001.

\bibitem{IEEEhowto:9}
C. Saradhi and S. Subramaniam, ``Physical layer impairment aware
routing (PLIAR) in WDM optical networks: Issues and challenges ,"
\emph{IEEE Communications Surveys \& Tutorials}, vol. 11, no. 4, pp.
109-130, December 2009.

\bibitem{IEEEhowto:10}
C. Kolias, G. Kambourakis, and S. Gritzalis, ``Attacks and
countermeasures on 802.16: Analysis and assessment," \emph{IEEE
Communications Surveys \& Tutorials}, vol. 15, no. 1, pp. 487-514,
February 2013.

\bibitem{IEEEhowto:11}
M. Stamp, \emph{Information security: Principles and practice}, 2nd Edition, John Wiley \& Sons, 2011.

\bibitem{IEEEhowto:12}
M. Whitman and H. Mattord, \emph{Principles of information security}, 4th Edition, Delmar Cengage Learning, 2012.

\bibitem{IEEEhowto:13}
Y. Xiao, H.-H. Chen, B. Sun, R. Wang, and S. Sethi, ``MAC security
and security overhead analysis in the IEEE 802.15.4 wireless sensor
networks," \emph{EURASIP Journal on Wireless Communications and
Networking}, DOI:10.1155/WCN/2006/ 93830, 2006.

\bibitem{IEEEhowto:14}
G. Apostolopoulos, V. Peris, P. Pradhan, and D. Saha, ``Securing
electronic commerce: Reducing the SSL overhead," \emph{IEEE
Network}, vol. 14, no. 4, pp. 8-16, July 2000.

\bibitem{IEEEhowto:15}
K. Wong, Y. Zheng, J. Cao, and S. Wang, ``A dynamic user
authentication scheme for wireless sensor networks,"
\emph{Proceedings of the 2006 IEEE International Conference on
Sensor Networks, Ubiquitous, and Trustworthy Computing}, vol. 14,
no. 4, Taichung, Taiwan, June 2006.

\bibitem{IEEEhowto:16}
A. Aziz and W. Diffie, ``Privacy and authentication for wireless
local area networks," \emph{IEEE Personal Communications}, vol. 1,
no. 1, pp. 25-31, August 2002.

\bibitem{IEEEhowto:17}
G. Raju and R. Akbani, ``Authentication in wireless networks,"
\emph{Proceedings of the 40th Annual Hawaii International Conference
on System Sciences}, Hawaii, USA, January 2007.

\bibitem{IEEEhowto:18}
L. Venkatraman and D. P. Agrawal, ``A novel authentication scheme
for ad hoc networks," \emph{Proceedings of the 2000 IEEE Wireless
Communications and Networking Confernce}, Chicago, USA, September
2000.

\bibitem{IEEEhowto:19}
A. H. Lashkari, K. Lumpur, M. Mansoor, and A. S. Danesh, ``Wired
equivalent privacy (WEP) versus Wi-Fi protected access (WPA),"
\emph{Proceedings of the 2009 International Conference on Signal
Processing Systems}, Singapore, May 2009.

\bibitem{IEEEhowto:20}
K. J. Hole, E. Dyrnes, and P. Thorsheim, ``Securing Wi-Fi networks,"
\emph{Computer}, vol. 38, no. 7, pp. 28-34, July 2005.

\bibitem{IEEEhowto:21}
RFC 5246, ``The transport layer security (TLS) protocol version
1.2," August 2008, available on-line at
https://tools.ietf.org/html/rfc5246.

\bibitem{IEEEhowto:22}
RFC 4346, ``The transport layer security (TLS) protocol version
1.1," April 2006, available on-line at
https://tools.ietf.org/html/rfc4346.

\bibitem{IEEEhowto:23}
RFC 2246, ``The transport layer security (TLS) protocol version
1.0," January 1999, available on-line at
https://tools.ietf.org/html/rfc2246.

\bibitem{IEEEhowto:24}
S. Lakshmanan, C. Tsao, R. Sivakumar, and K. Sundaresan, ``Securing
wireless data networks against eavesdropping using smart antennas,"
\emph{Proceedings of The 28th International Conference on
Distributed Computing Systems}, Beijing, China, June 2008.

\bibitem{IEEEhowto:25}
R. Raymond and S. Midkiff, ``Denial-of-service in wireless sensor
networks: Attacks and fefenses," \emph{IEEE Pervasive Computing},
vol. 7, no. 1, pp. 74-81, January 2008.

\bibitem{IEEEhowto:26}
B. Kannhavong, \emph{et al.}, ``A survey of routing attacks in
mobile ad hoc networks," \emph{IEEE Wireless Communications}, vol.
14, no. 5, pp. 85-91, December 2007.

\bibitem{IEEEhowto:27}
U. Meyer and S. Wetzel, ``A man-in-the-middle attack on UMTS,"
\emph{Proceedings of the 3rd ACM workshop on Wireless security},
Philadelphia, USA, October 2004.

\bibitem{IEEEhowto:28}
T. Ohigashi and M. Morii, ``A practical message falsification attack
on WPA," \emph{Proceedings of 2009 Joint Workshop on Information
Security}, Kaohsiung, Taiwan, August 2009.

\bibitem{IEEEhowto:29}
P. Christof, J. Pelzl, and B. Preneel, \emph{Understanding
cryptography: A textbook for students and practitioners}, NY:
Springer-Verlag, 2010.

\bibitem{IEEEhowto:30}
C. Elliott, ``Quantum cryptography," \emph{IEEE Security \& Privacy}, vol. 2, no. 4, pp. 57-61, April 2004.

\bibitem{IEEEhowto:31}
{{Q. Wang, K. Xu, and K. Ren, ``Cooperative secret key generation from phase estimation in narrowband fading channels," \emph{IEEE Journal on Selected Areas in Communications}, vol. 30, no. 9, pp. 1666-1674, September 2012.}}

\bibitem{IEEEhowto:32}
{{Y. Wei, K. Zengy and P. Mohapatra, ``Adaptive wireless channel probing for shared key generation," in \emph{Proceedings of The 30th Annual IEEE International Conference on Computer Communications (INFOCOM 2011)}, Shanghai, China, April 2011.}}

\bibitem{IEEEhowto:33}
A. D. Wyner, ``The wire-tap channel," \emph{Bell System Technical Journal}, vol. 54, no. 8, pp. 1355-1387, 1975.

\bibitem{IEEEhowto:34}
S. K. Leung-Yan-Cheong and M. E. Hellman, ``The Gaussian wiretap channel," \emph{IEEE Transactions Information Theory}, vol. 24, no. 7, pp. 451-456, July 1978.

\bibitem{IEEEhowto:35}
Y. Zou, X. Wang, and W. Shen, ``Intercept probability analysis of
cooperative wireless networks with best relay selection in the
presence of eavesdropping attack," in \emph{Proceedings of The 2013
IEEE Internation Conference on Communications}, Budapest, Hungary,
June 2013.

\bibitem{IEEEhowto:36}
Y. Zou, X. Wang, and W. Shen, ``Eavesdropping attack in
collaborative wireless networks: Security protocols and intercept
behavior," in \emph{Proceedings of The 17th IEEE International
Conference on Computer Supported Cooperative Work in Design},
Whistler, Canada, June 2013.

\bibitem{IEEEhowto:37}
O. Cepheli, T. Maslak, and G. Kurt, ``Analysis on the effects of
artificial noise on physical layer security," \emph{Proceedings of
The 2013 21st Signal Processing and Communications Applications
Conference}, Haspolat, Turkey, April 2013.

\bibitem{IEEEhowto:38}
A. Araujo, J. Blesa, E. Romero, and O. Nieto-Taladriz, ``Artificial
noise scheme to ensure secure communications in CWSN,"
\emph{Proceedings of The 2012 8th International Wireless
Communications and Mobile Computing Conference}, Limassol, Cyprus,
August 2012.

\bibitem{IEEEhowto:39}
N. Romero-Zurita, M. Ghogho, and D. McLernon, ``Outage probability
based power distribution between data and artificial noise for
physical layer security," \emph{IEEE Signal Processing Letters}, vol
19, no. 2, pp. 71-74, Febuaray 2012.

\bibitem{IEEEhowto:40}
J. Wu and J. Chen, ``Multiuser transmit security beamforming in
wireless multiple access channels," \emph{Proceedings of The 2012
IEEE International Conference on Communications}, Ottawa, Canada,
June 2012.

\bibitem{IEEEhowto:41}
H. Wang, Q. Yin, and X. Xia, ``Distributed beamforming for
physical-layer security of two-way relay networks," \emph{IEEE
Transactions on Signal Processing}, vol. 60, no. 7, pp. 3532-3545,
July 2012,

\bibitem{IEEEhowto:42}
Y. Zou, X. Wang, and W. Shen, ``Optimal relay selection for
physical-layer security in cooperative wireless networks,"
\emph{IEEE Journal on Selected Areas in Communications}, vol. 31, no. 10, pp. 2099-2111, October 2013.

\bibitem{IEEEhowto:43}
Y. Zou, X. Wang, and W. Shen, ``Physical-layer security with multiuser scheduling in cognitive radio networks," \emph{IEEE Transactions on Communications}, vol. 61, no. 12, pp. 5103-5113, December 2013.

\bibitem{IEEEhowto:44}
D. Ma and G. Tsudik, ``Security and privacy in emerging wireless networks," \emph{IEEE Wireless Communications}, vol. 17, no. 5, pp. 12-21, October 2010.

\bibitem{IEEEhowto:45}
H. Kumar, D. Sarma, and A. Kar, ``Security threats in wireless sensor networks," \emph{IEEE Aerospace and Electronic Systems
Magazine}, vol. 23, no. 6, pp. 39-45, June 2008.

\bibitem{IEEEhowto:46}
Y. Shiu, \emph{et al.}, ``Physical layer security in wireless
networks: A tutorial," \emph{IEEE Wireless Communications}, vol. 18,
no. 2, pp. 66-74, Apirl 2011.

\bibitem{IEEEhowto:47}
Y. Jiang, C. Lin, X. Shen, and M. Shi, ``Mutual authentication and
key exchange protocols for roaming services in wireless mobile
networks," \emph{IEEE Transactions on Wireless Communications}, vol.
5, no. 9, pp. 2569-2577, Septermber 2006.

\bibitem{IEEEhowto:48}
W. Stalling, \emph{Cryptography and network security: Principles and
Practices}, Third Edition, NJ: Prentice¨CHall, January 2010.

\bibitem{IEEEhowto:49}
X. Chen, K. Makki, K. Yen, and N. Pissinou, ``Sensor network
security: A survey," \emph{IEEE Communications Surveys \&
Tutorials}, vol. 11, no. 2, pp. 52-73, April 2009.

\bibitem{IEEEhowto:50}
D. Dzung, M. Naedele, T. Von Hoff, and M. Crevatin, ``Security for
industrial communications systems," \emph{Proceedings of the IEEE},
vol. 93, no. 6, pp. 1152-1177, June 2005.

\bibitem{IEEEhowto:51}
E. Shi amd A. Perrig, ``Designing secure sensor networks,"
\emph{IEEE Wireless Communications}, vol. 11, no. 6, pp. 38-43,
December 2004.

\bibitem{IEEEhowto:52}
X. Lin, ``CAT: Building couples to early detect node compromise
attack in wireless sensor networks," \emph{Proceedings of The 2009
IEEE Global Telecommunications Conference}, Honolulu, USA, December
2009.

\bibitem{IEEEhowto:53}
A. D. Wood and J. A. Stankovic, ``Denial of service in sensor
networks," \emph{IEEE Computer}, vol. 35, no. 10, pp. 54-62, October
2002.

\bibitem{IEEEhowto:54}
H. Huang, N. Ahmed, P. Karthik, ``On a new type of denial of service
attack in wireless networks: The distributed jammer network,"
\emph{IEEE Transactions on Wireless Communications}, vol. 10, no. 7,
pp. 2316-2324, July 2011.

\bibitem{IEEEhowto:55}
J. Ren amd T. Li, ``CDMA physical layer built-in security
enhancement," \emph{Proceedings of The 2003 IEEE 58th Vehicular
Technology Conference}, Orlando, USA, October 2003.

\bibitem{IEEEhowto:56}
A. Mpitziopoulos,G. Pantziou, and C. Konstantopoulos, ``Defending
wireless sensor networks from jamming attacks," \emph{Proceedings of
the IEEE 18th International Symposium on Personal, Indoor and Mobile
Radio Communications}, Athens, Greece, January 2013.

\bibitem{IEEEhowto:57}
K. Pelechrinis, C. Koufogiannakis, and S. Krishnamurthy, ``On the
efficacy of frequency hopping in coping with jamming attacks in
802.11 networks," \emph{IEEE Transactions on Wireless
Communications}, vol. 9, no. 10, pp. 3258-3271, October 2010.

\bibitem{IEEEhowto:58}
IEEE 802.15.1 Working Group, \emph{IEEE standard for
telecommunications and information exchange between systems -
LAN/MAN - specific requirements - part 15.1: Wireless medium access
control (MAC) and physical layer (PHY) specifications for wireless
personal area networks (WPANs)}, 2005.

\bibitem{IEEEhowto:59}
IEEE 802.11 Working Group, \emph{IEEE standard for information
technology - telecommunications and information exchange between
systems - local and metropolitan area networks - specific
requirements - part 11: Wireless LAN medium access control (MAC) and
physical layer (PHY) specifications}, 2007.

\bibitem{IEEEhowto:60}
IEEE 802.16m Working Group, \emph{IEEE standard for local and
metropolitan area networks part 16: Air interface for broadband
wireless access systems amendment 3: Advanced air interface}, 2011.

\bibitem{IEEEhowto:61}
A. Ghosh, \emph{et al.}, ``LTE-advanced: Next-generation wireless
broadband technology," \emph{IEEE Wireless Communications}, vol. 17,
no. 3, pp. 10-22, June 2010.

\bibitem{IEEEhowto:62}
B. Haibo, L. Sohraby, and C. Wang, ``Future internet services and
applications," \emph{IEEE Network}, vol. 24, no. 4, pp.  4-5, April
2010.

\bibitem{IEEEhowto:63}
R. Bruno and M. Conti, ``Throughput analysis and measurements in
IEEE 802.11 WLANs with TCP and UDP traffic flows," \emph{IEEE
Transactions on Mobile Computing}, vol. 7, no. 2, pp. 171-186,
Febuary 2008.

\bibitem{IEEEhowto:64}
S. Lee, G. Ahn, and A. Campbell, ``Improving UDP and TCP performance
in mobile ad hoc networks with INSIGNIA," \emph{IEEE Communications
Magazine}, vol. 39, no. 6, pp. 156-165, June 2001.

\bibitem{IEEEhowto:65}
C. Labovitz, A. Ahuja, and F. Jahanian, ``Delayed Internet routing
convergence," \emph{IEEE/ACM Transactions on Networking}, vol. 9,
no. 3, pp. 293-306, August 2008.

\bibitem{IEEEhowto:66}
R. Derryberry, \emph{et al.}, ``Transmit diversity in 3G CDMA
systems," \emph{IEEE Communications Magazine}, vol. 40, no. 4, pp.
68-75, April 2002.

\bibitem{IEEEhowto:67}
D. Bai, \emph{et al.}, ``LTE-advanced modem design: Challenges and
perspectives," \emph{IEEE Communications Magazine}, vol. 50, no. 2,
pp. 178-186, February 2012.

\bibitem{IEEEhowto:68}
S. M. Bellovin, ``Security problems in the TCP/IP protocol suite,"
\emph{ACM SIGCOMM Computer communications Review}, vol. 19, no. 2,
pp. 32-48, April 1989.

\bibitem{IEEEhowto:69}
G. Zargar and P. Kabiri, ``Identification of effective network
features to detect Smurf attacks," \emph{Proceedings of The 2009
IEEE Student Conference on Research and Development}, UPM Serdang,
November 2009.

\bibitem{IEEEhowto:70}
T. Shon and W. Choi, ``An analysis of mobile WiMAX security:
Vulnerabilities and solutions," \emph{Lecture Notes in Computer
Science}, vol. 4658, pp. 88-97,  2007.

\bibitem{IEEEhowto:71}
A. Perrig, J. Stankovic, and D. Wagner, ``Security in wireless
sensor networks," \emph{Communications of The ACM}, vol. 47, no. 6,
pp. 53-57, June 2004.

\bibitem{IEEEhowto:72}
A. Mpitziopoulos, ``A survey on jamming attacks and countermeasures
in WSNs," \emph{IEEE Communications Surveys \& Tutorials}, vol. 11,
no. 4, pp. 42-56, December 2009.

\bibitem{IEEEhowto:73}
V. Nagarajan and D. Huang, ``Using power hopping to counter MAC
spoof attacks in WLAN," \emph{Proceedings of The 2010 IEEE Consumer
Communications and Networking Conference}, Las Vegas, USA, January
2010.

\bibitem{IEEEhowto:74}
W. Zhou, A. Marshall, and Q. Gu, ``A novel classification scheme for
802.11 WLAN active attacking traffic patterns," \emph{Proceedings of
The 2006 IEEE Wireless Communications and Networking Conference},
Las Vegas, USA, April 2006.

\bibitem{IEEEhowto:75}
J. Park and S. Kasera, ``Securing Ad Hoc wireless networks against
data injection attacks using firewalls," \emph{Proceedings of The
2007 IEEE Wireless Communications and Networking Conference},
Honkong, China, April 2007.

\bibitem{IEEEhowto:76}
Computer Emergency Response Team (CERT), ``CERT advisory: IP
spoofing attacks and hijacked terminal connections," January 1995,
available on-line at http://www.cert.org/advisories/CA-1995-01.html.

\bibitem{IEEEhowto:77}
N. Hastings and P. McLean, ``TCP/IP spoofing fundamentals,"
\emph{Proceedings of The 1996 IEEE Fifteenth Annual International
Phoenix Conference on Computers and Communications}, Arizona, USA,
March 1996.

\bibitem{IEEEhowto:78}
B. Harrisa and R. Hunt, ``TCP/IP security threats and attack
methods," \emph{Computer Communications}, vol. 22, no. 10, pp.
885-897, June 1999.

\bibitem{IEEEhowto:79}
F. El-Moussa, N. Linge, and M. Hope, ``Active router approach to
defeating denial-of-service attacks in networks," \emph{IET
Communications}, vol. 1, no. 1, pp. 55-63, February 2007.

\bibitem{IEEEhowto:80}
C. Schuba, \emph{et al.}, ``Analysis of a denial of service attack
on TCP," \emph{Proceedings of The 1997 IEEE Symposium on Security
and Privacy}, Oakland, USA, May 1997.

\bibitem{IEEEhowto:81}
A. Kuzmanovic and E. W. Knightly, ``Low-rate TCP-targeted denial of
service attacks and counter strategies," \emph{IEEE/ACM Transactions
on Networking}, vol. 14, no. 4, pp. 683-696, August 2006.

\bibitem{IEEEhowto:82}
R. Chang, ``Defending against flooding-based distributed
denial-of-service attacks: A tutorial," \emph{IEEE Communications
Magazine}, vol. 40, no. 10, pp. 42-51, October 2002.

\bibitem{IEEEhowto:83}
RFC 2577, ``FTP security considerations," May 1999, available
on-line at http://tools.ietf.org/html/rfc2577.

\bibitem{IEEEhowto:84}
T. Bass, A. Freyre, D. Gruber, and G. Watt, ``E-mail bombs and
countermeasures: Cyber attacks on availability and brand integrity,"
\emph{IEEE Network}, vol. 12, no. 2, pp. 10-17, March 1998.

\bibitem{IEEEhowto:85}
A. Kieyzun, P. Guo, K. Jayaraman,and M. Ernst, ``Automatic creation
of SQL injection and cross-site scripting attacks,"
\emph{Proceedings of The IEEE 31st International Conference on
Software Engineering}, Vancouver, Canada, May 2009.

\bibitem{IEEEhowto:86}
C. Stevenson, \emph{et al.}, ``IEEE 802.22: The first cognitive
radio wireless regional area network standard," \emph{IEEE
Communications Magazine}, vol. 47, no. 1, pp. 130-138, January 2009.

\bibitem{IEEEhowto:87}
E. Ferro and F. Potorti, ``Bluetooth and Wi-Fi wireless protocols: A
survey and a comparison," \emph{IEEE Wireless Communications}, vol.
12, no. 1, pp. 12-26, February 2005.

\bibitem{IEEEhowto:88}
M. Polla, F. Martinelli, and D. Sgandurra, ``A Survey on security
for mobile devices," \emph{IEEE Communications Surveys \&
Tutorials}, vol. 15, no. 1, pp. 446-471, March 2013.

\bibitem{IEEEhowto:89}
D. Pareit, I. Moerman, and P. Demeester, ``The history of WiMAX: A
complete survey of the evolution in certification and
standardization for IEEE 802.16 and WiMAX," \emph{IEEE
Communications Surveys \& Tutorials}, vol. 14, no. 4, pp. 1183-1211,
October 2012.

\bibitem{IEEEhowto:90}
J. Cao, H. Ma, H. Li, and Y. Zhang, ``A survey on security aspects
for LTE and LTE-A networks," \emph{IEEE Communications Surveys \&
Tutorials}, vol. 15, no. 2, April 2013.

\bibitem{IEEEhowto:91}
T. Muller, ``Bluetooth security architecture," July 1999, available
on-line at http://www.afn.org/~afn48922/downs/wireless/1c11600.pdf.

\bibitem{IEEEhowto:92}
M, Kui and X. Cuo, ``Research of Bluetooth security manager,"
\emph{Proceedings of The 2003 IEEE International Conference on
Neural Networks \& Signal Processing}, Nanjing, China, December,
2003.

\bibitem{IEEEhowto:93}
P. Toengel, ``Bluetooth: Authentication, authorisation and
encryption," July 1999, available on-line at
http://www.toengel.net/studium/mm\_and\_sec/bluetooth.pdf

\bibitem{IEEEhowto:94}
A. Lashkari, M. Danesh,and B. Samadi, ``A survey on wireless
security protocols (WEP, WPA and WPA2/802.11i)," \emph{Proceedings
of The 2nd IEEE International Conference on Computer Science and
Information Technology}, Beijing, China, August 2009.

\bibitem{IEEEhowto:95}
A. Lashkari,M. Mansoor, and A. Danesh, ``Wired equivalent privacy
(WEP) versus Wi-Fi protected access (WPA)," \emph{Proceedings of The
2009 International Conference on Signal Processing Systems},
Singapore, May 2009.

\bibitem{IEEEhowto:96}
J. Lee and C. Fan, ``Efficient low-latency RC4 architecture designs
for IEEE 802.11i WEP/TKIP," \emph{Proceedings of The 2007
International Symposium on Intelligent Signal Processing and
Communications Systems}, Xiamen, China, November 2007.

\bibitem{IEEEhowto:97}
{{A. Stubbleleld, J. Ioannidis, and A. D. Rubin, ``A key recovery attack on the 802.11b wired equivalent privacy protocol (WEP)," \emph{ACM Transactions on Information and System Security}, vol. 7, no. 2, pp.319-332, May 2004.}}

\bibitem{IEEEhowto:98}
{{E. Tews, R.-P. Weinmann, and A. Pyshkin, ``Breaking 104 bit wep in less than 60 seconds," \emph{Lecture Notes in Computer Science}, vol. 4867, pp. 188-202, 2007.}}

\bibitem{IEEEhowto:97}
J. Lin, Y. Kao, and C. Yang, ``Secure enhanced wireless transfer
protocol," \emph{Proceedings of The First International Conference
on Availability, Reliability and Security}, Vienna, Austria, April
2006.

\bibitem{IEEEhowto:98}
C. Nancy, H. Russ, W. David, and W. Jesse, ``Security flaws in
802.11 data link protocols," \emph{Communications of The ACM}, vol.
46, no. 5, pp. 35-39, May 2003.

\bibitem{IEEEhowto:101}
{{E. Tews and M. Beck, ``Practical attacks against WEP and WPA," in \emph{Proceedings of the Second ACM Conference on Wireless Network Security (ACM WiSec)}, Zurich, Switzerland, March 2009.}}

\bibitem{IEEEhowto:99}
Q. Li, X. Lin, J. Zhang, and W. Roh, ``Advancement of MIMO
technology in WiMAX: From IEEE 802.16d/e/j to 802.16m," \emph{IEEE
Communications Magazine}, vol. 47, no. 6, pp. 100-107, June 2009.

\bibitem{IEEEhowto:100}
T. Han, \emph{et al.}, ``Analysis of mobile WiMAX security:
Vulnerabilities and solutions," \emph{Proceedings of The Fifth IEEE
International Conference on Mobile Ad Hoc and Sensor Systems},
Atlanta, USA, September 2008.

\bibitem{IEEEhowto:101}
F. Yang, ``Comparative analysis on TEK exchange between PKMv1 and
PKMV2 for WiMAX," \emph{Proceedings of The Seventh International
Conference on Wireless Communications, Networking and Mobile
Computing}, Wuhan, China, September 2011.

\bibitem{IEEEhowto:102}
D. Johnston and J. Walker, ``Overview of IEEE 802.16 security,"
\emph{IEEE Security \& Privacy}, vol. 2, no. 3, pp. 40-48, June
2004.

\bibitem{IEEEhowto:103}
S. Adibi, \emph{et al.}, ``Authentication authorization and
accounting (AAA) schemes in WiMAX," \emph{Proceedings of The 2006
IEEE International Conference on Electro\/information Technology},
Michigan, USA, May 2006.

\bibitem{IEEEhowto:104}
E. Biham, ``A fast new DES implementation in software,"
\emph{Proceedings of The Fourth International Workshop on Fast
software Encryption}, Haifa, Israel, January 1997.

\bibitem{IEEEhowto:105}
S. Ahson and M. Ilyas, \emph{WiMAX standards and security}, Florida:
CRC Press, 2007.

\bibitem{IEEEhowto:106}
D. Astely, \emph{et al.}, ``LTE: The evolution of mobile broadband,"
\emph{IEEE Communications Magazine}, vol.47, no.4, pp.44-51, April
2009.

\bibitem{IEEEhowto:107}
The 3rd Generation Partnership Project (3GPP), \emph{Technical
specification group services and system aspects: Service
requirements for home node B (HNB) and home eNode B (HeNB) (Rel
11)}, September 2012.

\bibitem{IEEEhowto:108}
M. Zhang and Y. Fang, ``Security analysis and enhancements of 3GPP
authentication and key agreement protocol," \emph{IEEE Transactions
on Wireless Communications}, vol. 4, no. 2, pp. 734-742, March 2005.

\bibitem{IEEEhowto:109}
G.M. Koien, ``Mutual entity authentication for LTE,"
\emph{Proceedings of The Seventh International Wireless
Communications and Mobile Computing Conference}, Istanbul, Turkey,
July 2011.

\bibitem{IEEEhowto:110}
The 3rd Generation Partnership Project (3GPP), \emph{Technical
specification group services and system aspects: 3G security,
specification of the 3GPP confidentiality and integrity algorithms,
Document 2: KASUMI specification}, 2001.

\bibitem{IEEEhowto:111}
A. Kircanski and A. Youssef, ``On the sliding property of SNOW 3G and SNOW 2.0," \emph{IET Information Security}, vol. 5, no. 4, pp.
199-206, December 2011.

\bibitem{IEEEhowto:112}
B. Sklar, ``Rayleigh fading channels in mobile digital communication systems - part I: Characterization," \emph{IEEE Communications Magazine}, vol. 35, no. 7, pp. 90-100, July 1997.

\bibitem{IEEEhowto:113}
A. Abdi, C. Tepedelenlioglu, M. Kaveh, and G. Giannakis, ``On the estimation of the K parameter for the Rice fading distribution", \emph{IEEE Communications Letters}, vol. 5, no. 3, pp. 92-94, March 2001.

\bibitem{IEEEhowto:114}
M. D. Yacoub, J. E. V. Bautista, and L. Guerra de Rezende Guedes, ``On higher order statistics of the Nakagami-m distribution", \emph{IEEE Transactions on Vehicular Technology}, vol. 48, no. 3, pp. 790-794, May 1999.

\bibitem{IEEEhowto:115}
{{S. Goel and R. Negi, ``Secret communication in presence of colluding eavesdroppers," in \emph{Proceedings of IEEE Military Communications Conference}, Atlantic, NJ, 2005.}}

\bibitem{IEEEhowto:116}
C. E. Shannon, ``Communications theory of secrecy systems,"
\emph{Bell System Technical Journal}, vol. 28, pp. 656-715, 1949.


\bibitem{IEEEhowto:118}
A. Khisti, G. Womell, A. Wiesel, and Y. Eldar, ``On the Gaussian MIMO wiretap channel," \emph{Proceedings of The 2007 IEEE International Symposium on Information Theory}, Nice, France, June 2007.

\bibitem{IEEEhowto:119}
A. Khisti and G. W. Wornell, ``Secure transmission with multiple antennas: The MISOME wiretap channel," \emph{IEEE Transactions on Information Theory}, vol. 56, no. 7, pp. 3088-3104, July 2010.

\bibitem{IEEEhowto:120}
A. Khisti and G. W. Wornell, ``Secure transmission with multiple antennas - part II: The MIMOME wiretap channel," \emph{IEEE Transactions on Information Theory}, vol. 56, no. 11, pp. 5515-5532, November 2010.

\bibitem{IEEEhowto:120}
T. Chrysikos, T. Dagiuklas, and S. Kotsopoulos, ``A closed-form expression for outage secrecy capacity in wireless information-theoretic security," \emph{Proceedings of The First International ICST Workshop on Security in Emerging Wireless Communication and Networking Systems (SEWCN 2009)}, Athens, Greece, September 2009.

\bibitem{IEEEhowto:121}
F. Oggier and B. Hassibi, ``The secrecy capacity of the MIMO wiretap channel," \emph{IEEE Transactions on Information Theory}, vol. 57, no. 8, pp. 4961-4972, August 2011.

\bibitem{IEEEhowto:122}
X. He, A. Khisti, and A. Yener, ``MIMO broadcast channel with arbitrarily varying eavesdropper channel: Secrecy degrees of freedom," in \emph{Proceedings of The 2011 IEEE Global Telecommunications Conference}, Houston, USA, December 2011.

\bibitem{IEEEhowto:123}
S. Goel and R. Negi, ``Guaranteeing secrecy using artificial noise,"
\emph{IEEE Transactions on Wireless Communications}, vol. 7, no. 6,
pp. 2180-2189, July 2008.

\bibitem{IEEEhowto:124}
X. Zhou and M. McKay, ``Secure transmission with artificial noise
over fading channels: Achievable rate and optimal power allocation,"
\emph{IEEE Transactions on Vehicular Technology}, vol. 59, no. 8,
pp. 3831-3842, August 2010.

\bibitem{IEEEhowto:125}
W. Liao, T. Chang, W. Ma, and C. Chi, ``QoS-based transmit
beamforming in the presence of eavesdroppers: An optimized
artificial-noise-aided approach," \emph{IEEE Transactions on
Vehicular Technology}, vol. 59, no. 3, pp. 1202-1216, March 2011.

\bibitem{IEEEhowto:126}
Q. Li and W. Ma, ``A robust artificial noise aided transmit design
for MISO secrecy," \emph{Proceedings of The 2011 IEEE International
Conference on Acoustics, Speech and Signal Processing}, Prague,
Czech Republic, May 2011.

\bibitem{IEEEhowto:127}
D. Goeckel, \emph{et al.}, ``Artificial noise generation from
cooperative relays for everlasting secrecy in two-hop wireless
networks," \emph{IEEE Journal on Selected Areas in Communications},
vol. 29, no. 10, pp. 2067-2076, October 2011.

\bibitem{IEEEhowto:128}
J. Zhang and M. Gursoy, ``Collaborative relay beamforming for
secrecy," \emph{Proceedings of The 2010 IEEE International
Conference on Communications}, Cape Town, South Africa, May 2010.

\bibitem{IEEEhowto:129}
A. Mukherjee amd A. Swindlehurst, ``Robust beamforming for security
in MIMO wiretap channels with imperfect CSI," \emph{IEEE
Transactions on Signal Processing}, vol. 59, no. 1, pp. 351-361,
January 2011.

\bibitem{IEEEhowto:130}
C. Jeong, I. Kim, and K. Dong, ``Joint secure beamforming design at
the source and the relay for an amplify-and-forward MIMO untrusted
relay system," \emph{IEEE Transactions on Signal Processing}, vol.
60, no. 1, pp. 310-325, January 2012.

\bibitem{IEEEhowto:131}
{{N. Anand, S.-J. Lee, and E. W. Knightly, ``Strobe: Actively securing wireless communications using zero-forcing beamforming," in \emph{Proceedings of  31st IEEE INFOCOM}, Orlando, FL, March 2012.}}

\bibitem{IEEEhowto:131}
H. Qin, \emph{et al.}, ``Optimal power allocation for joint beamforming and artificial noise design in secure wireless
communications," \emph{Proceedings of The 2011 IEEE International Conference on Communications Workshops}, Kyoto, Japan, June 2011.

\bibitem{IEEEhowto:132}
N. Romero-Zurita, M. Ghogho, and D. McLernon, ``Outage probability based power distribution between data and artificial noise for physical layer security," \emph{IEEE Signal Processing Letters}, vol. 19, no. 2, pp. 71-74, February 2012.

\bibitem{IEEEhowto:133}
Y. Zou, J. Zhu, X. Wang, and V. Leung, ``Improving diversity for physical-layer security in wireless communications," \emph{IEEE Network}, vol. 29, no. 1, pp. 42-48, Jan. 2015.

\bibitem{IEEEhowto:134}
{{L. Zheng and D. Tse, ``Diversity and multiplexing: A fundamental tradeoff in multiple-antenna channels", \emph{IEEE Transactions on Information Theory}, vol. 49, no. 5, pp. 1073-1096, May 2003.}}

\bibitem{IEEEhowto:135}
{{S. Alamouti, ``A simple transmitter diversity scheme for wireless communications", \emph{IEEE Journal on Selected Areas in Communications}, vol. 16, no. 8, pp. 1451-1458, October 1998.}}

\bibitem{IEEEhowto:136}
{{S. Srikanth, P. A. Murugesa, and X. Fernando, ``Orthogonal frequency division multiple access in WiMAX and LTE: A comparison", \emph{IEEE Communications Magazine}, vol. 50, no. 9, pp. 153-161, September 2012.}}

\bibitem{IEEEhowto:137}
{{J. Laiho, K. Raivio, P. Lehtimaki, K. Hatonen, and O. Simula, ``Advanced analysis methods for 3G cellular networks", \emph{IEEE Transactions on Wireless Communications}, vol. 4, no. 3, pp. 930-942 , June 2005.}}

\bibitem{IEEEhowto:138}
{{Y. Zou, X. Wang, and W. Shen, ``Physical-layer security with multiuser scheduling in cognitive radio networks", \emph{IEEE Transactions on Communications}, vol. 61, no. 12, pp. 5103 - 5113, December 2013.}}


\bibitem{IEEEhowto:140}
{{A. Sendonaris, E. Erkip, and B. Aazhang, ``User cooperation diversity - Part I: System description," \emph{IEEE Transactions Communications}, vol. 51, no. 11, pp. 1927-1938, November 2003.}}

\bibitem{IEEEhowto:141}
{{Y. Zou, Y.-D. Yao, and B. Zheng, ``Opportunistic distributed space-time coding for decode-and-forward cooperation systems," \emph{IEEE Transactions on Signal Processing}, vol. 60, no. 4, pp. 1766-1781, April 2012.}}

\bibitem{IEEEhowto:142}
{{Y. Zou, X. Li, and Y.-C. Liang, ``Secrecy outage and diversity analysis of cognitive radio systems," \emph{IEEE Journal on Selected Areas in Communications}, vol. 32, no. 11, pp. 2222-2236, November 2014.}}

\bibitem{IEEEhowto:143}
{{K. Ren, H. Su, and Q. Wang, ``Secret key generation exploiting channel characteristics in wireless communications," \emph{IEEE Wireless Communications}, vol. 18, no. 4, pp. 6-12, August 2011.}}

\bibitem{IEEEhowto:144}
{{Y. Li, L. J. Cimini, and N. R. Sollenberger, ``Robust channel estimation for OFDM systems with rapid dispersive fading channels," \emph{IEEE Transactions on Communications}, vol. 46, no. 7, pp. 902-915, August 2002.}}

\bibitem{IEEEhowto:145}
{{B. Muquet, M. Courville, and P. Duhamel, ``Subspace-based blind and semi-blind channel estimation for OFDM systems," \emph{IEEE Transactions on Signal Processing}, vol. 50, no. 7, pp. 1699-1712 , July 2002.}}

\bibitem{IEEEhowto:146}
{{M. Steiner, G. Tsudik, and M. Waidner, ``Diffie-Hellman key distribution extended to group communication," in \emph{Proceedings of The 3rd ACM Conference on Computer and Communications Security}, New Delhi, India, March 1996.}}

\bibitem{IEEEhowto:147}
{{J. E. Hershey, A. A. Hassan, and R. Yarlagadda, ``Unconventional cryptographic keying variable management," \emph{IEEE Transactions on Communications}, vol. 43, no. 1, pp. 3-6, January 1995.}}

\bibitem{IEEEhowto:148}
{{A. A. Hassan, W. E. Stark, J. E. Hershey, and S. Chennakeshu, ``Cryptographic key agreement for mobile radio," \emph{Digital Signal Processing}, vol. 6, no. 10, pp. 207-212, October 1996.}}

\bibitem{IEEEhowto:149}
{{S. Mathur, W. Trappe, N. Mandayam, C. Ye, and A. Reznik, ``Radio-telepathy: Extracting a secret key from an unauthenticated wireless channel," in \emph{Proceedings of The 14th Annual International Conference on Mobile Computing and Networking (MobiCom 2008)}, California, USA, September 2008.}}

\bibitem{IEEEhowto:150}
S. Jana, \emph{et al.}, ``On the effectiveness of secret key extraction from wireless signal strength in real environments," in \emph{Proceedings of The 15th Annual International Conference on Mobile Computing and Networking (MobiCom 2009)}, Beijing, China, September 2009.

\bibitem{IEEEhowto:152}
{{S. Gollakota and D. Katabi, ``Physical layer wireless security made fast and channel independent," in \emph{Proceedings of The 30th Annual IEEE International Conference on Computer Communications (INFOCOM 2011)}, Shanghai, China, April 2011.}}

\bibitem{IEEEhowto:151}
{{H. Liu, J. Yang, Y. Wang, and Y. Chen, ``Collaborative secret key extraction leveraging received signal strength in mobile wireless networks," in \emph{Proceedings of The 31st Annual IEEE International Conference on Computer Communications (INFOCOM 2012)}, Florida, USA, March 2012.}}

\bibitem{IEEEhowto:152}
{{Y. Shehadeh, O. Alfandi, K. Tout, and D. Hogrefe, ``Intelligent mechanisms for key generation from multipath wireless channels," in \emph{Proceedings of The 2011 Wireless Telecommunications Symposium (WTS)}, NY, USA, April 2011.}}

\bibitem{IEEEhowto:153}
{{Q. Wang, H. Su, K. Ren, and K. Kim, ``Fast and scalable secret key generation exploiting channel phase randomness in wireless networks," in \emph{Proceedings of The 30th Annual IEEE International Conference on Computer Communications (INFOCOM 2011)}, Shanghai, China, March 2011.}}

\bibitem{IEEEhowto:154}
J. W. Wallace, C. Chen, and M. A. Jensen, ``Key generation exploiting MIMO channel evolution: Algorithm and theoretical limits," in \emph{Proceedings of The 3rd European Conference on Antennas and Propagation}, Berlin, Germany, March 2009.

\bibitem{IEEEhowto:155}
K. Zeng, D. Wu, A. Chan, and P. Mohapatra, ``Exploiting multiple-antenna diversity for shared secret key generation in wireless networks," in \emph{Proceedings of The 29th Annual IEEE International Conference on Computer Communications (INFOCOM 2010)}, California, USA, March 2010.

\bibitem{IEEEhowto:156}
T. Shimizu, H. Iwai, and H. Sasaoka, ``Physical-layer secret key agreement in two-way wireless relaying systems," \emph{IEEE Transactions on Information Forensics and Security}, vol. 6, no. 3, pp. 650-660, September 2011.

\bibitem{IEEEhowto:156}
{{M. Edman, A. Kiayias, and B. Yener, ``On passive inference attacks against physical-layer key extraction?," in \emph{Proceedings of the Fourth European Workshop on System Security}, Salzburg, Austria, April 2011.}}

\bibitem{IEEEhowto:157}
{{S. Eberz, M. Strohmeier, M. Wilhelm, and I. Martinovic, ``A practical man-in-the-middle attack on signal-based key generation protocols," \emph{Computer Security}, pp. 235-252, 2012.}}

\bibitem{IEEEhowto:158}
{{L. Shi, \emph{et al.}, ``ASK-BAN: Authenticated secret key extraction utilizing channel characteristics for body area networks," in \emph{Proceedings of the sixth ACM conference on Security and Pprivacy in Wireless and Mobile Networks (ACM WiSec'13)}, Budapest, Hungary, Apirl 2013.}}


\bibitem{IEEEhowto:158}
K. Pelechrinis, M. Iliofotou, and S. V., Krishnamurthy, ``Denial of service attacks in wireless networks: The case of jammers," \emph{IEEE Communications Surveys \& Tutorials}, vol. 13, no. 2, pp. 245-257, May 2011.

\bibitem{IEEEhowto:159}
T. X. Brown, J. E. James, and A.Sethi, ``Jamming and sensing of encrypted wireless ad hoc networks," in \emph{Proceedings of The 7th ACM International Symposium on Mobile Ad Hoc Networking and Computing (MobiHoc 2006)}, Florence, Italy, May 2006.


\bibitem{IEEEhowto:161}
{{W. Xu, W. Trappe, Y. Zhang, and T. Wood, ``The feasibility of launching and detecting jamming attacks in wireless networks," in \emph{Proceedings of The 6th ACM International Symposium on Mobile Ad Hoc Networking and Computing (MobiHoc 2005)}, Illinois, USA, May 2005.}}

\bibitem{IEEEhowto:162}
{{D. J. Torrieri, ``Frequency hopping with multiple frequency-shift keying and hard decisions," \emph{IEEE Transactions on Communications}, vol. 32, no. 5, pp. 574-582, May 1984.}}

\bibitem{IEEEhowto:163}
{{V. Navda, A. Bohra, S. Ganguly, and D. Rubenstein, ``Using channel hopping to increase 802.11 resilience to jamming attacks," in \emph{Proceedings of The 26th Annual IEEE International Conference on Computer Communications (INFOCOM 2007)}, Alaska, USA, May 2007.}}

\bibitem{IEEEhowto:164}
{{R. Gummadi, D. Wetheral, B. Greenstein, and S. Seshan, ``Understanding and mitigating the impact of RF interference on 802.11 networks," in \emph{Proceedings of ACM SIGCOMM 2007}, Kyoto, Japan, August 2007.}}

\bibitem{IEEEhowto:165}
{{O. Besson, P. Stoica, and Y. Kamiya, ``Direction finding in the presence of an intermittent interference," \emph{IEEE Transactions on Signal Processing}, vol. 50, no. 7, pp. 1554-1564, July 2002.}}

\bibitem{IEEEhowto:166}
{{Y. Liu and P. Ning, ``BitTrickle: Defending against broadband and high-power reactive jamming attacks," in \emph{Proceedings of The 31st Annual IEEE International Conference on Computer Communications (INFOCOM 2012)}, Orlando, USA, March 2012.}}

\bibitem{IEEEhowto:167}
{{E. Lance and G. K. Kaleh, ``A diversity scheme for a phase-coherent frequency-hopping spread-spectrum system," \emph{IEEE Transactions on Communications}, vol. 45, no. 9, pp. 1123-1129, September 1997.}}

\bibitem{IEEEhowto:168}
{{L. Freitag, M. Stojanovic, S. Singh, and M. Johnson, ``Analysis of channel effects on direct-sequence and frequency-hopped spread-spectrum acoustic communication," \emph{IEEE Journal of Oceanic Engineering}, vol. 26, no. 4, pp. 586-593, October 2001.}}

\bibitem{IEEEhowto:169}
{{J. Jeung, S. Jeong, and J. Lim, ``Adaptive rapid channel-hopping scheme mitigating smart jammer attacks in secure WLAN," in \emph{Proceedings of The 2011 Military Communications Conference}, Baltimore, USA, November 2011.}}

\bibitem{IEEEhowto:170}
{{Z. Liu, H. Liu, W. Xu, and Y. Chen, ``An error-minimizing framework for localizing jammers in wireless networks," \emph{IEEE Transactions on Parallel and Distributed Systems}, vol. 25, no. 2, pp. 508-517, Febuary 2014.}}

\bibitem{IEEEhowto:171}
{{B. P. Crow, I. Widjaja, J. G. Kim, and P. T. Sakai, ``IEEE 802.11 wireless local area networks," \emph{IEEE Communications Magazine}, vol. 35, no. 9, pp. 116-126, September 1997.}}

\bibitem{IEEEhowto:172}
{{G. Bianchi, ``Performance analysis of the IEEE 802.11 distributed coordination function," \emph{IEEE Journal on Selected Areas in Communications}, vol. 18, no. 3, pp. 535-547, March 2003.}}

\bibitem{IEEEhowto:178}
X. Liu, G. Noubir, R. Sundaram, and S. Tan, ``SPREAD: Foiling smart jammers using multi-layer agility," in \emph{Proceedings of The 26th Annual IEEE International Conference on Computer Communications (INFOCOM 2007)}, Alaska, USA, March 2007.


\bibitem{IEEEhowto:178}
{{Y. Abdallah, M. A. Latif, M. Youssef, A. Sultan, and H. E. Gamal, ``Keys through ARQ: Theory and practice," \emph{IEEE Transactions on Information Forensics and Security}, vol. 6, no. 3, pp. 737-751, Sept. 2011.}}

\bibitem{IEEEhowto:179}
{{D. Loh, C. Cho, C. Tan, and R. Lee, ``Identifying unique devices through wireless fingerprinting," in \emph{Proceedings of the 1st ACM Conference on Wireless Network Security (WiSec'08)}, Alexandria, USA, March 2008.}}

\bibitem{IEEEhowto:181}
{{V. Brik, S. Banerjee, M. Gruteser, and S. Oh, ``Wireless device indentification with radiometric signatures," in \emph{Proceedings of the 14th ACM International Conference on Mobile Computing and Networking (ACM MobiCom'08)}, New York, USA, September 2008.}}


\bibitem{IEEEhowto:182}
{{W. Hou, X. Wang, J.-Y. Chouinard, and A. Refaey, ``Physical layer authentication for mobile systems with time-varying carrier frequency offsets," \emph{IEEE Transactions on Communications}, vol. 62, no. 5, pp. 1658-1667, May 2014.}}

\bibitem{IEEEhowto:180}
{{L. Xiao, L. Greenstein, N. Mandayam, and W. Trappe, ``Using the physical layer for wireless authentication in time-variant channels," \emph{IEEE Transations on Wireless Communications}, vol. 7, no. 7, pp. 2571-2579, July 2008.}}


\bibitem{IEEEhowto:183}
{{Y. Liu and P. Ning, ``Enhanced wireless channel authentication using time-synched link signature," in \emph{Proceedings of 2012 IEEE International Conference on Computer Communications (IEEE INFOCOM'12)}, Shanghai, China, March 2012.}}

\bibitem{IEEEhowto:184}
{{J. Xiong and K. Jamieson, ``SecureArray: Improving WiFi security with fine-grained physical-layer information," in \emph{Proceedings of the 19th ACM International Conference on Mobile Computing and Networking (ACM MobiCom'13)}, Miami, USA, Septermber 2013.}}

\bibitem{IEEEhowto:185}
{{X. Du, D. Shan, K. Zeng and L. Huie, ``Physical layer challenge-response authentication in wireless networks with relay," in \emph{Proceedings of 2014 IEEE International Conference on Computer Communications (IEEE INFOCOM'14)}, March 2014.}}

\bibitem{IEEEhowto:186}
{{P. Yu, J. Baras, and B. Sadler, ``Physical-layer authentication," \emph{IEEE Transactions on Information Forensics and Security}, vol. 3, no. 1, pp. 38-51, March 2008.}}

\bibitem{IEEEhowto:187}
{{P. Yu and B. Sadler ``MIMO authentication via deliberate fingerprinting at the physical layer," \emph{IEEE Transactions on Information Forensics and Security}, vol. 6, no. 3, pp. 606-615, March 2011.}}

\bibitem{IEEEhowto:188}
{{P. Yu, G. Verma, and B. Sadler, ``Wireless physical layer authentication via fingerprint embedding," \emph{IEEE Communications Magazine}, vol. 53, no. 6, pp. 48-53, June 2015.}}

\bibitem{IEEEhowto:190}
{{A. Menezes, P. Oorschot, and S. Vanstone, \emph{Handbook of applied cryptography}, CRC Press, 2001.}}


\bibitem{IEEEhowto:191}
{{M. Latif, A. Sultan, and H. Gamal, ``ARQ-based secret key sharing," in \emph{Proceedings of the 2009 IEEE Internation Conference on Communications (IEEE ICC'09)}, Dresden, Germany, June 2009.}}

\bibitem{IEEEhowto:192}
{{S. Xiao, W. Gong, and D. Towsley, ``Secure wireless communication with dynamic secrets," in \emph{Proceedings of the IEEE INFOCOM 2010}, San Diego, CA, March 2010.}}

\bibitem{IEEEhowto:193}
{{Y. Khiabani and S. Wei, ``ARQ-based symmetrickey generation over correlated erasure channels," \emph{IEEE Transactions on Information Forensics and Security}, vol. 8, no. 7, pp. 1152-1161, July 2013.}}

\bibitem{IEEEhowto:194}
C. B. Sankaran, ``Network access security in next generation 3GPP systems: A tutorial," \emph{IEEE Communications Magazine}, vol. 47, no. 2, pp. 84-91, February 2009.

\bibitem{IEEEhowto:195}
Y. Zou, X. Wang, W. Shen, and L. Hanzo, ``Security versus reliability analysis for opportunistic relaying," \emph{IEEE Transactions on Vehicular Technology}, vol. 63, no. 6, pp. 2653-2661, July 2014.

\bibitem{IEEEhowto:196}
N. Yang, \emph{et al.}, ``Safeguarding 5G wireless communication networks using physical layer security," \emph{IEEE Communications Magazine}, vol. 53, no. 4, pp. 20-27, April 2015.

\bibitem{IEEEhowto:197}
W. Roh \emph{et al.}, ``Millimeter-wave beamforming as an enabling technology for 5G cellular communications: Theoretical feasibility and prototype results," \emph{IEEE Communications Magazine}, vol. 52, no. 2, pp. 106-113, Feb. 2014.

\bibitem{IEEEhowto:198}
J. Andrews, \emph{et al.},``What will 5G be?," \emph{IEEE Journal on Selected Areas in Communications}, vol. 32, no. 6, pp. 1065-1082, June 2014.

\bibitem{IEEEhowto:199}
C.-X. Wang, \emph{et al.}, ``Cellular architecture and key technologies for 5G wireless communication networks," \emph{IEEE Communications Magazine}, vol. 52, no. 2, pp. 122-130, Febuary 2014.

\bibitem{IEEEhowto:200}
P. Rost, \emph{et al.},``Cloud technologies for flexible 5G radio access networks," \emph{IEEE Communications Magazine}, vol. 52, no. 5, pp. 68-76, May 2014.

\bibitem{IEEEhowto:201}
Y. Zou, J. Zhu, L. Yang, Y.-.C. Liang, and Y.-D. Yao, ``Securing physical-layer communications for cognitive radio networks," \emph{IEEE Communications Magazine}, vol. 53, no. 9, pp. 48-54, September 2015.

\bibitem{IEEEhowto:202}
Y. Zou, J. Zhu, X. Li, and L. Hanzo, ``Relay selection for wireless communications against eavesdropping: A security-reliability tradeoff perspective," \emph{IEEE Network}, accepted to appear.

\end{thebibliography}
\end{document}